\documentclass[sn-basic,Numbered]{sn-jnl}

\usepackage{graphicx}%
\usepackage{multirow}%
\usepackage{amsmath,amssymb,amsfonts}%
\usepackage{amsthm}%
\usepackage{mathrsfs}%
\usepackage[title]{appendix}%
\usepackage{xcolor}%
\usepackage{textcomp}%
\usepackage{manyfoot}%
\usepackage{booktabs}%
\usepackage{algorithm}%
\usepackage{algorithmicx}%
\usepackage{algpseudocode}%
\usepackage{listings}%
\usepackage{subcaption}
\usepackage{natbib}
\usepackage{float}
\usepackage{glossaries}
\usepackage{longtable}
\usepackage{supertabular}
\usepackage{chngcntr}
\usepackage{soul}
\usepackage[normalem]{ulem}

\theoremstyle{thmstyleone}%
\theoremstyle{thmstyletwo}%

\theoremstyle{thmstylethree}%

\begin{document}

\title[]{Efficient Inverse Design Optimization through Multi-fidelity Simulations, Machine Learning, and Boundary Refinement Strategies}


\author*[1]{\fnm{Luka} \sur{Grbcic}}\email{lgrbcic@lbl.gov}

\author[2]{\fnm{Juliane} \sur{M\"uller}}\email{juliane.mueller@nrel.gov }

\author[1]{\fnm{Wibe Albert} \sur{de Jong}}\email{wadejong@lbl.gov}

\affil*[1]{\orgdiv{Applied Mathematics and Computational Research Division}, \orgname{Lawrence Berkeley National Laboratory}, \orgaddress{\street{1 Cyclotron Rd}, \city{Berkeley}, \postcode{94720}, \state{California}, \country{USA}}}

\affil[2]{\orgdiv{Computational Science Center}, \orgname{National Renewable Energy Laboratory}, \orgaddress{\street{15013 Denver West Parkway}, \city{Golden}, \postcode{80401}, \state{Colorado}, \country{USA}}}


\abstract{This paper introduces a methodology designed to augment the inverse design optimization process in scenarios constrained by limited compute, through the strategic synergy of multi-fidelity evaluations, machine learning models, and optimization algorithms. The proposed methodology is analyzed on two distinct engineering inverse design problems: airfoil inverse design and the scalar field reconstruction problem. It leverages a machine learning model trained with low-fidelity simulation data, in each optimization cycle, thereby proficiently predicting a target variable and discerning whether a high-fidelity simulation is necessitated, which notably conserves computational resources. Additionally, the machine learning model is strategically deployed prior to optimization to compress the design space boundaries, thereby further accelerating convergence toward the optimal solution. The methodology has been employed to enhance two optimization algorithms, namely Differential Evolution and Particle Swarm Optimization. Comparative analyses illustrate performance improvements across both algorithms. Notably, this method is adaptable across any inverse design application, facilitating a synergy between a representative low-fidelity ML model, and high-fidelity simulation, and can be seamlessly applied across any variety of population-based optimization algorithms.}


\maketitle

\section*{Article Highlights}
\begin{itemize}
	\item Low-fidelity machine learning (ML) models streamline optimization, speeding up processes and cutting computational costs.
	\item Low-fidelity ML models narrow the optimization boundaries, quickening identification of optimal solutions.
	\item A combined ML and optimization framework, shows wide applicability in solving various complex engineering problems.
\end{itemize}

\section{Introduction}\label{sec1}

\label{sec:introduction}

Inverse design problems represent a frontier in the field of engineering and science, where the objective is to discover the necessary system inputs to achieve a desired known output. Rather than following the traditional forward design process--which starts with given parameters and attempts to predict the outcome--inverse design turns the procedure on its head, beginning with the desired outcome and working backward to determine the optimal parameters to realize it. Particularly in scenarios with computationally expensive or hierarchical simulations, multi-fidelity evaluations play a pivotal role, offering a trade-off between accuracy and computational cost.

Multi-fidelity (MF) methods, that range from faster and approximate or low-fidelity (LF) objective function evaluations to detailed--high fidelity (HF), computationally intensive ones have been explored in-depth for optimization purposes \cite{forrester2008engineering, peherstorfer2018survey, fernandez2016review, beran2020comparison}. In the context of inverse design optimization, which is the focus of this work, coupled with a multi-fidelity approach, an additional layer of complexity is introduced when the target output is a distribution. Bayesian and surrogate-based optimization methods have provided great insights in this specific domain, especially when the inverse design problem is rooted in uncertainty or when prior knowledge is available \cite{eldred2006formulations, marzouk2009stochastic, sarkar2019multifidelity}. However, due to the curse of dimensionality, these optimization approaches can encounter computational challenges. 

Variable-fidelity methods have further enhanced the efficiency of inverse design optimization by adapting the fidelity level dynamically based on the current stage of the optimization process \cite{fernandez2016review}. This ensures a balance between computational efficiency and solution accuracy, leading to faster convergence rates and reduced computational costs. The success of variable-fidelity methods is evident in their widespread application across various engineering disciplines \cite{beran2020comparison, fusi2015multifidelity, guo2021design}.
These methods usually either employ LF models during the initial exploratory phases and gradually transition to HF models as the solution converges or they have an adaptive mechanism for fidelity selection. These mechanics include monitoring convergence conditions \cite{koziel2013surrogate, leifsson2010multi}, correction techniques \cite{fischer2016multi, demange2016multifidelity}, space mapping \cite{robinson2006multifidelity}, model error monitoring \cite{mehmani2014managing, jo2016adaptive}, etc. Furthermore, the dominant surrogate model algorithms used in variable-fidelity optimization approaches are Kriging (or Gaussian Process Regression), Co-Kriging, Polynomial Chaos Expansion (PCE), and Moving Least Squares \cite{fernandez2016review}. Additionally, Deep Neural Networks (DNN) are commonly used for multi-fidelity inverse design as a surrogate model for optimization purposes (\cite{du2021rapid, deng2023fast}), but not as a part of the variable-fidelity optimization mechanism.

LF warm-starting optimization techniques have also demonstrated their efficacy in improving the convergence of optimization algorithms \cite{habibi2023actually, poloczek2016warm}. By initializing the optimization process with solutions generated with LF machine learning (ML) models, warm starting leverages prior knowledge to reduce the number of evaluations required to reach optimal solutions. This approach is particularly useful in scenarios where optimization problems or targets share similarities, such as in iterative design processes or when dealing with parametric variations \cite{chen2020airfoil, chen2022inverse, kudyshev2020machine}.

The essence of MF simulations is to harmoniously integrate models of varying accuracy and computational expense. By leveraging the strengths of both HF simulations and LF ML models, it is possible to achieve accurate solutions while conserving computational resources. This is especially pivotal in scenarios where computational budgets are limited, but the accuracy cannot be compromised. Kriging, Co-Kriging, and PCE have the major benefit of having reliable uncertainty estimates, however, they do not scale well with an increase of data without an increase in computational complexity (\cite{lederer2020real, hensman2013gaussian}), and they require retraining when additional data is available. Furthermore, most methods switch between LF and HF simulations, however, the execution time of the LF simulation could also be non-trivial.

Hence, in this paper, in order to tackle the aforementioned issues, an innovative inverse design framework is presented and investigated. The framework integrates metaheuristic algorithms with a pre-trained LF ML model used for design approximation and decision making in order to discern whether there is a need for a HF simulation. This decision is achieved by comparing the discrepancy between its predicted approximation and the inverse design target value. This innovation stands out from prior research by facilitating the predictive power of the ML model regarding the necessity for HF simulations, leveraging LF data. Additionally, the other, equally important task of the LF ML model is its capability of design space boundary refinement before the optimization process is started. This can be considered as a form of an optimization process warm starting, and the purpose is to enhance the rate of convergence of the inverse design optimization algorithms. Two different strategies for boundary refinement are investigated and separately applied to two different problems. Finally, the ultimate benefit of this framework is that the LF ML model can enhance the optimization for any target design within its applicable domain, thereby substantially extending its reach and impact. This is possible since the ML models that are investigated are DNNs, and Gradient Boosting (GB) algorithms. Both of these algorithm types can be continually trained (provided there is not a large data distribution shift), and scale well with additional data. 

The metaheuristic algorithms used within the inverse design framework are: Particle Swarm Optimization (PSO) and Differential Evolution (DE). Even though any kind of optimization algorithm could be incorporated into the framework, PSO and DE were chosen since they generally require a high number of evaluations, and they've been previously used for similar tasks \cite{pehlivanoglu2019efficient, chakraborty2017surrogate}. To the best of the authors' knowledge, no approaches in the literature combine metaheuristic algorithms with ML techniques for both boundary refinement techniques and MF optimization within a single framework.

The importance of this research is emphasized by its application to airfoil inverse design (AID) and scalar field reconstruction (SFR) challenges. The AID problem is chosen since it occupies a pivotal role in engineering, particularly in the realm of aeronautics \cite{bartoli2019adaptive, han2018aerodynamic} and wind energy generation \cite{sharma2021recent}, and it has been extensively studied in the field of multi-fidelity inverse design optimization \cite{du2019aerodynamic, pehlivanoglu2019efficient, leifsson2011inverse, koziel2013surrogate, zhu2020proper, lei2021deep, han2018aerodynamic, tandis2017inverse}. The SFR problem emerges as an inverse design challenge across various scientific and engineering domains, representing a specific variant of the inverse boundary value problem \cite{tarantola2005inverse}. This problem centers on deducing the distribution of a scalar field from sparse measurements \cite{aster2018parameter, hanna2005monte, xu2016bayesian, gordillo2020gradient, wang2022reconstruction, mohasseb2017novel, winter2023multi, ren2021efficient}. Solving the SFR problem utilizing optimization algorithms has been of research interest \cite{chen2018identification, chen2018identification2, slota2008solving, karr2000solving}. 

Finally, the goals of this research are to: (i) show that the proposed framework can accelerate the rate of convergence of both optimization algorithms, and on both inverse design tasks, (ii) show that the LF ML models can be re-used when the inverse design target is changed, without retraining, and (iii) quantify the amount of data needed for the LF ML models to be of use through detailed analysis.

The manuscript is structured to offer a thorough understanding of the research. Following the introduction, Sect. \ref{sub:ml_framework} delves into the ML-enhanced inverse design framework. Sect. \ref{sec:airfoil_opt} and Sect. \ref{sec:boundary_value_opt} provide an in-depth examination of the AID and the SFR problems, respectively, as well as their boundary refinement strategies. Finally, Sect. \ref{sec:results} presents a comprehensive discussion of the results of the ML model, techniques for boundary refinement, and a meticulous analysis of the ML-enhanced framework, contrasting it with traditional optimization algorithms.

\section{ML-Enhanced Inverse Design Framework}
\label{sub:ml_framework}

\begin{figure}[!h]
	\centering
	\includegraphics[width=0.75\linewidth]{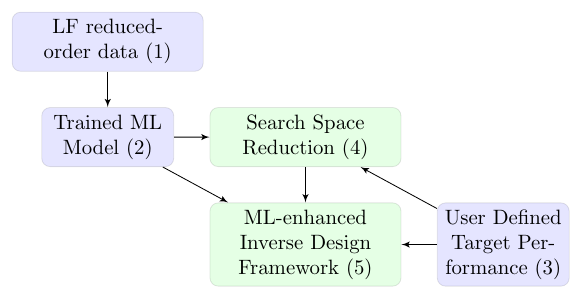}
	\caption{The creation of the ML model commences with the generation of LF data (1) used for training the ML model (2). Once the model is trained and its accuracy is determined, it enables the boundary refinement (4) and the ML-enhanced optimization methodology (5). The inverse design procedure requires the specification of a target performance or vector (3). Blocks highlighted in light green (4) and (5) denote stages involving an optimization process.
	}
	\label{fig:method_flowchart}
\end{figure}

In this section we introduce our ML-enhanced inverse design framework. The methodology consists of two primary stages: training an ML model and applying it to refine the boundaries of optimization problem thus enabling  acceleration and the rate of convergence improvement, and finally, executing the ML-enhanced optimization process to find the design corresponding to the target performance. 
The framework uniquely combines LF simulation data for ML model training with HF simulations for optimization, creating a versatile MF system. Once trained, the ML model can be utilized to augment the inverse design for a given problem. This, however, is applicable to the solutions that lie within the boundaries of the dataset used for training the model. The general workflow of the inverse design framework and the components is displayed in Fig. \ref{fig:method_flowchart}.

\newpage
\subsection{Inverse Design Optimization and Objective Function Definition}

The inverse design problem can be mathematically articulated as
\begin{equation}
	\mathbf{x} = f^{-1}(y)
	\label{eqn:inverse_design_formulation}
\end{equation}

here, \(\mathbf{x} \in \mathbb{R}^m\) represents the design parameters and \(y \in \mathbb{R}^q\) is the known target value. The objective is to ascertain the design parameters \(\mathbf{x}\) that generate \(y\) when evaluated with the inverse of the objective function \(f: \mathbb{R}^m \to \mathbb{R}^q\). This is in stark contrast to forward problems, where \(y\) is typically unknown. Inverse design problems tend to be ill-posed, commonly encountering the problem of multiple viable solutions, which complicates the process of identifying a unique and optimal solution.

In Eq. (\ref{eqn:general_inverse_opt}) the inverse design optimization problem is defined as

\begin{equation}
	\begin{aligned}
		& \underset{\mathbf{x}}{\text{minimize}}
		& & \varepsilon(f(\mathbf{x}), y) \\
		& \text{subject to}
		& & \mathbf{x}_{lb} \leq \mathbf{x} \leq \mathbf{x}_{ub}
		\label{eqn:general_inverse_opt}
	\end{aligned}
\end{equation}

here, \(\varepsilon: \mathbb{R}^q \times \mathbb{R}^q \to \mathbb{R}\) is the error-based objective function that returns a scalar value in \(\mathbb{R}\). The goal of this optimization problem is to minimize the discrepancy between the desired output \(y\) and the outcome derived from the proposed design \(\mathbf{x}\). The design parameters \(\mathbf{x}\) are constrained within a compact design space, defined by the lower and upper boundaries \(\mathbf{x}_{lb} \in \mathbb{R}^m\) and \(\mathbf{x}_{ub} \in \mathbb{R}^m\) respectively, which represent the feasible range of the design variables.

In this study, the root mean square error is used as $\varepsilon$:

\begin{equation}
	\begin{aligned}
		& \underset{\mathbf{x}}{\text{minimize}}
		& & \varepsilon(\mathbf{x}) = \sqrt{\frac{1}{q} \lVert \mathbf{P}^{C}(\mathbf{x}) - \mathbf{T}\rVert_2^2} \\
		& \text{subject to}
		& & \mathbf{x}_{lb} \leq \mathbf{x} \leq \mathbf{x}_{ub},
	\end{aligned}
	\label{eqn:obj_function}
\end{equation}

where \(\mathbf{x} = (x_1, \ldots, x_m)^T\) is the optimization design vector in the decision space \(\mathbb{R}^m\), \(m\) is the dimension of the general optimization design vector, \(\mathbf{P}^{C}(\mathbf{x}) = (P_1^C(\mathbf{x}), \ldots, P_q^C(\mathbf{x}))^T \in \mathbb{R}^q\) denotes the computed performance vector based on the design vector \(\mathbf{x}\), while \(\mathbf{T} = (T_1, \ldots, T_q)^T \in \mathbb{R}^q\) signifies the user-defined target performance vector. Both \(\mathbf{P}^{C}(\mathbf{x})\) and \(\mathbf{T}\) are of dimension \(q\). Ideally, an exact match between these values would result in an objective function value of zero.

\subsection{ML-Enhanced Optimization}
\label{sub:ml_enh_opt}

As shown in Fig. \ref{fig:method_flowchart}, the requirement for the ML-enhanced optimization process is to train a ML model using simulation data generated with Latin Hypercube Sampling (LHS). Each simulation yields an input design vector and a simulation result vector. These design vectors, which are of the same size and within the same bounds as those evaluated during the optimization process, are used as inputs to the ML model. The outputs are statistical measures (mean, minimum, maximum, etc.) derived from the simulation result vectors. Once trained, this ML model can be reused within the inverse design framework when the target performance changes. The details of the ML model can be found in Sect. \ref{sec:mlmodels}.

The ML model has two main tasks: (i) refine the lower and upper optimization boundaries (denoted as $\mathbf{lb}_{R}$ and $\mathbf{ub}_{R}$, respectively), and (ii) decide whether to run a HF simulation based on a design $\mathbf{x}$ being evaluated during the optimization process. The details of the boundary refinement procedure are given in Sect. \ref{sub:ml_search_space}. Furthermore, the pseudo-code of the ML-enhanced optimization process or task (ii), and all the necessary parameters are detailed in Alg. \ref{alg:main_opt_algorithm}.

When the ML model is trained, the optimization process begins by defining a target vector (\(\mathbf{T}\)) and deriving a target scalar value (\(\mathit{T}_{info} \in \mathbb{R}\)), where \(\mathit{T}_{info} = \frac{1}{q} \sum_{i=1}^{q} T_i\) (the mean of \(\mathbf{T}\)) or \(\mathit{T}_{info} = \max(\mathbf{T})\) (the maximum of \(\mathbf{T}\)), depending on the application case in this study. During each evaluation of the objective function $\varepsilon$ (defined in Eq. \ref{eqn:general_inverse_opt}), the pre-trained ML model ($M(\mathbf{x})$) predicts the value (\(\mathit{M}_{info} \in \mathbb{R}\)) for a given optimization design vector $\mathbf{x}$. The scalar values \(\mathit{M}_{info}\) and \(\mathit{T}_{info}\) must represent the same statistically derived quantities in the space \(\mathbb{R}\), ensuring consistency in the comparison of predicted and target performance metrics. Subsequently, the absolute error ($\Delta$) between the ML predicted value ($\mathit{M}_{info}$) and the target scalar value ($\mathit{T}_{info}$) is then computed. If $\Delta$ exceeds a pre-established threshold ($\omega$), the objective function ($\varepsilon$) is assigned the value $\lambda \cdot e^\Delta$ ($\lambda = 2$). If $\Delta$ is less than or equal to $\omega$, $\varepsilon$ is evaluated with a HF simulation and the result is compared with $\mathbf{T}$ through a discrepancy metric.

The threshold parameter ($\omega$) is calculated using a user-defined scaling factor ($c$) and the error of the ML model ($\epsilon_\mathit{M}$), such as root mean square error or mean absolute error obtained through ML model analysis. The initial design vector \(\mathbf{x}_{init} \in \mathbb{R}^m\) is a vector randomly initialized within the optimization boundaries ($\mathbf{lb}_{R}$ and $\mathbf{ub}_{R}$) by the optimization algorithms. The remaining budget (RB) denotes the remaining HF simulation budget, which is used as a comparison metric with unenhanced optimization algorithms. A higher RB value indicates enhanced performance, reflecting increased computational efficiency and savings. The optimization process stops when the simulation budget is exceeded.

The proposed ML-enhanced inverse design framework can be used in conjunction with any population-based global optimization algorithm. In this study, it is investigated how the ML model enhances two population-based  algorithms, namely, DE and PSO. The fundamental goal of this framework is to enhance the robustness and efficiency of the optimization algorithms through the use of ML-generated boundary refinement and ML-guided evaluation of HF simulations.

\begin{algorithm}
\begin{algorithmic}[1]
	\Procedure{Optimize}{$\varepsilon, \mathbf{x_{\text{init}}}, \mathbf{T}, \mathit{T}_{info}, \mathbf{lb}_R, \mathbf{ub}_R, \mathit{M}(\textbf{x}), {\epsilon}_\mathit{M}, c$}
	\State $\mathit{RB} \gets 0$
	\State $\mathbf{x} \gets \mathbf{x}_{\text{init}}$
	\While{budget not exceeded}
	\State $\mathbf{x} \gets \text{OptimizerStep}(\varepsilon, \mathbf{x}, \mathbf{lb}_{R}, \mathbf{ub}_{R})$
	\State \textbf{inside OptimizerStep:}
	\State $M_{info} \gets \mathit{M}(\mathbf{x})$
	\State $\Delta \gets  | M_{info} - T_{info} |$
	\State $\omega \gets \mathit{c} \cdot {\epsilon}_\mathit{M}$ 
	\If{$\Delta > \text{$\omega$}$}
	\State $\varepsilon \gets \lambda \cdot e^\text{$\Delta$}$
	\State $\mathit{RB} \gets \mathit{RB} + 1$
	\Else
	\State $\varepsilon$ \Comment{Run HF simulation, compare with $\mathbf{T}$}
	\EndIf
	\EndWhile
	\State \textbf{return} $\mathbf{x}$ \Comment{Ensure \textbf{x} is within the bounds $\mathbf{lb}_R$ and $\mathbf{ub}_{R}$}
	\EndProcedure
\end{algorithmic}
\caption{Pseudo-code of the ML-Enhanced optimization algorithm within the inverse design framework.}
\label{alg:main_opt_algorithm}
\end{algorithm}

\subsection{ML-Generated Boundary Refinement}
\label{sub:ml_search_space}

The ML model is used to narrow down the optimization boundaries through a boundary refinement method, shown in Alg.~\ref{alg:boundary_compression},. This approach aims to significantly minimize the demand for computational resources, an essential factor when operating within a stringent computational budget. The requirements for the boundary refinement are $\mathit{T}_{info}$ and $\mathit{M}_{info}$ values. The objective of each of the $\mathit{N}$ optimization runs in the algorithm is to minimize the absolute difference between these two values. The $\mathit{N}$ value is predefined by the user and ultimately will determine the number of optimization solutions that will be used to refine the boundaries for the ML-enhanced inverse design framework. More specifically, to narrow down the boundaries through with the boundary refinement method, it is necessary to determine the optimal solution defined as in Eq. (\ref{eqn:search_space_opt}):

\begin{equation}
	\begin{aligned}
		& \mathbf{x}^* = \underset{\mathbf{x}}{\text{argmin}} \, 
		|\mathit{M}(\textbf{x}) - \mathit{T}_{info}| \\
		& \text{subject to} \quad 
		\mathbf{x}_{lb} \leq \mathbf{x} \leq \mathbf{x}_{ub}
	\end{aligned}
	\label{eqn:search_space_opt}
\end{equation}

The solution vector $\mathbf{x}^* \in \mathbb{R}^m$ represents an optimized design based on the absolute difference between $\mathit{T}_{info}$ and $\mathit{M}_{info}$ (that is predicted by $\mathit{M}(\textbf{x})$). 

Given the inherent ill-posedness of most inverse design problems, this optimization procedure is repeatedly executed, resulting in a matrix of optimal solutions $\mathbf{S}$. Repeating the optimization process $N$ times yields various solutions due to the multi-modal landscape and the stochastic nature of the used optimizer (DE), which converges to different local optima. The condition in Eq. (\ref{eqn:search_space_opt}) is especially sensitive to this because it relies on partial information (ML prediction of a single scalar value instead of a complete array), further reducing the fidelity of the ML model trained with LF simulation data.

More specifically, each row of $\mathbf{S}$ represents one of the optimal solutions $N$, while each column is one of the design variables $m$ as defined in:

\begin{equation}
	\mathbf{S} = 
	\begin{bmatrix}
		\mathbf{x}^{*}_{1} \\
		\mathbf{x}^{*}_{2} \\
		\vdots \\
		\mathbf{x}^{*}_{N}
	\end{bmatrix}
	=
	\begin{bmatrix}
		x^{*}_{1,1} & x^{*}_{1,2} & \cdots & x^{*}_{1,m} \\
		x^{*}_{2,1} & x^{*}_{2,2} & \cdots & x^{*}_{2,m} \\
		\vdots & \vdots & \ddots & \vdots \\
		x^{*}_{N,1} & x^{*}_{N,2} & \cdots & x^{*}_{N,m}
	\end{bmatrix}
	\label{eqn:sequence_sa_matrices}
\end{equation}

where $x^*_{N,m}$ is the design point in dimension $m$ of the optimized solution $N$, and $\mathbf{x}^{*}_{N}$ is the $N^{th}$ solution vector.

\begin{algorithm}
\begin{algorithmic}[1]
\Procedure{BoundaryRefinement}{$\mathit{T}_{info}, \mathbf{x}_{lb}, \mathbf{x}_{ub},\mathit{M}(\textbf{x}), N$}
\State Initialize an empty matrix \( \mathbf{S} \)
\State $\mathbf{S} = []$
\For{\( n = 1 \) to \( N \)}
\State $\mathit{M}_{info}$ \( \gets \) $\mathit{M}(\textbf{x})$
\State \( f \) \( \gets \) \( |\mathit{M}_{info} - \mathit{T}_{info}| \)
\State Find \( \mathbf{x}^* \) that minimizes \( f \) 
\State subject to: \( \mathbf{x}_{lb} \leq \mathbf{x} \leq \mathbf{x}_{ub} \)
\State Add \( \mathbf{x}^* \) to \( \mathbf{S} \)
\EndFor
\State $\mathbf{S} = \ [\mathbf{x}^*_1, \mathbf{x}^*_2, \ldots, \mathbf{x}^*_N] \ $
\State \Return \( \mathbf{S}^T \)
\EndProcedure
\end{algorithmic}
\caption{The boundary refinement algorithm.}
\label{alg:boundary_compression}
\end{algorithm}

The obtained solutions within the matrix  $\mathbf{S}$ are then subjected to statistical processing that depends on the specific inverse design problem being solved (see Sect. \ref{sec:airfoil_opt} and Sect. \ref{sec:boundary_value_opt}), which yields the compressed lower and upper boundaries $\mathbf{lb}_{R}$ and $\mathbf{ub}_{R}$, respectively.

\subsection{ML Model}
\label{sec:mlmodels}

The ML model $\mathit{M}(\textbf{x})$ takes as input the optimization design vector $\mathbf{x}$ and maps it to $\mathit{M}_{info}$ which is then compared with the $\mathit{T}_{info}$ value, thereby minimizing the necessity for HF simulations. The performance of three different ML algorithms is analyzed within this methodology -- A DNN, and two different GB algorithms -- LightGBM (LGB) \cite{ke2017lightgbm} and XGBoost (XGB) \cite{chen2016xgboost}. 
XGBoost or eXtreme Gradient Boosting (XGB) is a scalable tree boosting framework that effectively integrates a sparsity-aware algorithm alongside a weighted quantile sketch, thereby facilitating an approximate tree learning process. The combination of cache access patterns, elevated data compression, and sharding empowers XGBoost to construct an efficient and powerful tree boosting system. 

LightGBM (LGB) is a robust and efficient gradient boosting framework aimed at enhanced performance and speed. It incorporates innovative strategies such as gradient-based one-side sampling and exclusive feature bundling to expedite processing and improve efficiency. LightGBM has a unique leaf-wise tree growth strategy, which deviates from the conventional level-wise approach seen in other boosting algorithms, and contributes to improved model accuracy by minimizing loss, thereby achieving faster convergence. 

A DNN configured as an MLP is fundamentally composed of three distinct types of layers: the input, hidden, and output layers. These layers are constituted by artificial neuron nodes. The MLP model can incorporate multiple hidden layers as part of its neural architecture. Each neuron residing within the hidden and output layers utilizes a nonlinear activation function, echoing the complex processing mechanisms observed in the human brain \citep{nielsen2015neural}. This structure effectively facilitates the MLP's ability to model and solve intricate nonlinear problems.

The accuracy of all trained ML models was assessed using the $RMSE$ (Eq. (\ref{eqn:$RMSE$})) since it is used to evaluate the $\omega$ value within the ML-enhanced framework (as shown in Alg. \ref{alg:main_opt_algorithm}).

\begin{equation}
	\text{$RMSE$}  = \sqrt{\frac{\sum_{l=1}^{L} (y_{l} - \hat{y}_{l})^2}{L}}
	\label{eqn:$RMSE$}
\end{equation}

The variables $y_l$, $\hat{y}_l$, and $L$  represent the $l^{th}$ actual value, the $l^{th}$ ML model prediction, and the test set size, respectively. More specifically, the variable $y_l$ represents the $l^{th}$ data point that is the result of the $l^{th}$ LF simulation, and $y_l$ must represent the same statistically derived information as the T$_{info}$ value. The K-Fold cross-validation procedure ($k=5$) was used to evaluate the accuracy and uncertainty of all three investigated algorithms. For the K-Fold analysis of the ML model, the test set size $L$ is varied as 500, 1000, 5000, and 15000.

\subsection{Metaheuristic Optimization Algorithms}
\label{sub:metaheuristics}

Two distinct metaheuristic optimization algorithms will be compared: Particle Swarm Optimization (PSO) and Differential Evolution (DE). Both  algorithms belong to  the broader categories of swarm intelligence and evolutionary algorithms. Fundamentally, these categories rely on populations of agents that abide by specific rules to identify optimal solutions. 
Using both PSO and DE will demonstrate the general applicability of the ML-enhancement.

PSO is a population-based stochastic optimization algorithm, inspired by the social behavior of bird flocking or fish schooling \cite{kennedy1995particle}. In PSO, each individual particle in the swarm population represents a solution in the design space. Every particle updates its position based on its local best position, as well as the global best solution of the swarm. This cooperative search process, conducted through the iterative adjustment of velocities and positions increases the rate of convergence of the swarm towards the local or global optimum.

DE is a population-based stochastic search technique, commonly used for global optimization problems over continuous optimization design vectors \cite{storn1997differential}. In DE, the potential solutions are evolved over time via a simple arithmetic operation:  a combination of mutation, crossover, and selection operations. Each individual in the population is a potential solution, and the evolution of these individuals is performed based on the differences between randomly sampled pairs of individuals within the population. The differential evolution of the population ensures a good rate of convergence, however, converging to a global optimum is not guaranteed.
The success-history-based parameter adaptation (SHADE) variant of DE is used in this investigation. In the SHADE variant, the scaling factor and crossover rate are adaptively adjusted for each individual in the population based on a history of successful parameters. This dynamic adaptation allows for more effective exploration and exploitation of the design space, potentially improving the performance of the algorithm. 

For the investigated problems, the swarm size and the population size parameters for the PSO and DE algorithms were both set to 10. Both DE and PSO implementations in the Indago 0.4.5 Python module for numerical optimization were used \cite{Indago}. For the PSO algorithm, the inertia parameter was set at 0.8. The cognitive and social rates of the swarm were standardized to 1, mainly based on the default recommendations with the Python module. For DE, the key hyperparameters, such as the archive size factor, historical memory size, and mutation rate, were  configured to 2.6, 4, and 0.11, respectively, following recommendations from the utilized Python module.

\section{Airfoil Inverse Design}
\label{sec:airfoil_opt}

This section defines the AID problem through the optimization design vector, constraints, and the boundary refinement strategy.  

\subsection{AID Problem Description}

The goal of the AID problem is to determine the optimal geometry of an airfoil given a set of target pressure coefficients on the surface of the airfoil. The parameters used for the AID and the boundary refinement in the context of Eq. (\ref{eqn:obj_function}) and Eq. (\ref{eqn:search_space_opt}) are presented in Table \ref{tab:airfoil_parameters}. Each evaluation of the function $\varepsilon$ (Eq. (\ref{eqn:obj_function})) necessitates executing a flow simulation over a generated design.

\begin{table}[h]
	\caption{Mapping of problem-specific parameters for the AID to their corresponding general parameters used in the objective function and the boundary refinement process.}\label{tab:airfoil_parameters}
	\begin{tabular}{@{}ll@{}}
		\toprule
		General Parameter & Problem Specific Parameter \\
		\midrule
		$\mathit{T}_{info}$ & $C^T_{p_{min}}$  \\
		$\mathbf{T}$ & $\mathbf{C}_{p}^{T}$ \\
		$\mathbf{P}^{C}(\textbf{x})$ & $\mathbf{C}_{p}^{C}(\mathbf{x})$  \\
		$\mathit{M}_{info}$ &  $C_{p_{min}}$\\
		$m$ & $N_c$ \\
		\botrule
	\end{tabular}

\end{table}

$\mathbf{C}_{p}^{C}(\mathbf{x}) \in \mathbb{R}^q$ denotes the computed pressure distribution around an airfoil based on the design vector $\mathbf{x}$, while $\mathbf{C}_{p}^{T} \in \mathbb{R}^q$ signifies the user-defined target pressure coefficient distribution measured at the same locations on the surface of the airfoil. For all target cases, $q = 300$. For each evaluation of $\mathbf{x} \in \mathbb{R}^{N_c}$, the computed pressure distribution $\mathbf{C}_{p}^{C}(\mathbf{x})$ is linearly interpolated to match the target pressure distribution in both the size and airfoil surface location for each individual component $q$. $C^T_{p_{min}} \in \mathbb{R}$ denotes the target minimum pressure coefficient obtained from $\mathbf{C}_{p}^{T}$, and since the minimum pressure coefficient is used, the ML model's task is to map the optimization design vector $\mathbf{x}$ to the minimum pressure coefficient $C_{p_{min}} \in \mathbb{R}$ measured on the surface of the airfoil.

The AID optimization problem uses B-Spline approximation of the $NACA0012$ airfoil geometry for the lower and upper boundary definition, as this method offers superior shape parametrization \cite{rajnarayan2018universal}. Utilizing the \textit{splrep} function from the \textit{scipy 1.9.1} module \cite{virtanen2020scipy}, the B-Splines are defined with coefficients, knots, a maximum degree of 5, and a smoothness parameter set at 0.00001. The knots generated by the \textit{splrep} function as well as the degree and the smoothness of the splines remain unchanged throughout the optimization procedure. 

The optimization design vector for the AID problem $\mathbf{x_A}$ is defined as: 

\begin{equation}
	\mathbf{x_A} = [\mathbf{c_l}, \mathbf{c_u}]^T \in \mathbb{R}^{N_c}, \quad \begin{cases} 
		\gamma \cdot (\mathbf{c_{L0012}} - 1\cdot10^{-5}) \leq \mathbf{c_l} \leq \mathbf{0} \\
		\mathbf{0} \leq \mathbf{c_u} \leq \gamma \cdot (\mathbf{c_{U0012}} + 1\cdot10^{-5}) \\
	\end{cases}
	\label{eqn:bspline_variables}
\end{equation}

where it is defined as being comprised of the 15 lower and 15 upper surface B-spline coefficient vectors ($\mathbf{c_l}$ and $\mathbf{c_u}$), meaning that $N_c=30$. In detail, $\mathbf{x_A}$ with its individual components ($c_{li}$ and $c_{ui}$) is defined as:

\begin{equation}
	\begin{aligned}
		\mathbf{x_A} &= \begin{bmatrix}
			c_{l1}, & c_{l2}, & \ldots, & c_{l\frac{N_c}{2}}, & c_{u\frac{N_c}{2} + 1}, & c_{u\frac{N_c}{2} + 2}, & \ldots, & c_{uN_c}
		\end{bmatrix}^T \in \mathbb{R}^{N_c}, \\
		&\begin{cases} 
			\gamma \cdot (c_{L0012_i} - 1 \cdot 10^{-5}) \leq c_{li} \leq 0 & \text{for} \quad i = 1, 2, \ldots, \frac{N_c}{2} \\
			0 \leq c_{ui} \leq \gamma \cdot (c_{U0012_i} + 1 \cdot 10^{-5}) & \text{for} \quad i = \frac{N_c}{2} + 1, \frac{N_c}{2} + 2, \ldots, N_c 
		\end{cases}
	\end{aligned}
	\label{eqn:bspline_variables_xa}
\end{equation}

The bounds of the lower and upper B-spline coefficients are determined by the $NACA0012$ lower and upper B-Spline coefficient vectors  denoted as $\mathbf{c_{L0012}}$ and $\mathbf{c_{U0012}}$, respectively, scaled by a multiplication factor $\gamma=3$. The extracted lower boundary B-Spline coefficients in $\mathbf{c_{L0012}}$ are negative. Subsequently, there is an overlap in the initial lower and upper B-Spline coefficients ($\mathbf{c_{L0012}}$ and $\mathbf{c_{U0012}}$) generated by B-Spline interpolation of the NACA0012 geometry, where some coefficients in these vectors converge to zero. To address this potential overlap and to enable optimization in the near-zero space, we adjust the values of $\mathbf{c_{L0012}}$ and $\mathbf{c_{U0012}}$ by subtracting and adding a small value of $1 \times 10^{-5}$ to these vectors, respectively. 

\begin{figure}[!h]
	\centering
	\includegraphics[width=0.75\linewidth]{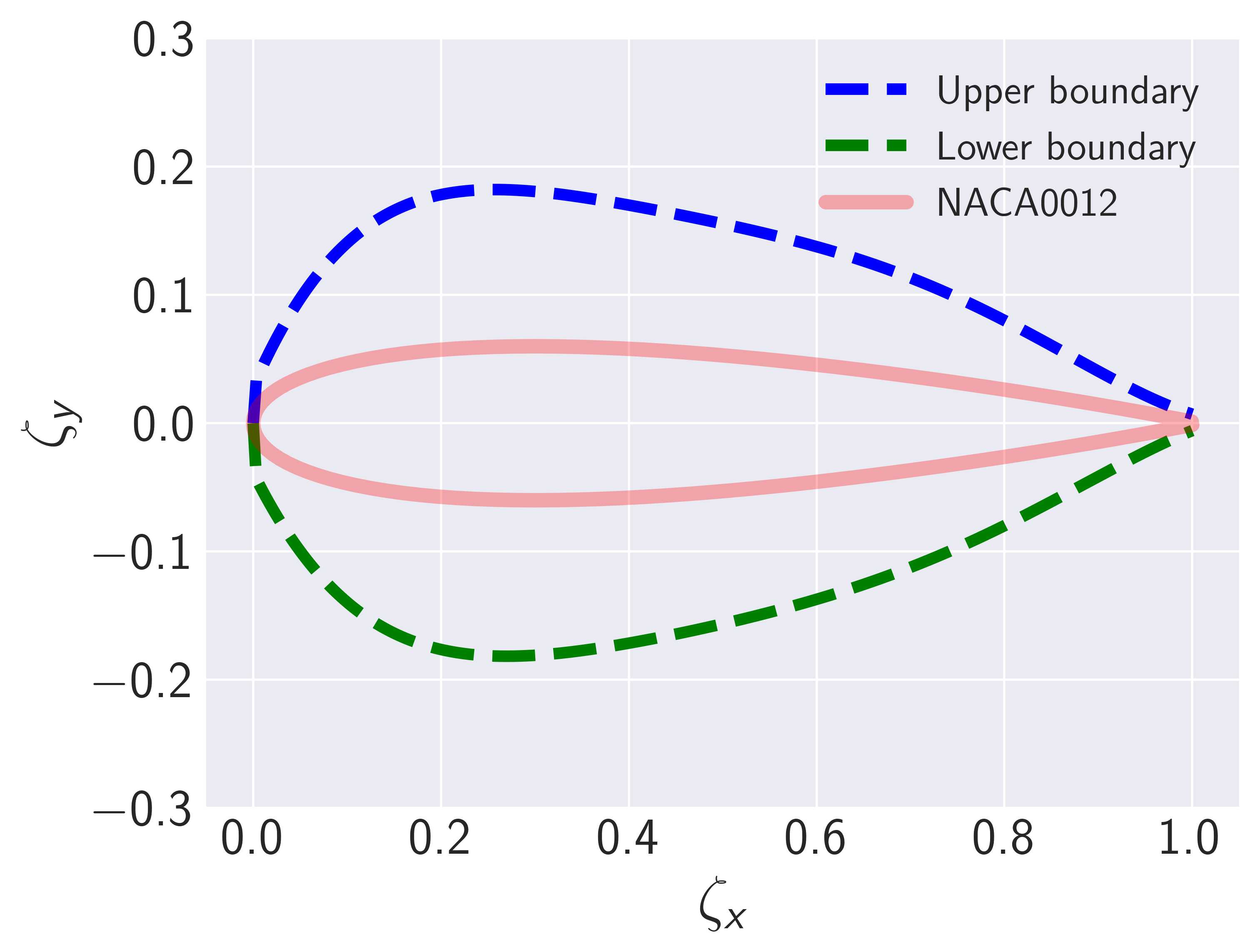}
	\caption{Optimization boundaries generated by multiplying the $NACA0012$ baseline B-Spline coefficients by a factor of 3.  The $\zeta_x$ and $\zeta_y$ axis are the chord-length normalized $x$ and $y$ coordinates.}
	\label{fig:naca0012_constraints}
\end{figure}

The constraints on $\mathbf{c_l}$ and $\mathbf{c_u}$ in Eq.~(\ref{eqn:bspline_variables}) represent the initial lower and upper bounds ($\mathbf{x}_{lb}$ and $\mathbf{x}_{ub}$) used for the ML-generated boundary refinement (Eq. (\ref{eqn:search_space_opt})) as well as the lower and upper boundaries used by the unenhanced optimization algorithms (stand-alone DE and PSO). The lower and upper boundaries of the optimization design vector based on the scale factor $\gamma$ and the B-Splines of $NACA0012$ are shown in Fig. \ref{fig:naca0012_constraints}. Details of the airfoil targets, flow simulation solver, and the dataset used to train the ML models are provided in the Appendix \ref{app:airfoil_experiments_dataset}.

\subsection{AID Boundary Refinement}

After the boundary refinement technique is applied, a whole matrix of solutions $\mathbf{S}$ is obtained, as detailed in the Sect. \ref{sub:ml_search_space}. By statistically analyzing the matrix of optimal solutions derived from the ML model, the column-averaged values of the solution matrix $\mathbf{S}$ provide a meaningful representation of the refined design space. More specifically, the matrix of solutions for the AID problem $\mathbf{S}_A$ is defined as:

\begin{equation}
	\mathbf{S}_A = 
	\begin{bmatrix}
		\mathbf{x}^{opt}_{A_{1}} \\
		\mathbf{x}^{opt}_{A_{2}} \\
		\vdots \\
		\mathbf{x}^{opt}_{A_{N}}
	\end{bmatrix}
	=
	\begin{bmatrix}
		c_{l1}^{opt_1} & c_{l2}^{opt_1} & \ldots & c_{l\frac{N_c}{2}}^{opt_1} & c_{u\frac{N_c}{2} + 1}^{opt_1} & c_{u\frac{N_c}{2} + 2}^{opt_1} & \ldots & c_{uN_c}^{opt_1} \\
		c_{l1}^{opt_2} & c_{l2}^{opt_2} & \ldots & c_{l\frac{N_c}{2}}^{opt_2} & c_{u\frac{N_c}{2} + 1}^{opt_2} & c_{u\frac{N_c}{2} + 2}^{opt_2} & \ldots & c_{uN_c}^{opt_2} \\
		\vdots & \vdots & \ddots & \vdots & \vdots & \vdots & \ddots & \vdots \\
		c_{l1}^{opt_N} & c_{l2}^{opt_N} & \ldots & c_{l\frac{N_c}{2}}^{opt_N} & c_{u\frac{N_c}{2} + 1}^{opt_N} & c_{u\frac{N_c}{2} + 2}^{opt_N} & \ldots & c_{uN_c}^{opt_N}
	\end{bmatrix}
	\label{eqn:matrixSa}
\end{equation}

where $\mathbf{x}^{opt}_{A_{N}}$ is the $N^{th}$ optimized solution of the AID problem, and it consists of optimal lower and upper B-Spline coefficients (Eq. (\ref{eqn:bspline_variables_xa})). Subsequently, the averaged design vector $\mathbf{\bar{x}}_A$ is defined by column-averaging the matrix $\mathbf{S}_A$:

\begin{equation}
	\mathbf{\bar{x}}_A =
	\begin{bmatrix}
		\frac{1}{N} \sum_{i=1}^{N} c_{l1}^{opt_i}, & 
		\ldots, & 
		\frac{1}{N} \sum_{i=1}^{N} c_{l\frac{N_c}{2}}^{opt_i}, & 
		\frac{1}{N} \sum_{i=1}^{N} c_{u\frac{N_c}{2} + 1}^{opt_i}, &
		\ldots, &
		& \frac{1}{N} \sum_{i=1}^{N} c_{uN_c}^{opt_i}
	\end{bmatrix}
	\label{eqn:column_averaged}
\end{equation}

Furthermore, the averaged design vector $\bar{\mathbf{x}}_A$ is scaled by a safety factor $\eta$ to ensure that the target design is within the new boundaries:

\begin{equation}
	\mathbf{x_{A_\eta}} = \eta \cdot \bar{\mathbf{x}}_A
	\label{eqn:x_a_zeta}
\end{equation}

Finally, the design vector $\mathbf{x_{A_\eta}}$ represents an airfoil shape itself, i.e. its design variables are the lower and upper B-Spline coefficients, hence, new lower and upper boundaries ($\mathbf{lb}_R$ and $\mathbf{ub}_R$) are constructed based on this design vector for the AID problem. Different values of hyperparameter $\eta$ are investigated to assess their impact on the performance of the ML enhanced optimization, more specifically, $\eta \in \{1, 1.1, 1.2, 1.3\}$. These values  showcase a range from less efficient to more efficient performance to provide a comprehensive view of the method's effectiveness.

\section{Scalar Field Reconstruction}
\label{sec:boundary_value_opt}

This section defines the SFR problem through the optimization design vector, constraints, and the boundary refinement strategy.

\subsection{SFR Problem Description}

The goal of the SFR problem is to determine the scalar boundary values  based on a set of target scalar measurements on a given domain. The essence  lies in optimizing the boundary conditions for a diffusion partial differential equation (PDE). This mathematical model describes how a scalar quantity spreads within a given domain. Instead of prescribing boundary conditions outright, the problem aims to find the ideal boundary conditions that, when applied to the diffusion PDE, result in a reconstructed scalar field that closely aligns with measured data. The diffusion PDE  is defined as:

\begin{equation}
	\frac{\partial s}{\partial t} = D \nabla^2 s \quad \text{in } \Omega, \quad t \in [0, t_{max}]
	\label{eqn:diffusion}
\end{equation}

where $s$ is the non-dimensional scalar value, $D$ is the diffusion coefficient set to 1 (m$^2$/s), $t$ denotes the time (s), while $t_{max}$ denotes the maximum or end time of the simulation, and $\Omega$ is the domain. For the purposes of demonstrating the ML-enhanced inverse design framework, 
the $t_{max}$ is set to 0.1 s and is treated as a converged state, i.e. the scalar diffusion is treated as a quasi-transient problem.

The parameters used for the SFR and the boundary refinement in the context of Eq. (\ref{eqn:obj_function}) and Eq. (\ref{eqn:search_space_opt}) are presented in Table \ref{tab:scalar_field_parameters}.

\begin{table}[h]
	\caption{Mapping of problem-specific parameters for the SFR problem to their corresponding general parameters used in the objective function and the boundary refinement process.}\label{tab:scalar_field_parameters}
	\begin{tabular}{@{}ll@{}}
		\toprule
		General Parameter & Problem Specific Parameter \\
		\midrule
	    $T_{info}$ & $s^T_{_{max}}$  \\
		$\mathbf{T}$ & $\mathbf{s}^{T}$ \\
		$\mathbf{P}^{C}(\textbf{x})$ & $\mathbf{s}^{C}(\mathbf{x})$  \\
		$M_{info}$ &  ${s_{max}}$ \\
		$m$ & $I$ \\
		\botrule
	\end{tabular}
	
\end{table}

$\mathbf{s}^{C}$($\mathbf{x}$) denotes the computed scalar distribution in $\mathbb{R}^q$ (with $q = 30$) on a given domain $\Omega$ based on the design vector $\mathbf{x}$ (denoted as $\mathbf{x_s}$ for the SFR problem) which is used to define the boundary condition, while $\mathbf{s}^{T}$, also in $\mathbb{R}^q$, signifies the user-defined target scalar field measured at the same locations. $s^T_{{max}}$ denotes the target maximum scalar value obtained from $\mathbf{s}^{T}$ in $\mathbb{R}$. The design vector constraint for the SFR problem is defined in Eq. (\ref{eqn:scalar_field_boundaries}). The ML model maps the design vector $\mathbf{x_s}$ to the maximum scalar value ${s_{max}}$ observed in the domain, with ${s_{max}}$ being in $\mathbb{R}$.

\begin{equation}
	\begin{aligned}
		\mathbf{x_s} = [s_1, s_2, \ldots, s_{I}]^T \in \mathbb{R}^{I}, \\
		\quad 0 \leq s_i \leq s_{ub}
		\quad \text{for} \quad i = 1, 2, \ldots, I.
		\label{eqn:scalar_field_boundaries}
	\end{aligned}
\end{equation}

The scalar design vector $\mathbf{x_s}$ consists of scalar values $s_i$ that collectively form a boundary condition for a given domain. The minimum scalar value at the boundary is defined as 0 and the maximum value $s_{ub}$ is set to 30.
For a HF simulation, the number of scalar values $i$ defined at the top of the domain $\Omega$ is 80 ($I=80$). However, with each evaluation of $\mathbf{x_s}$, it is necessary to obtain the $M_{info}$ value, and a discrepancy arises due to the ML model being trained with LF data where the number of scalar values was set to 20 ($I=20$). In order to evaluate the $\mathbf{x_s}$ with $I=80$ using the ML model, the $\mathbf{x_s}$ are linearly interpolated and the scalar values at the LF boundary points are extracted in order to be evaluated by the ML model in order to predict $s{_{max}}$. 
In accordance with the diffusion PDE (Eq. (\ref{eqn:diffusion})), the scalar boundary values are defined as the Dirichlet boundary condition (Eq. (\ref{eqn:diffusion_bc_top})) for the top part of the domain $\partial \Omega_{top}$:

\begin{equation}
	\begin{aligned}
		s = g(\mathbf{x_s}, t) \quad \text{on } \partial \Omega_{top}, \quad t \in [0, T].
		\label{eqn:diffusion_bc_top}
	\end{aligned}
\end{equation}

\begin{figure}[!h]
	\centering
	\includegraphics[width=0.75\linewidth]{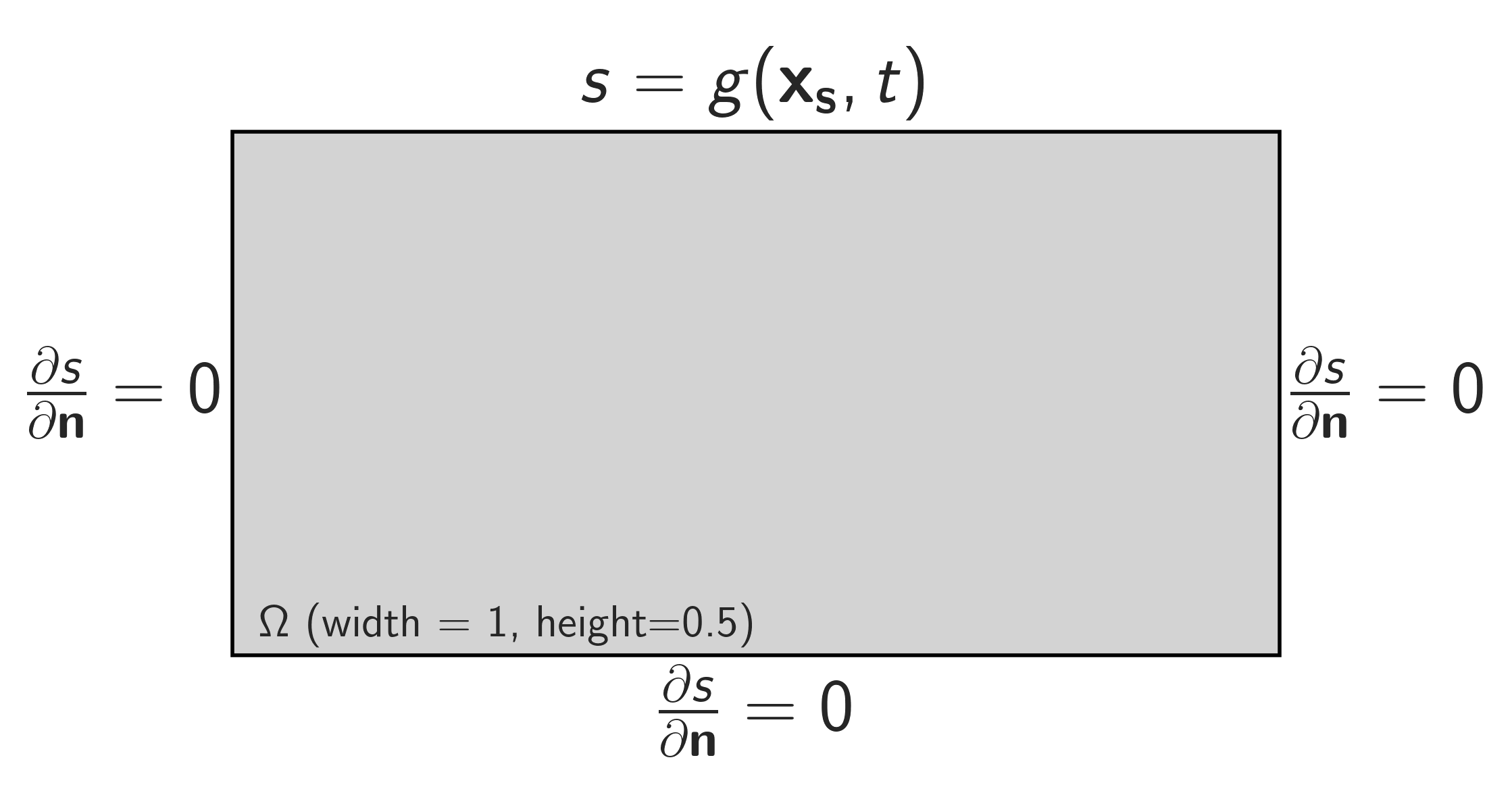}
	\caption{The mathematical domain $\Omega$ of the scalar reconstruction problem. The top boundary condition is the Dirichlet boundary condition where the optimization design vector $\mathbf{x_s}$ is set, while the bottom, left, and right parts are defined as the Neumann boundary condition. The height and width are presented in meters. }
	\label{fig:domain}
\end{figure}

Other parts of the domain $\partial \Omega_{other}$ (left, right, bottom) are defined as the Neumann boundary condition:

\begin{equation}
	\begin{aligned}
		\frac{\partial s}{\partial \mathbf{n}} = 0 \quad \text{on } \partial \Omega_{other}, \quad t \in [0, T],
		\label{eqn:diffusion_bc_other}
	\end{aligned}
\end{equation}

where $\mathbf{n}$ is the unit normal vector pointing outward from the domain. Finally, the mathematical domain $\Omega$ for the given SFR problem along with the appropriate boundary conditions are shown in Fig. \ref{fig:domain}. Details of the solver, inverse design target parameters for the SFR problem, and the dataset used to train the ML models are provided in the Appendix \ref{app:sfr_experiments_dataset}.

\subsection{SFR Boundary Refinement}
\label{sec:sfr_ssr}
In addressing the SFR problem, the design space size presents a significant challenge. Compared to the AID problem, the SFR problem is less constrained and more ill-posed. This means that, based on the objective function and the lower and upper boundaries of the design space, the optimization landscape is more multi-modal for the SFR problem. This difference requires a more robust strategy for boundary refinement. Firstly, the LF solution matrix $\mathbf{S}_{s_{LF}}$ of optimized design vectors (obtained through Eq.~(\ref{eqn:search_space_opt})) that contain boundary condition scalar values is defined as:

\begin{equation}
	\mathbf{S}_{s_{LF}} = 
	\begin{bmatrix}
		\mathbf{x}^{opt}_{s_{1}} \\
		\mathbf{x}^{opt}_{s_{2}} \\
		\vdots \\
		\mathbf{x}^{opt}_{s_{N}}
	\end{bmatrix}
	=
	\begin{bmatrix}
		s_{1}^{opt_1} & s_{2}^{opt_1} & \ldots & 	s_{20}^{opt_1} \\
		s_{1}^{opt_2} & s_{2}^{opt_2} & \ldots & 	s_{20}^{opt_2} \\
		\vdots & \vdots & \ddots & \vdots \\
		s_{1}^{opt_N} & s_{2}^{opt_N} & \ldots & 	s_{20}^{opt_N}
	\end{bmatrix}
	\label{eqn:matrixSa}
\end{equation}

where $\mathbf{x}^{opt}_{s_{N}}$ is the $N^{th}$ optimized design vector. Each row of $\mathbf{S}_{s_{LF}}$ contains an optimized scalar value for each point on the LF domain boundary ($I=20$). Subsequently, the design vectors in each row of matrix $\mathbf{S}_{s_{LF}}$ are subjected to regression model fitting. Since the shape of the BC is unknown, and considering the number of possible solutions, in order to cover a variety of BC shapes, this fitting utilizes polynomials of degree $d$ $\in \{1,2,3,4\}$ for each $n = 1,2,\ldots,N$ design vector, forming the new regression model solution matrix $\mathbf{S}_{sreg}$ as:

\begin{equation}
	\mathbf{S}_{sreg} = 
	\begin{bmatrix}
		\mathbf{P}_1(\mathbf{x}^{opt}_{s_1}) \\
		\mathbf{P}_2(\mathbf{x}^{opt}_{s_1}) \\
		\mathbf{P}_3(\mathbf{x}^{opt}_{s_1}) \\
		\mathbf{P}_4(\mathbf{x}^{opt}_{s_1}) \\
		\vdots \\
		\mathbf{P}_1(\mathbf{x}^{opt}_{s_N}) \\
		\mathbf{P}_2(\mathbf{x}^{opt}_{s_N}) \\
		\mathbf{P}_3(\mathbf{x}^{opt}_{s_N}) \\
		\mathbf{P}_4(\mathbf{x}^{opt}_{s_N})
	\end{bmatrix}
	=
    \begin{bmatrix}
		\sum_{k=0}^{1} a_{k,1}^{opt_1} \hat{s}^k \\
		\sum_{k=0}^{2} a_{k,2}^{opt_1} \hat{s}^k \\
		\sum_{k=0}^{3} a_{k,3}^{opt_1} \hat{s}^k \\
		\sum_{k=0}^{4} a_{k,4}^{opt_1} \hat{s}^k \\
		\vdots \\
		\sum_{k=0}^{1} a_{k,1}^{opt_N} \hat{s}^k \\
		\sum_{k=0}^{2} a_{k,2}^{opt_N} \hat{s}^k \\
		\sum_{k=0}^{3} a_{k,3}^{opt_N} \hat{s}^k \\
		\sum_{k=0}^{4} a_{k,4}^{opt_N} \hat{s}^k
	\end{bmatrix}
	\label{eqn:reg_x_s_matrix}
\end{equation}

where $\mathbf{P}_4(\mathbf{x}^{opt}_{s_N})$ is the 4$^{th}$ degree polynomial of the optimized $N^{th}$ vector $\mathbf{x}^{opt}_{s_{N}}$. More specifically, the term $\sum_{k=0}^{4} a_{k,4}^{opt_N} \hat{s}^k$ represents the 4$^{th}$ degree polynomial regression model for the $N^{th}$ optimized design vector, where $a_{k,4}^{opt_N}$ are the polynomial regression coefficients and $\hat{s}$ is the unknown variable. As the matrix $\mathbf{S}_{sreg}$ is defined, each regression model is utilized to evaluate the HF discretized space ($I=80$) with equally spaced points between 0 and 1, resulting in the final $\mathbf{S}_{s_{HF}}$ matrix:

\begin{equation}
	\mathbf{S}_{s_{HF}} = 
	\begin{bmatrix}
		s_{1,1}^{opt_1,1} & s_{1,2}^{opt_1,1} & \ldots & s_{1,80}^{opt_1,1} \\
		s_{1,1}^{opt_1,2} & s_{1,2}^{opt_1,2} & \ldots & s_{1,80}^{opt_1,2} \\
		s_{1,1}^{opt_1,3} & s_{1,2}^{opt_1,3} & \ldots & s_{1,80}^{opt_1,3} \\
		s_{1,1}^{opt_1,4} & s_{1,2}^{opt_1,4} & \ldots & s_{1,80}^{opt_1,4} \\
		\vdots & \vdots & \ddots & \vdots \\
		s_{N,1}^{opt_N,1} & s_{N,2}^{opt_N,1} & \ldots & s_{N,80}^{opt_N,1} \\
		s_{N,1}^{opt_N,2} & s_{N,2}^{opt_N,2} & \ldots & s_{N,80}^{opt_N,2} \\
		s_{N,1}^{opt_N,3} & s_{N,2}^{opt_N,3} & \ldots & s_{N,80}^{opt_N,3} \\
		s_{N,1}^{opt_N,4} & s_{N,2}^{opt_N,4} & \ldots & s_{N,80}^{opt_N,4}
	\end{bmatrix}
	\label{eqn:matrix_s_HF}
\end{equation}

where $s_{N,1}^{opt_N,4}$ is the first scalar value at the boundary of the HF domain obtained from the $N^{th}$ design vector using the 4$^{th}$ degree polynomial model, and $s_{N,80}^{opt_N,4}$ is the last scalar value obtained from the same regression model of the same degree. 

Polynomial regression coefficients are determined using the \textit{numpy 1.24.3} function \textit{polyfit} for 20 equally spaced points between 0 and 1 (which corresponds to the LF discretization), and this regression model, generated by the function \textit{poly1d}, is subsequently evaluated for the HF discretization ($I=80$) with equally spaced points between 0 and 1, resulting in the final $\mathbf{S}_{sreg}$ set.
Finally, to determine the new optimization boundaries for the SFR problem, the maximum scalar value is extracted from the flattened matrix $\mathbf{S}_{s_{HF}}$: $max(\mathbf{S}_{s_{HF}})$ defines the upper optimization boundary value for each dimension $I=80$, thus forming the new upper boundary $\mathbf{ub}_R$. For safety reasons, the lower boundary $\mathbf{lb}_R$ remains as specified in Eq. (\ref{eqn:scalar_field_boundaries}), i.e., $\mathbf{0}$.

\begin{figure}[htbp]
	\centering
	\includegraphics[width=0.75\linewidth]{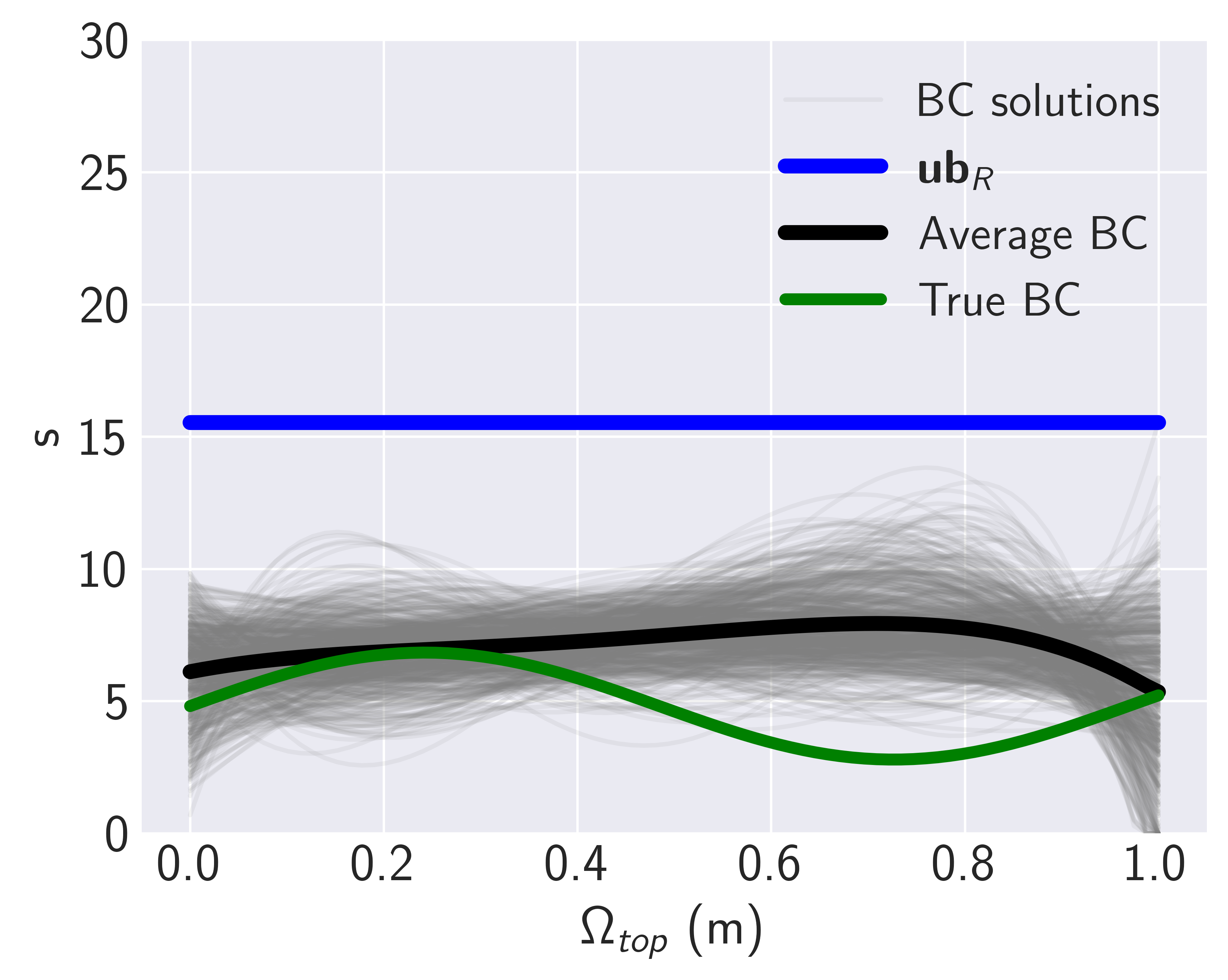}
	\caption{SFR boundary refinement example. The grey lines are the $N$ solutions within the matrix $\mathbf{S}_s$, the black line is the average value of the solutions in matrix $\mathbf{S}_s$, the green line is the true boundary condition that corresponds to the $\mathbf{T}$ field measurements, and  the blue line is the new upper boundary $\mathbf{ub}_R$  obtained with the boundary refinement procedure described in this section.}
	\label{fig:sfr_ssr_reduction_example}
\end{figure}

Fig. \ref{fig:sfr_ssr_reduction_example} illustrates an instance of the boundary refinement process. The green line represents the true solution, the blue line shows the reduction of the design space (approximately 50\% pruned) as described above and the black line represents the average of $N$ optimized boundary conditions for comparison. The green curve lies below the black curve, suggesting that the boundary refinement methodology used for the AID problem might be similarly effective here. 

\section{Results and Discussion}
\label{sec:results}

In this section, the results and analyses for the ML model, boundary refinement, and ML-enhanced framework for both demonstration problems are detailed. An in-depth hyperparameter analysis of the ML-enhanced framework is showcased, followed by overarching recommendations for optimal utilization. The section concludes by highlighting the advantages and limitations of the proposed technique. The details of all ML model hyperparameters, the hyperparameter tuning procedure, and the Python modules used are given in Appendix \ref{app:ml_hyperparams}.

\subsection{ML Models Results}
\label{sec:ml_model_results}
In this subsection, the ML model results for both the AID and SFR problems are presented through the accuracy metrics given in Sect. \ref{sec:mlmodels}.

\subsubsection{AID ML Model Results}

\begin{figure}[!h]
	\centering
	\includegraphics[width=0.75\linewidth]{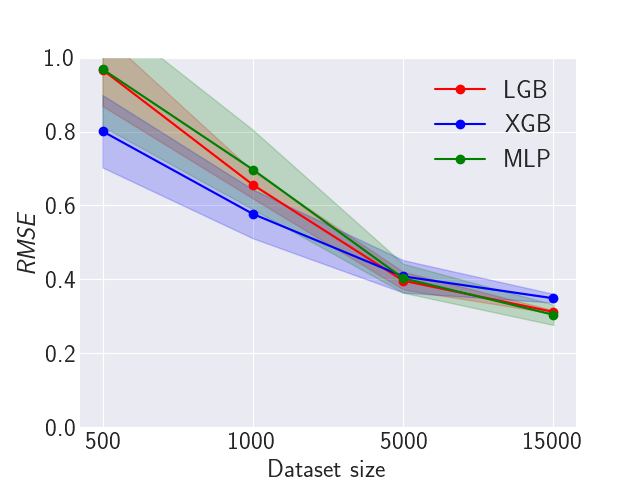}
	\caption{The mean (solid lines) and standard deviation (shaded areas) K-Fold $RMSE$ for the three investigated ML algorithms for the AID problem. Lower $RMSE$ values are better.}
	\label{fig:$RMSE$_aid}
\end{figure}

Fig. \ref{fig:$RMSE$_aid} presents the $RMSE$ scores for the three ML algorithms applied to the AID problem for varying dataset sizes. The dataset size was varied in order to assess the influence it has on the ML-enhanced framework, and to obtain the learning curve for each algorithm. All models show that the larger the dataset size, the better (lower) the resulting $RMSE$. It can be seen that for smaller datasets, XGB has  better  performance than LGB and MLP, but  its accuracy marginally lags when leveraging all 15000 data instances for training and cross-validation. Given its overall top performance, XGB (trained with four different dataset sizes) was selected as the ML model for the inverse design framework, and the K-fold cross-validation $RMSE$ values used for further analysis are presented in Table \ref{tab:$RMSE$_values_aid_omega}.

\begin{table}[h]
	\caption{XGB $RMSE$ values used for calculating the $\omega$ threshold parameter for each scenario and dataset size.}\label{tab:$RMSE$_values_aid_omega}
	\begin{tabular}{@{}ll@{}}
		\toprule
		Dataset size & $RMSE$ \\
		\midrule
		500  & 0.81 \\
		1000  & 0.61 \\
		5000  & 0.39 \\
		15000  & 0.34 \\
		\botrule
	\end{tabular}
	
\end{table}

\subsubsection{SFR ML Model Results}

\begin{figure}[!h]
	\centering
	\includegraphics[width=0.75\linewidth]{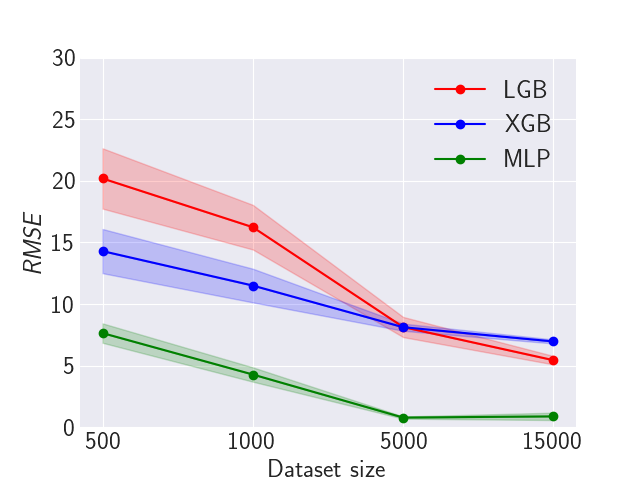}
	\caption{The mean (solid lines) and standard deviation (shaded areas)  K-Fold $RMSE$ for the three investigated ML algorithms for the SFR problem.}
	\label{fig:$RMSE$_sfr}
\end{figure}

Fig. \ref{fig:$RMSE$_sfr} shows the $RMSE$ values of the three ML algorithms when applied to the SFR problem using different  dataset sizes. The results  show the MLP's superior performance over both LGB and XGB across all dataset sizes. As a result, the MLP was selected  as the ML model within the inverse design framework for the SFR problem. The specific K-fold cross-validation $RMSE$ values for the MLP, which were used to compute the $\omega$ parameter for the SFR problem, are detailed in Table \ref{tab:$RMSE$_values_sfr_omega}. Given the minimal performance difference between the models trained with 5000 and 15000 instances, only three distinct models trained on three different amounts of data were compared  within the ML-enhanced framework.

\begin{table}[h]
	\caption{MLP $RMSE$ values used for calculating the $\omega$ threshold parameter for each of the two SFR BC scenarios and dataset size.}\label{tab:$RMSE$_values_sfr_omega}
	\begin{tabular}{@{}ll@{}}
		\toprule
		Dataset size & $RMSE$ \\
		\midrule
		500  & 7.63 \\
		1000  & 4.27 \\
		5000  & 0.78 \\
		\botrule
	\end{tabular}
	
\end{table}

\subsection{Boundary Refinement with the ML Model}
\label{sub:search_space_results}

In this subsection, the results of the boundary refinement technique for both investigated problems are presented. To generate the new boundaries $\mathbf{lb}_R$ and $\mathbf{ub}_R$, models trained on different dataset sizes  were compared. The DE was employed to solve Eq. (\ref{eqn:search_space_opt}) 150 times ($N=150$ solutions). The DE algorithm was configured with a maximum of 800 function evaluations, and the population size was set to equal the dimensionality of the optimization vector i.e. 20 for SFR and 30 for AID. A comparative analysis of different $N$ values is provided in Appendix \ref{app:ssr_convergence}.

\subsubsection{AID Boundary Refinement}

\begin{figure*}[]
	\centering
	\begin{subfigure}{0.45\textwidth}
		\centering
		\includegraphics[width=\linewidth]{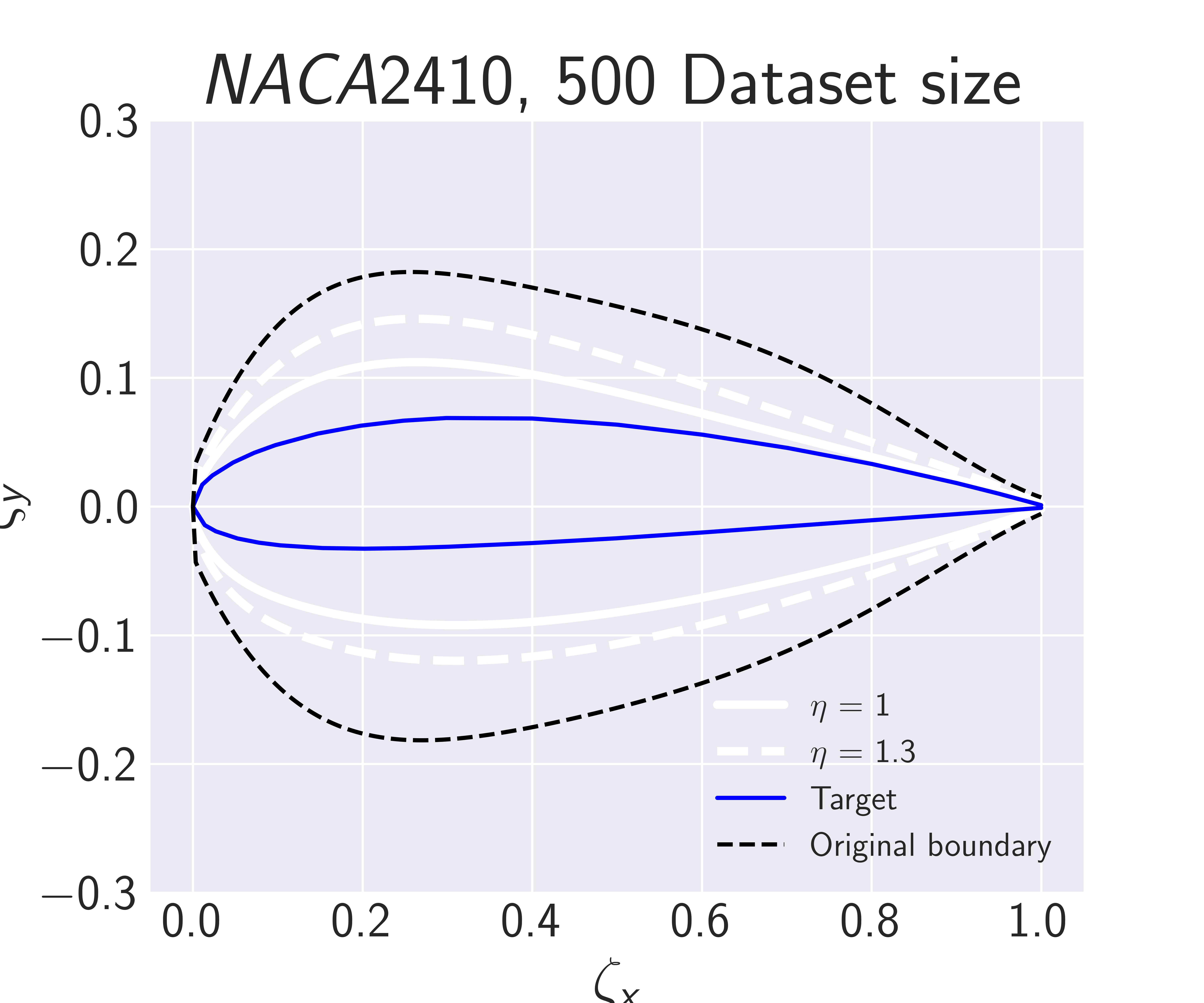}
		\caption{}
		\label{fig:naca2410_500ssr}
	\end{subfigure}%
	\hfill
	\begin{subfigure}{0.45\textwidth}
		\centering
		\includegraphics[width=\linewidth]{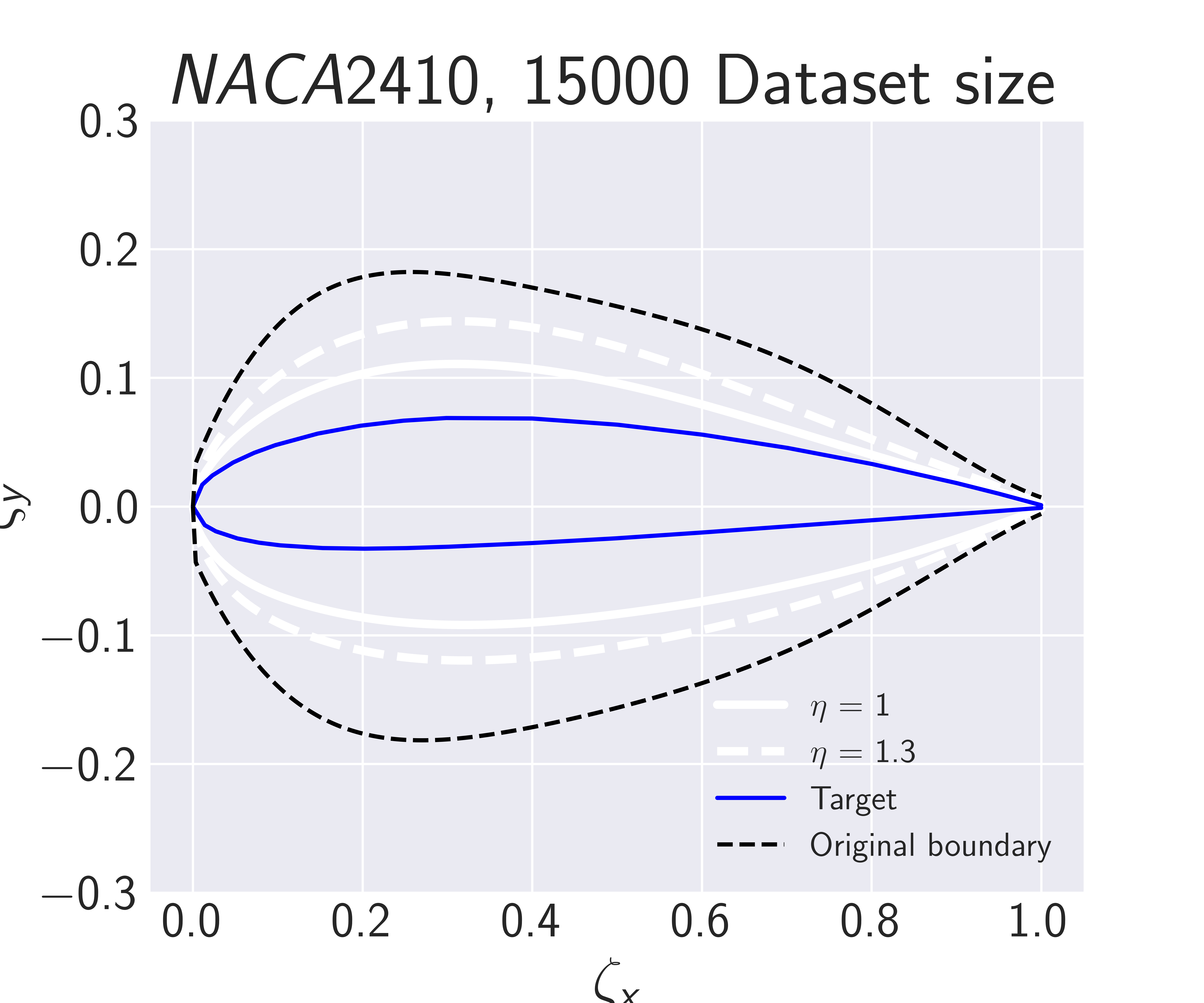}
		\caption{}
		\label{fig:naca2410_15000ssr}
	\end{subfigure}
	
	\begin{subfigure}{0.45\textwidth}
		\centering
		\includegraphics[width=\linewidth]{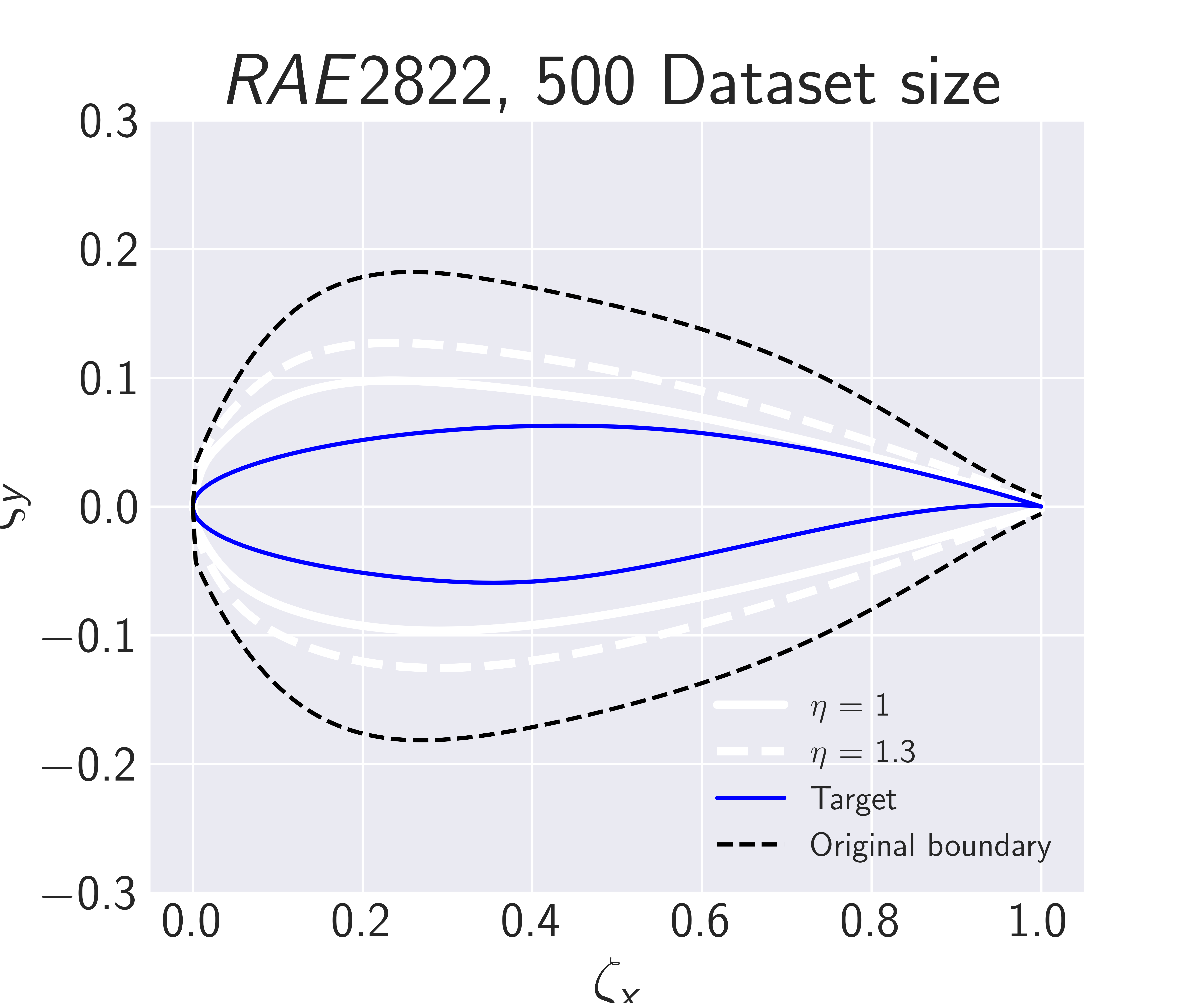}
		\caption{}
		\label{fig:rae_2822_500ssr}
	\end{subfigure}%
	\hfill
	\begin{subfigure}{0.45\textwidth}
		\centering
		\includegraphics[width=\linewidth]{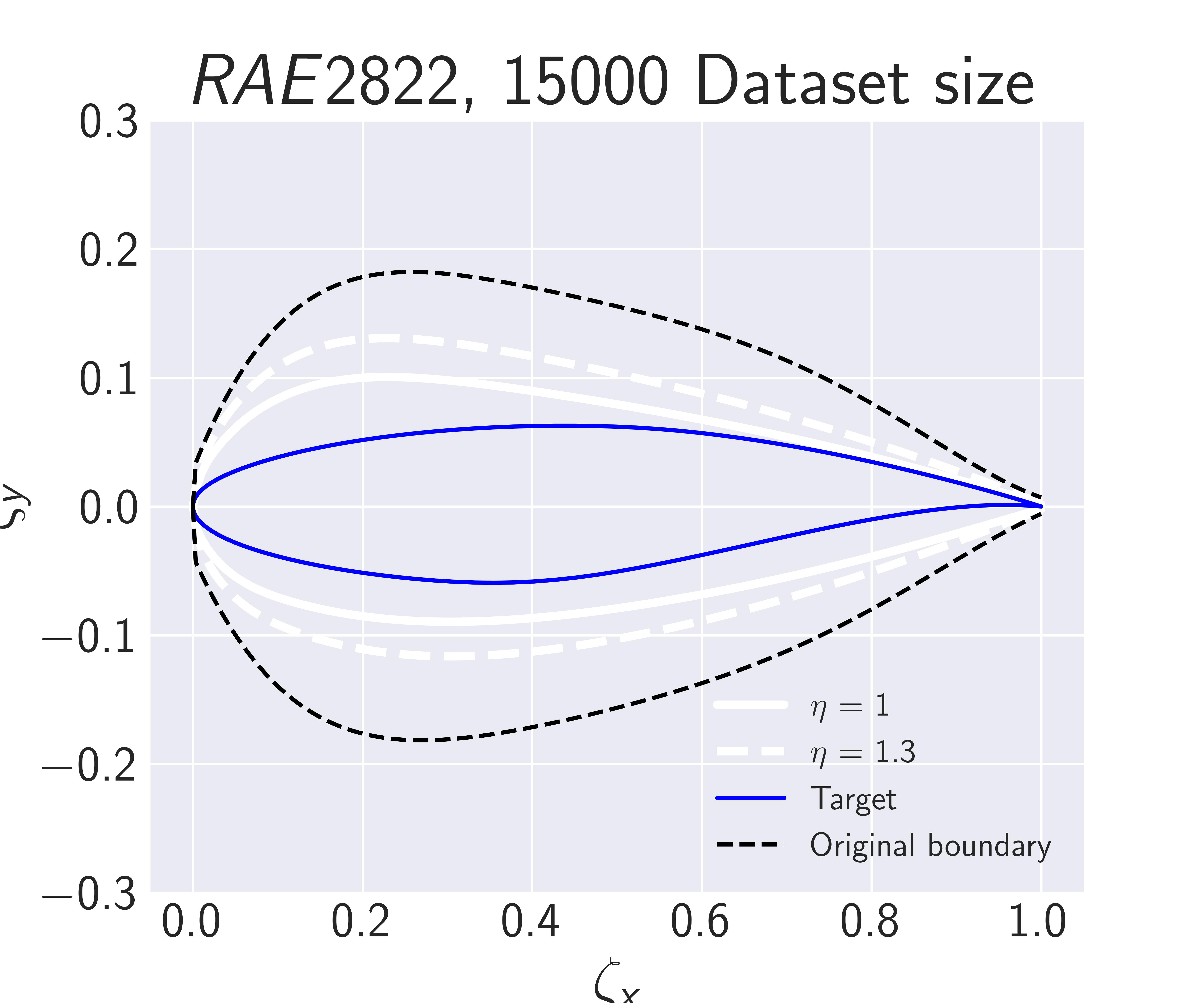}
		\caption{}
		\label{fig:rae_2822_15000ssr}
	\end{subfigure}
	
	\caption{Boundary refinement technique results generated by the XGB model: (a) $NACA2410$ with  dataset size 500 (b) $NACA2410$ with  dataset size 15000 (c) $RAE2822$ with  dataset size 500 (d) $RAE2822$ with dataset size 15000. The white lines indicate the narrowed boundaries, i.e. $\mathbf{lb}_R$ and $\mathbf{ub}_R$ for different values of $\eta$. There is no noticeable difference between the white lines when the dataset size is increased, indicating that a significant boundary compression is achievable even with an  ML model trained on a small  dataset.}
	\label{fig:airfoils_ssr_comparison}
\end{figure*}

In Sect. \ref{sec:ml_model_results}, the choice of the XGB algorithm was justified by its marginal superiority over other algorithms, especially when various training data sizes are taken into account. For the purpose of boundary refinement, the XGB was trained using data instances of sizes 500, 1000, 5000, and 15000. The results of the edge cases of the XGB-produced boundaries $\mathbf{lb}_R$ and $\mathbf{ub}_R$ are illustrated in Fig. \ref{fig:airfoils_ssr_comparison}. Since every solution in the matrix $\mathbf{S}_A$ represents an airfoil itself with lower and upper shape coefficients, the new lower and upper boundaries were derived solely by averaging the solution matrix $\mathbf{S}_A$ where $\eta=1$ encompass the genuine target designs. A notable overlap is observed in a section of the upper trailing edge between the target and the new boundary ($\zeta_x > 0.8$). When the safety factor is increased to its maximum investigated value of $\eta=1.3$, this overlap at $\zeta_x > 0.8$ significantly diminishes, and a noticeable distinction is achieved between the new and the original boundaries (Fig. \ref{fig:naca2410_15000ssr} and Fig. \ref{fig:rae_2822_15000ssr}).

When training the XGB model with different numbers of instances, only minor variations in results emerge. This suggests that the ML model trained with a small dataset suffices to prune a segment of the design space for such problems. This observation holds for both $NACA2410$ and $RAE2822$ boundary refinement procedures, as illustrated in Fig. \ref{fig:airfoils_ssr_comparison}. Finally, an analysis of how the number of solutions $N$ effects the change in the airfoil shape and the boundary refinement is shown in Fig. \ref{fig:ssr_convergence_airfoil} (Appendix \ref{app:ssr_convergence}).

\subsubsection{SFR Boundary Refinement}

\begin{figure}[H]
	\centering
	\begin{subfigure}[b]{0.45\textwidth}
		\includegraphics[width=1\linewidth]{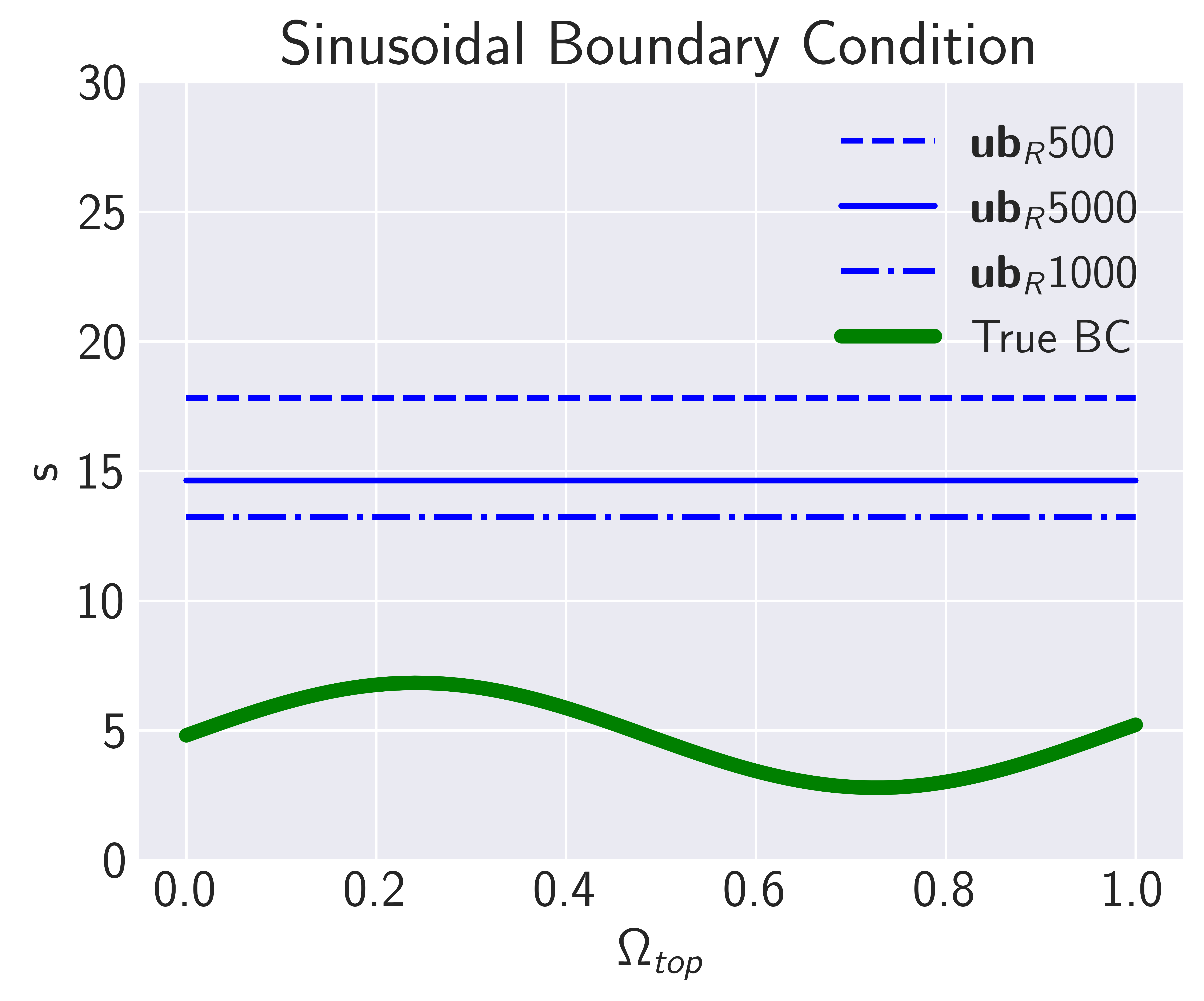}
		\caption{}
		\label{fig:bc1_ssr} 
	\end{subfigure}
	\begin{subfigure}[b]{0.45\textwidth}
		\includegraphics[width=1\linewidth]{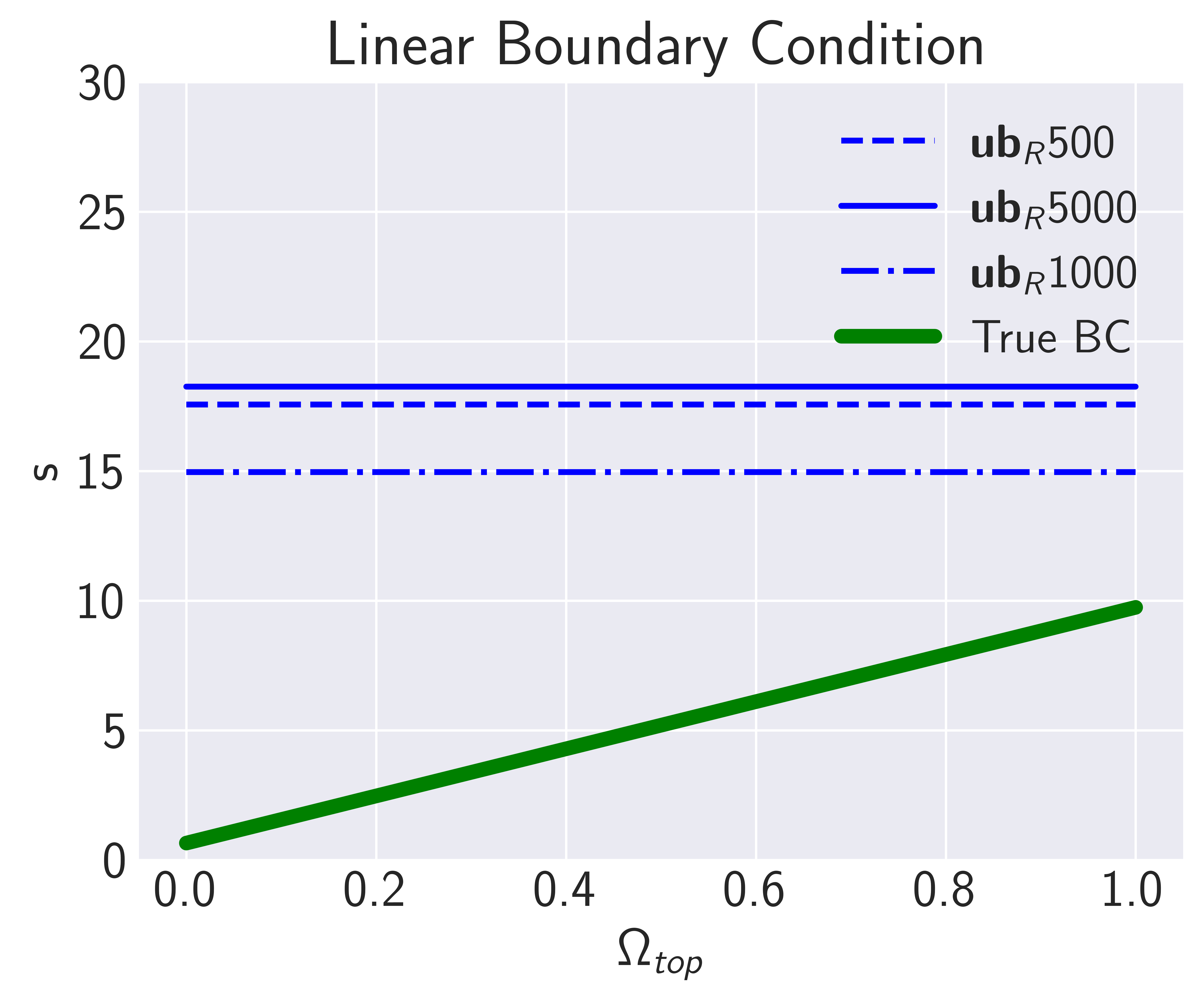}
		\caption{}
		\label{fig:bc2_ssr}
	\end{subfigure}
	\caption[Two bcs]{The boundary refinement technique developed for the SFR problem implemented on two distinct cases: (a) Sinusoidal BC (b) Linear BC. The green line represents the BC utilized to derive the HF target solution, while the blue lines indicate the upper boundaries of the design space produced using varied dataset sizes for training the MLP.}
	\label{fig:ssr_for_bcs}
\end{figure}

For the SFR problem, the MLP outperformed the other investigated algorithms in modeling $s_{max}$. Fig. \ref{fig:ssr_for_bcs} displays the MLP results of the boundary refinement. The MLP was trained with dataset sizes  500, 1000, and 5000. Across both BC scenarios, all three MLP models significantly reduce the size of the design space, confining the $\mathbf{ub}_R$ value between $s=13$ and $s=18$ (43\% to 60\% of the design space pruned). Compared to the airfoil problem, these newly produced boundaries exhibit greater sensitivity to changes in dataset size, but  all three can be reliably incorporated into the ML-enhanced inverse design framework without losing the true solutions. Additionally, Fig. \ref{fig:ssr_convergence_bcs} (Appendix \ref{app:ssr_convergence}) presents an analysis of how the number of solutions, $N$, impacts both cases of the SFR problem.

\subsection{ML-Enhanced Inverse Design Framework Results}

This section provides a comprehensive analysis of the ML-enhanced inverse design framework, detailing the hyperparameters ($c$ values, $\eta$, ML dataset size) for both problem categories. Following this, a meticulous comparison is presented between the conventional inverse design approach which employs classic optimization algorithms like DE and PSO, and the ML-enhanced optimization algorithms. Both strategies aim to minimize the objective function described in Eq. (\ref{eqn:obj_function})  subject to the  constraints specified in Eq. (\ref{eqn:bspline_variables}) and Eq. (\ref{eqn:scalar_field_boundaries}) for the AID and SFR challenges, respectively. Given the stringent computational budget, both methodologies are restricted to 200 HF simulations for each problem. Furthermore, to account for the inherent randomness of the population-based algorithms in use, all hyperparameter combinations are subjected to 30 runs, facilitating a robust uncertainty analysis. The term fitness is introduced to align with the conventions of PSO and DE, and it is equal to $RMSE$, which is the optimization objective used to evaluate the quality of solutions.

Within the ML-enhanced inverse design framework, the boundaries resulting from the boundary refinement techniques are utilized, i.e.,  when the ML model is trained with a particular dataset size,  the $\mathbf{lb}_R$ and $\mathbf{ub}_R$ corresponding to  that  ML model are applied. The user-defined hyperparameter $c$ is utilized to scale the K-fold cross-validation $RMSE$ values of the ML models. For the AID problem, the explored  values are $c \in \{1, 2, 4, 6, 8\}$, while for the SFR problem, they are $c \in \{0.25, 0.5, 1, 2, 4\}$. The differing ranges for $c$ between the two problems arise from the variance in magnitude of their $RMSE$ values. However, there is an overlap in the sets, which aids in formulating a generalized recommendation. The $RMSE$ metric of the ML models was used to calculate the $\omega$ threshold as defined in Sect. \ref{sub:ml_framework}. This decision is motivated by the intuitiveness and interpretability offered by the $RMSE$ value. By reflecting the degree of discrepancy in the model's predictions, it provides a clear and meaningful measure of the model's performance.

\subsubsection{AID Results}
\label{sec:aid_results}
\begin{figure*}[!htb]
	\centering
	\begin{subfigure}{0.45\textwidth}
		\centering
		\includegraphics[width=\linewidth]{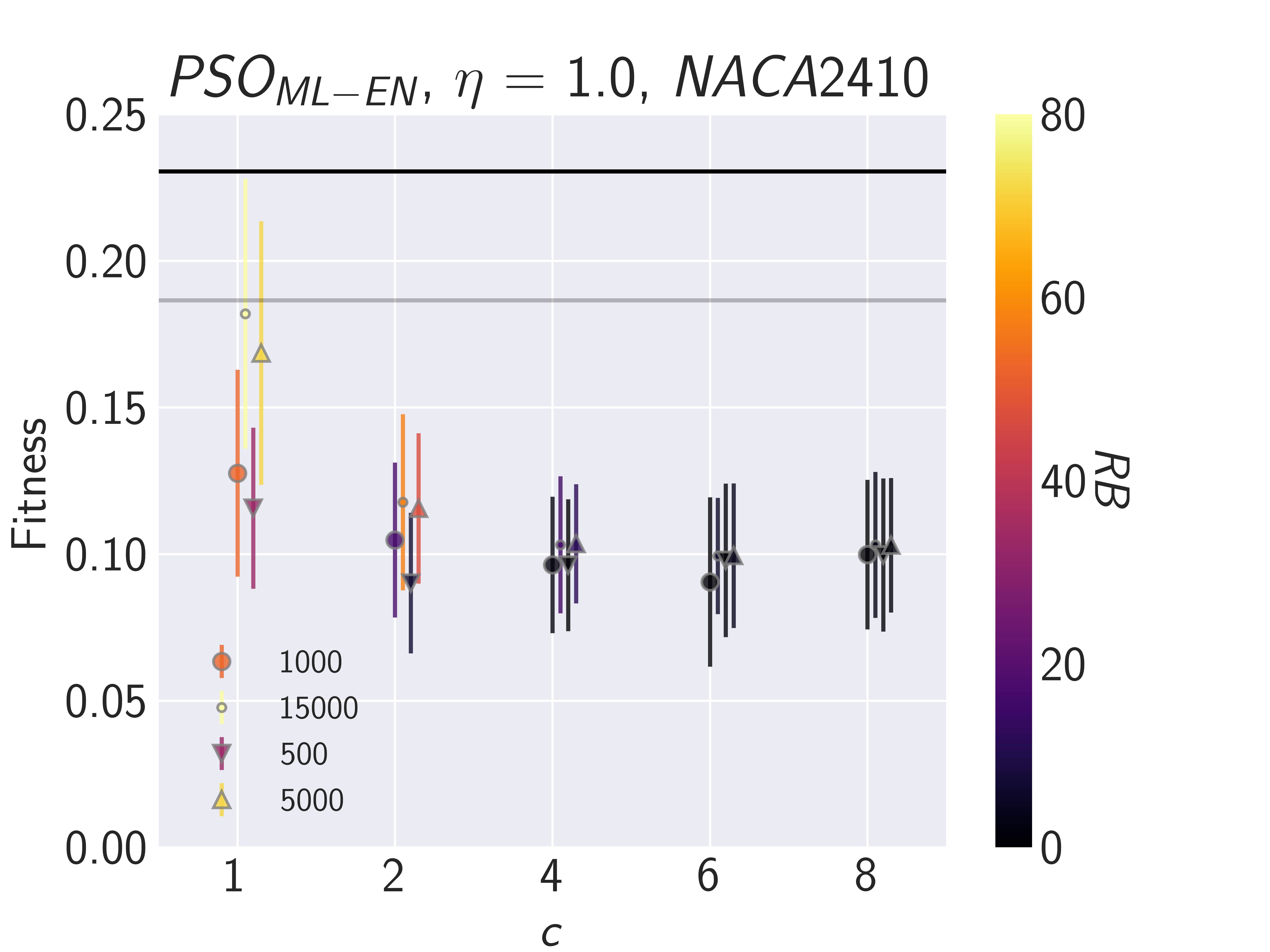}
		\caption{}
		\label{fig:naca2410_1_pso}
	\end{subfigure}%
	\hfill
	\begin{subfigure}{0.45\textwidth}
		\centering
		\includegraphics[width=\linewidth]{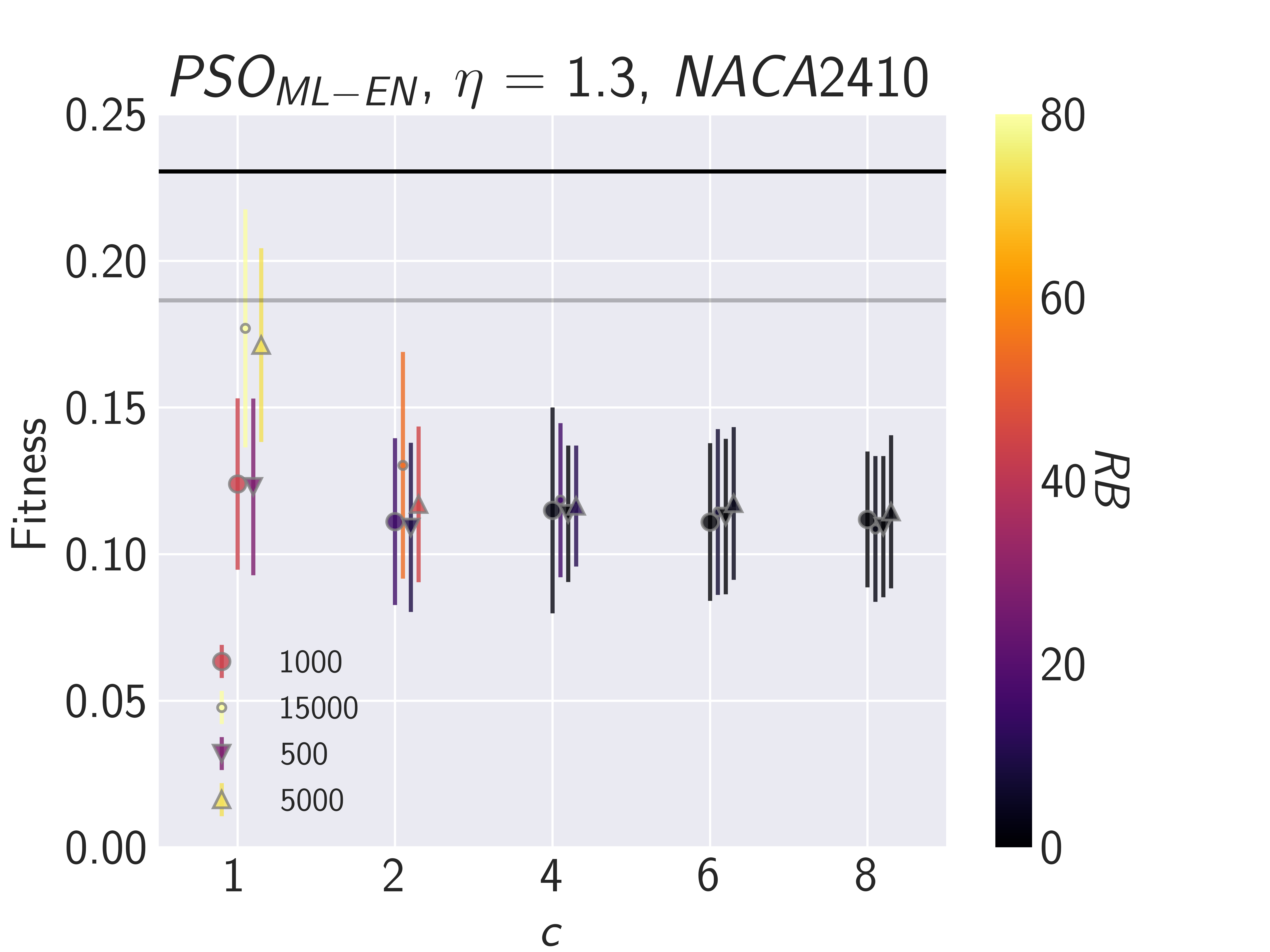}
		\caption{}
		\label{fig:naca2410_1_3_pso}
	\end{subfigure}
	
	\begin{subfigure}{0.45\textwidth}
		\centering
		\includegraphics[width=\linewidth]{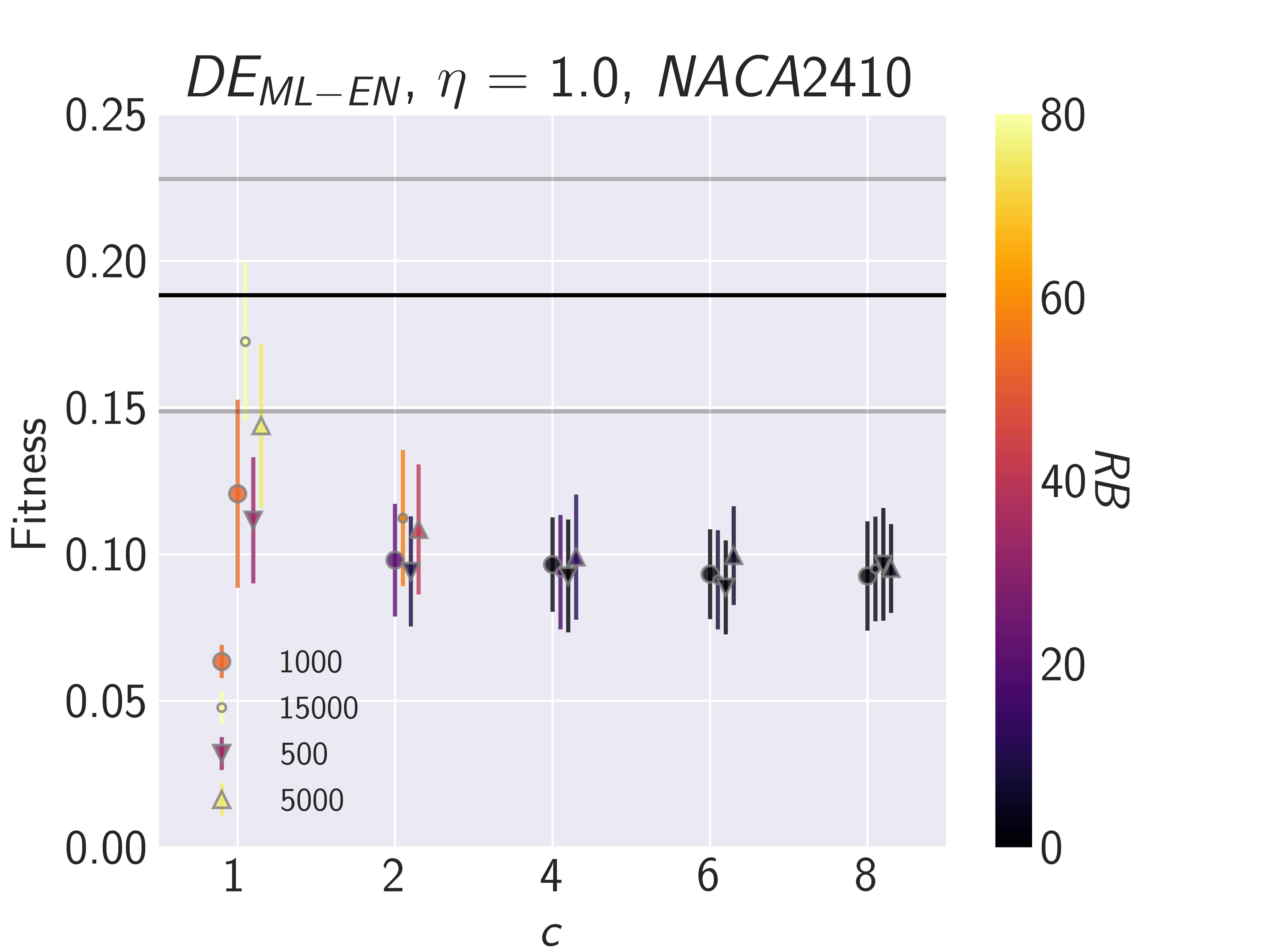}
		\caption{}
		\label{fig:naca2410_1_de}
	\end{subfigure}%
	\hfill
	\begin{subfigure}{0.45\textwidth}
		\centering
		\includegraphics[width=\linewidth]{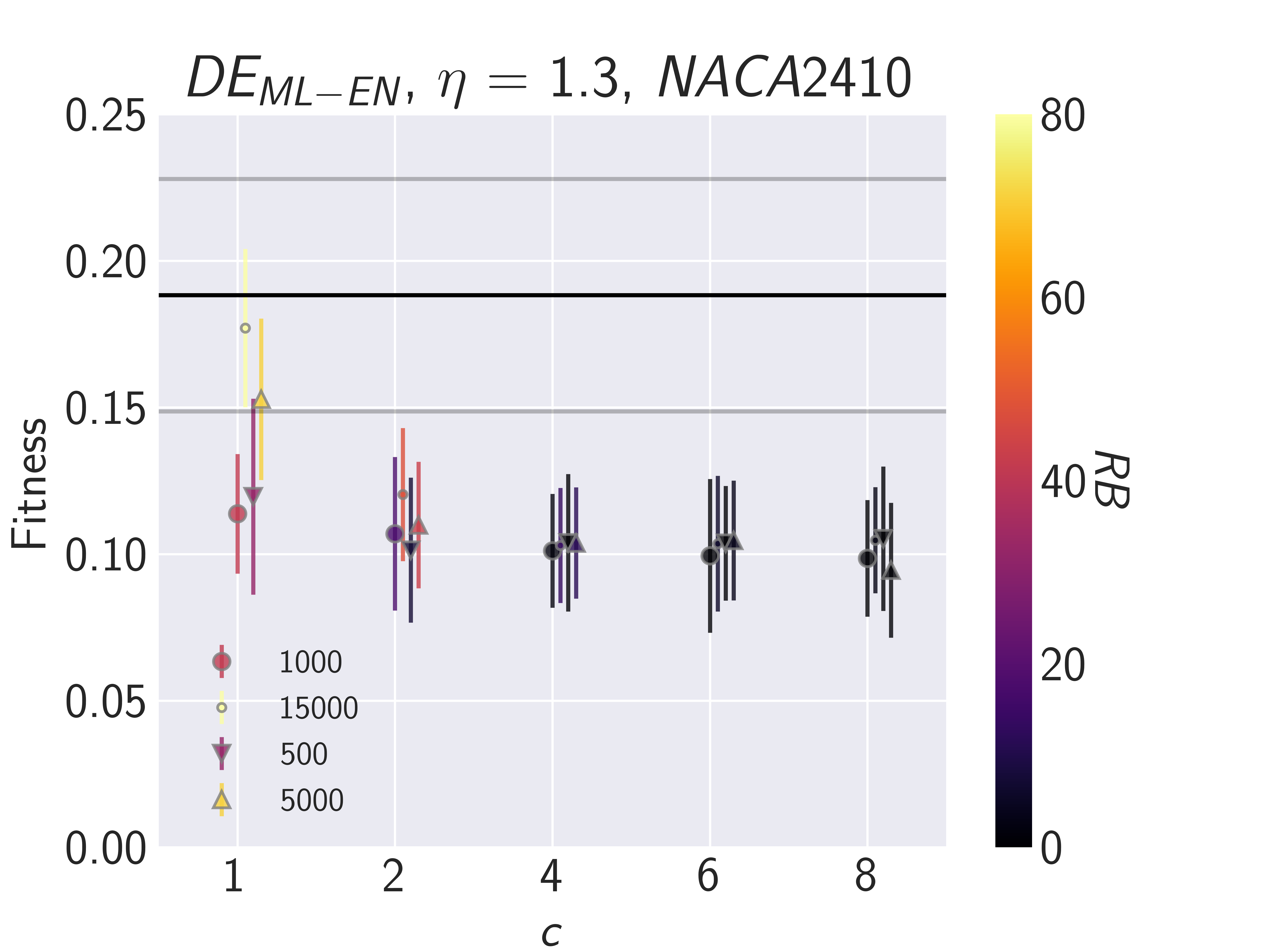}
		\caption{}
		\label{fig:naca2410_1_3_de}
	\end{subfigure}
	
	\caption{Results from the ML-enhanced framework for $NACA2410$ that include the boundary refinement for the following configurations: (a) $PSO_{ML-EN}$, $\eta$ = 1 (b) $PSO_{ML-EN}$, $\eta$ = 1.3 (c) $DE_{ML-EN}$, $\eta$ = 1
		(d) $DE_{ML-EN}$, $\eta$ = 1.3. The markers denote the different dataset sizes used to train the ML model, while the coloring of the markers represents the remaining simulation budget ($RB$) values. A higher $RB$ signifies greater savings in the computational budget (less requirements for HF simulations), while low fitness values imply a better approximation of the target performance. The markers are slightly offset for each $c$ value to improve visibility.}
	\label{fig:naca2410_pso_de_hyperparameters}
\end{figure*}

\begin{figure*}[!htb]
	\centering
	\begin{subfigure}{0.45\textwidth}
		\centering
		\includegraphics[width=\linewidth]{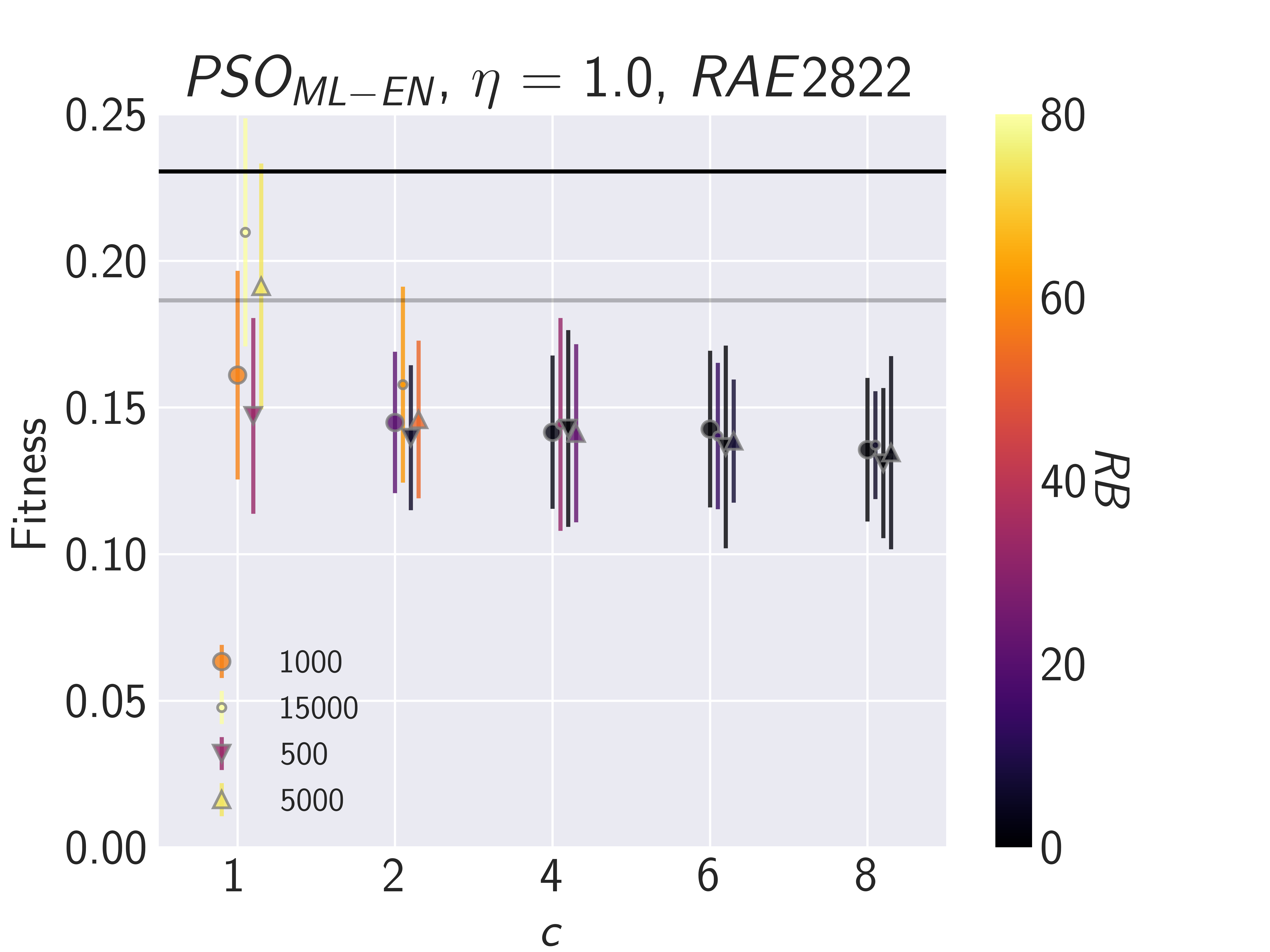}
		\caption{}
		\label{fig:rae2822_1_pso}
	\end{subfigure}%
	\hfill
	\begin{subfigure}{0.45\textwidth}
		\centering
		\includegraphics[width=\linewidth]{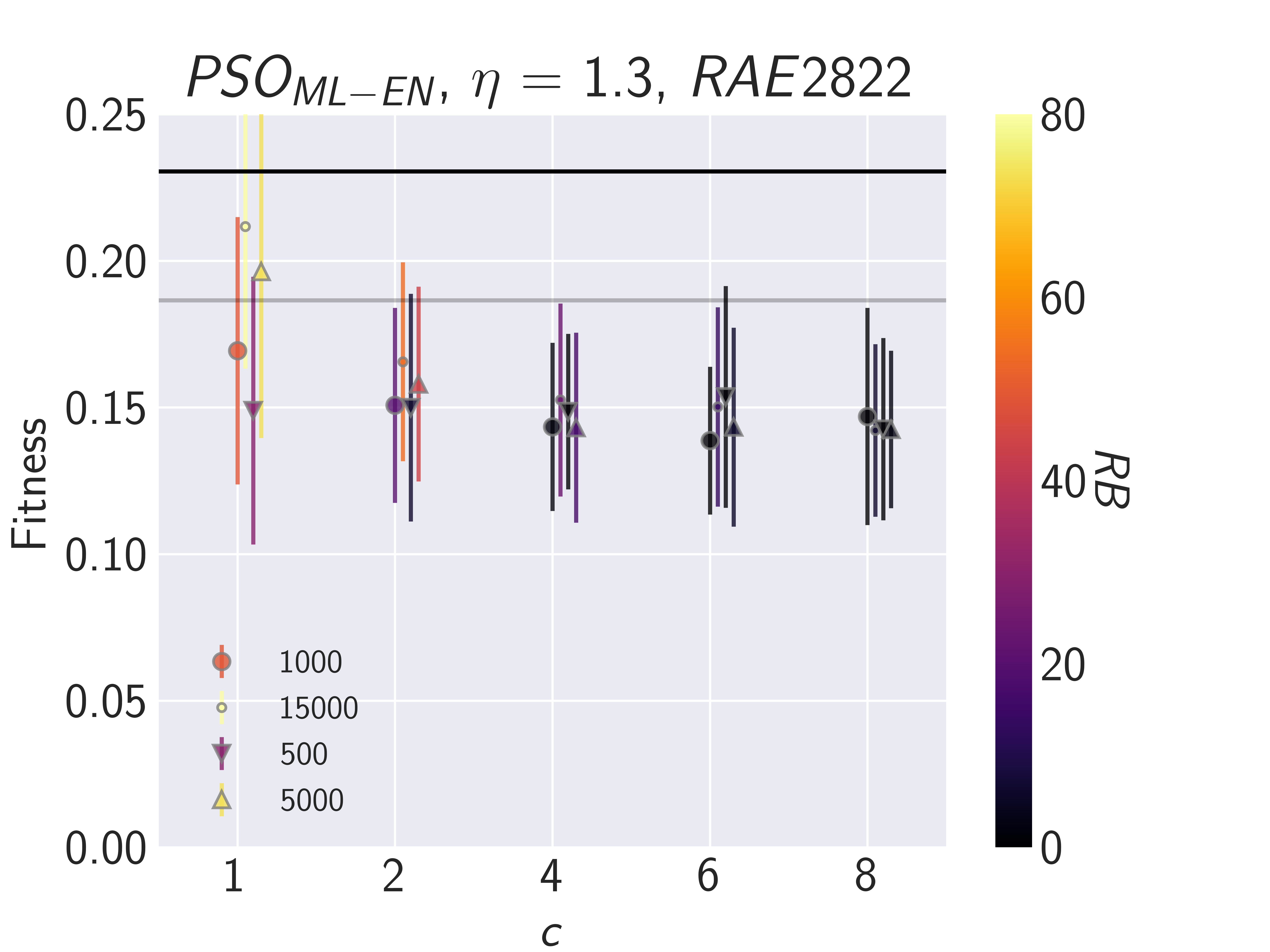}
		\caption{}
		\label{fig:rae2822_1_3_pso}
	\end{subfigure}
	
	\begin{subfigure}{0.45\textwidth}
		\centering
		\includegraphics[width=\linewidth]{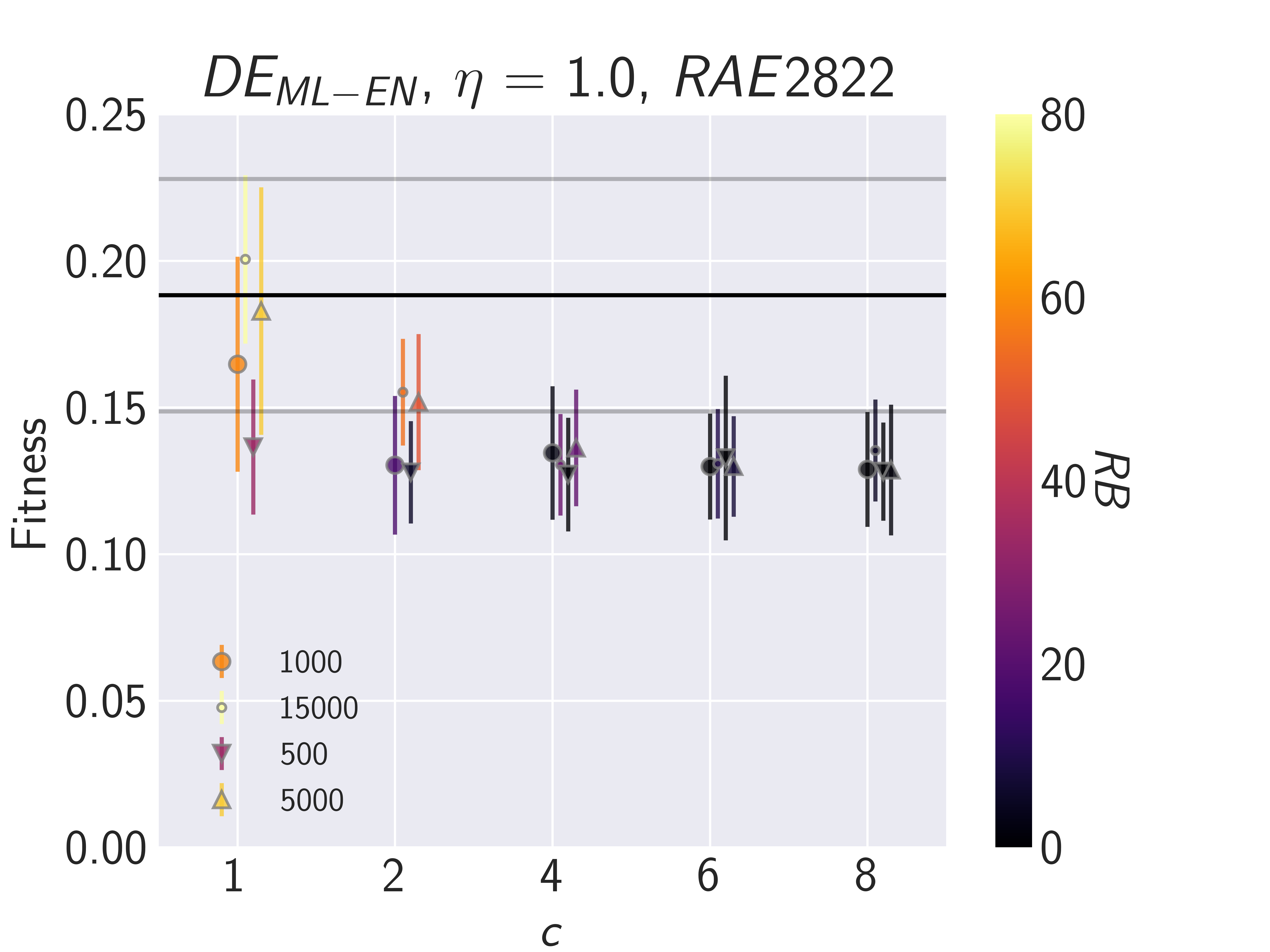}
		\caption{}
		\label{fig:rae2822_1_de}
	\end{subfigure}%
	\hfill
	\begin{subfigure}{0.45\textwidth}
		\centering
		\includegraphics[width=\linewidth]{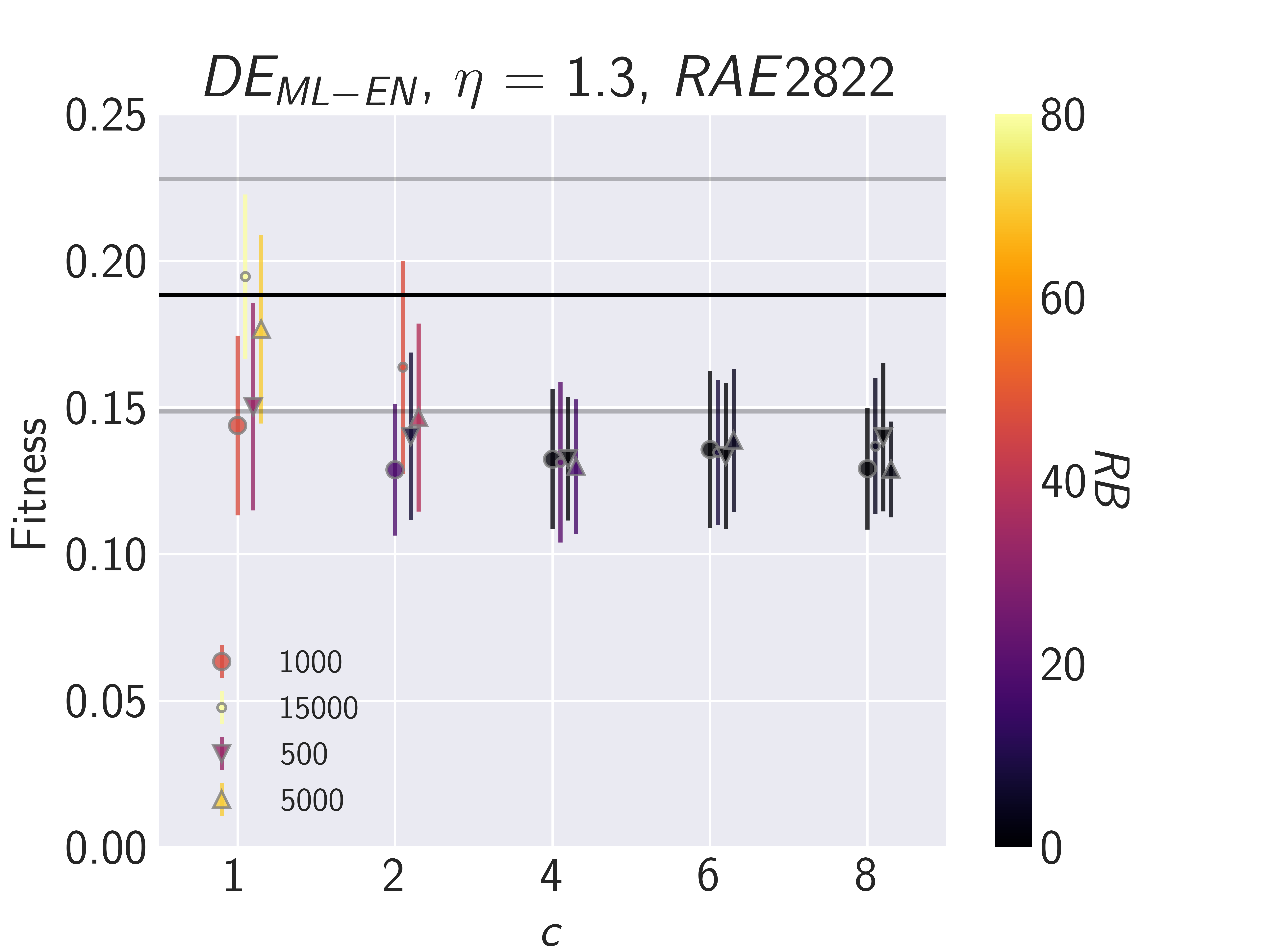}
		\caption{}
		\label{fig:rae2822_1_3_de}
	\end{subfigure}
	
	\caption{Results from the ML-enhanced framework for $RAE2822$ that include the boundary refinement for the following configurations: (a) $PSO_{ML-EN}$, $\eta$ = 1 (b) $PSO_{ML-EN}$, $\eta$ = 1.3 (c) $DE_{ML-EN}$, $\eta$ = 1
		(d) $DE_{ML-EN}$, $\eta$ = 1.3. The markers denote the different dataset sizes used to train the ML model, while the coloring of the markers represents the remaining simulation budget ($RB$) values. A higher $RB$ signifies greater savings in the computational budget, while low fitness values imply a better approximation of the target performance. The markers are slightly offset for each $c$ value to improve visibility.}
	\label{fig:rae2822_pso_de_hyperparameters}
\end{figure*}

The results of the ML-enhanced inverse design framework utilizing the XGB model and the boundary refinement technique ($\eta$ = 1 and $\eta$ = 1.3) applied to the AID problem for the $NACA2410$ and $RAE2822$ airfoils are shown in Fig. \ref{fig:naca2410_pso_de_hyperparameters} and Fig.  \ref{fig:rae2822_pso_de_hyperparameters}. $PSO_{ML-EN}$ and $DE_{ML-EN}$ denote the ML-enhanced versions of the optimization algorithms. For comparison, the average and standard deviation over the results of 30 runs  of the unehannced PSO and DE algorithms are shown as horizontal black and grey lines, respectively. The markers indicate the dataset size used to train the XGB model, which was then incorporated into the ML-enhanced optimization algorithm and used to form $\mathbf{lb}_R$ and $\mathbf{ub}_R$. These markers are color-coded based on the $RB$ values, implying that the fitness values were obtained from a number of HF simulations defined as $TSB - RB$, where $TSB$ represents the total simulation budget, specifically set at 200. The full results which include the $\mathbf{lb}_R$ and $\mathbf{ub}_R$ formed with $\eta$  = 1.1 and $\eta$ = 1.2 are given in Fig. \ref{fig:naca2410_pso_de_hyperparameters_appendix} and Fig. \ref{fig:rae2822_pso_de_hyperparameters_appendix} (Appendix \ref{app:aid_full_analysis}), respectively.

First, a clear observation is that the DE algorithm, in both its unenhanced and ML-enhanced forms, outperforms the PSO algorithm. Moreover, across most tested hyperparameters, airfoil types, and optimization algorithms, the ML-enhanced variant consistently surpasses the performance of its unenhanced counterpart. There are a few instances where DE or PSO exhibit competitive performance in terms of raw fitness ($RMSE$) value, particularly when the $c$ value is set to 1 and the XGB models trained with dataset sizes of 5000 and 15000 are employed. However, note that both $PSO_{ML-EN}$ and $DE_{ML-EN}$ have consumed only about 60\% of their HF simulation budgets (remaining budget $RB\sim70-80$), whereas their unenhanced versions have fully exhausted theirs.

Once the user defined $RMSE$ scaling parameter $c$ value reaches and exceeds 4, the $RB$ value becomes zero for most dataset sizes and $\eta$ values. Given that the $RB$ value is zero, it indicates that only HF simulations were utilized for assessing the design vector. Consequently, it can be inferred that, in this particular scenario, employing unenhanced algorithms alongside the refined boundaries would yield equivalent results. ML-enhanced algorithms, especially when employing models trained on dataset sizes of 5000 and 15000 and when $c$ = 2 (observable in Fig. \ref{fig:naca2410_pso_de_hyperparameters} and Fig. \ref{fig:rae2822_pso_de_hyperparameters}), not only converge to a better solution but also economize on the total HF computational budget ($RB\sim30-50$) when compared with the unenhanced versions. 

For a general recommendation on the use of ML-enhanced optimization algorithms for the AID problem within a limited HF computational budget, any of the investigated $\eta$ factors can be employed. However, to ensure the target design falls within the refined boundaries, an $\eta$ value of 1.3 is preferable. This choice allows for  convergence  across all configurations. In terms of achieving optimal fitness and conserving the computational budget, the $c$ value of 2 appears to be the best across all dataset sizes and algorithm combinations. Furthermore, a $c$ value of 1 can be considered for exploratory inverse designs, as it requires fewer HF simulation runs to attain comparable or superior results to the unenhanced algorithms.

The ML models trained on smaller datasets  (500 and 1000) suffice to expedite the  inverse design process, achieving  ($RB\sim20-50$) for $c$ = 1 and $c$ = 2. These ML models also lead to effective boundary refinement when  the entire simulation budget is used up in pursuit of the optimal design.

\begin{figure*}[ht]
	\centering
	\begin{subfigure}{0.3\textwidth}
		\centering
		\includegraphics[width=\linewidth]{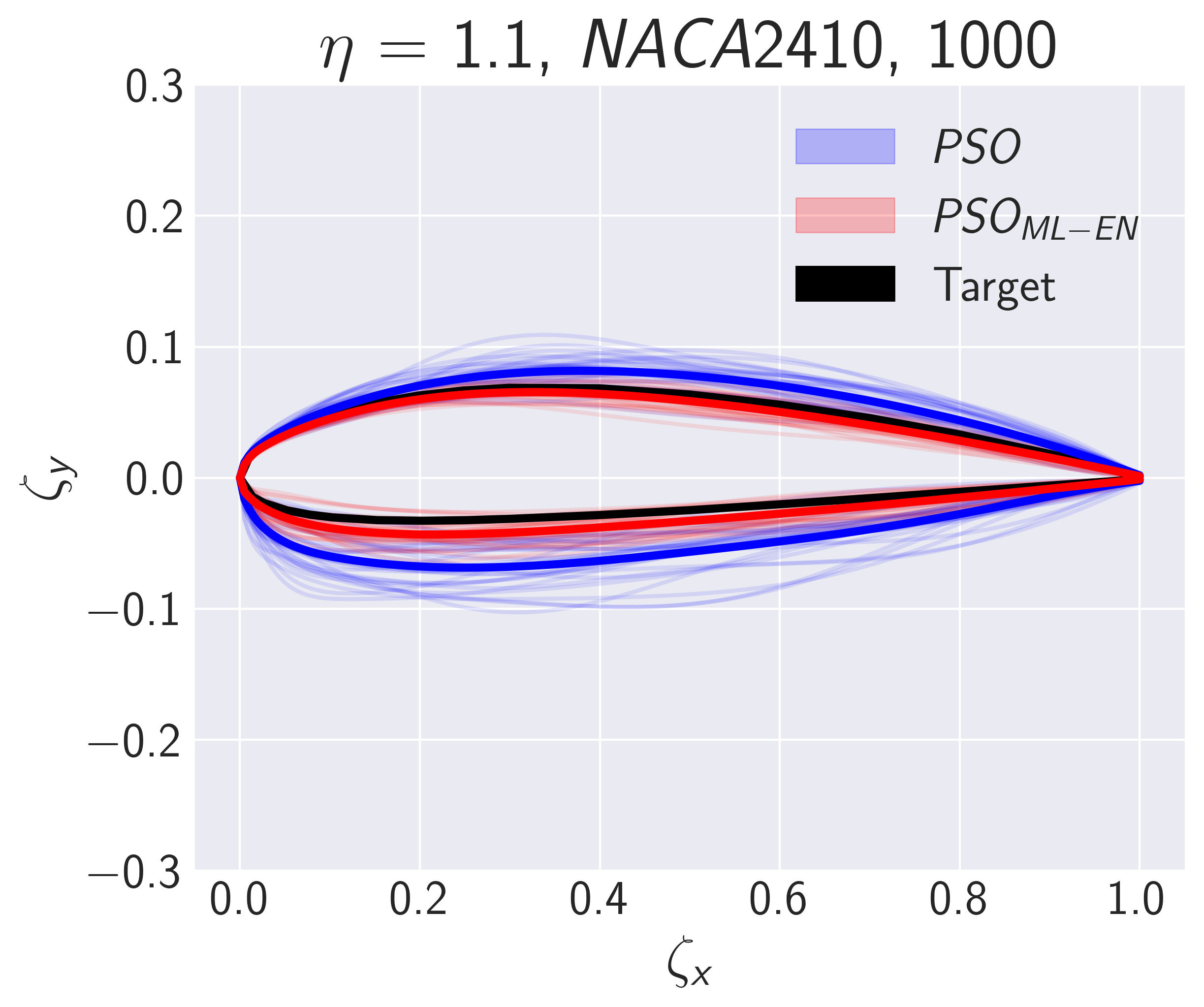}
		\caption{}
		\label{fig:enhpso_naca_best_design_comparison}
	\end{subfigure}%
	\hfill
	\begin{subfigure}{0.3\textwidth}
		\centering
		\includegraphics[width=\linewidth]{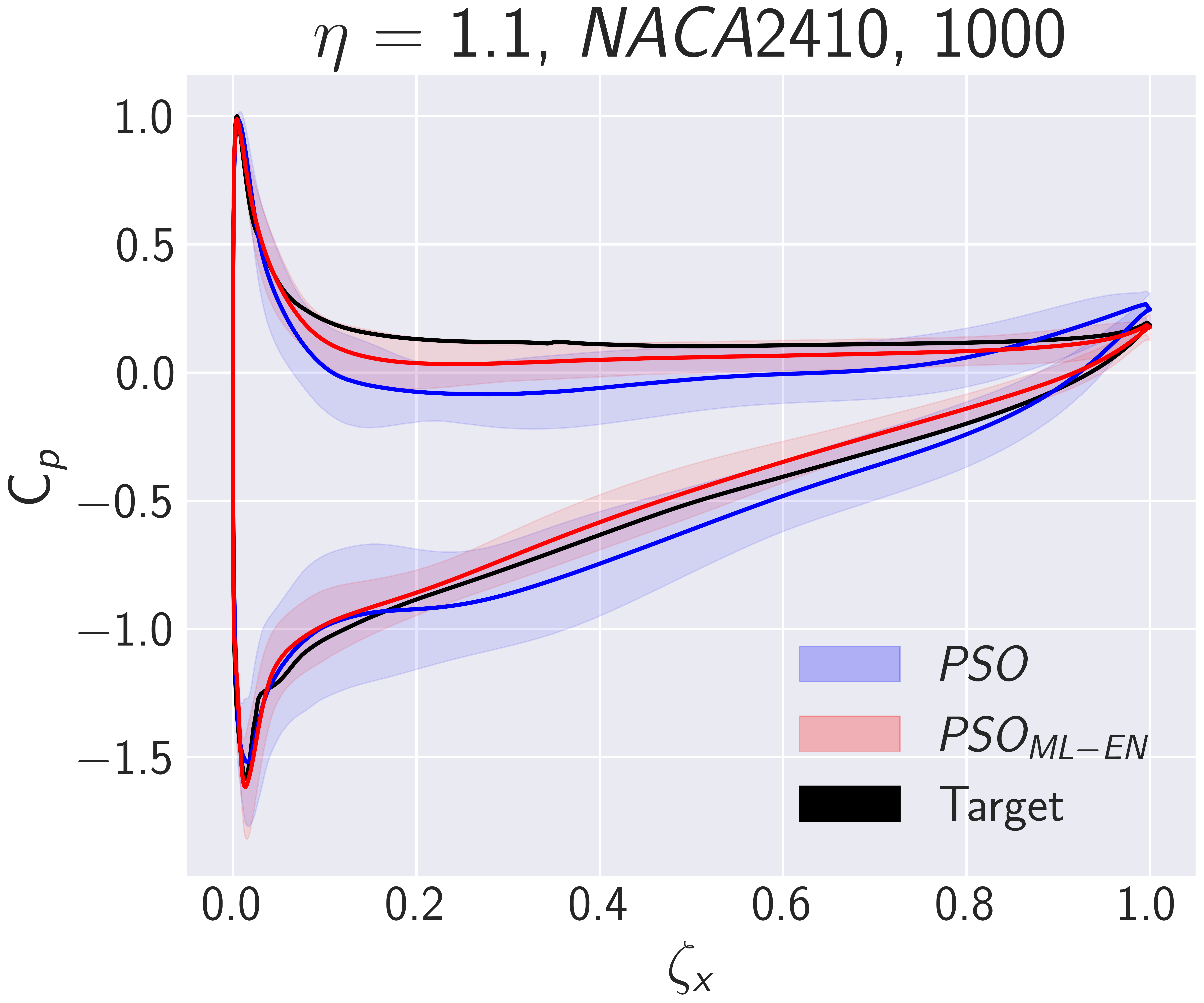}
		\caption{}
		\label{fig:enhpso_naca_pressure_comparison}
	\end{subfigure}%
	\hfill
	\begin{subfigure}{0.3\textwidth}
		\centering
		\includegraphics[width=\linewidth]{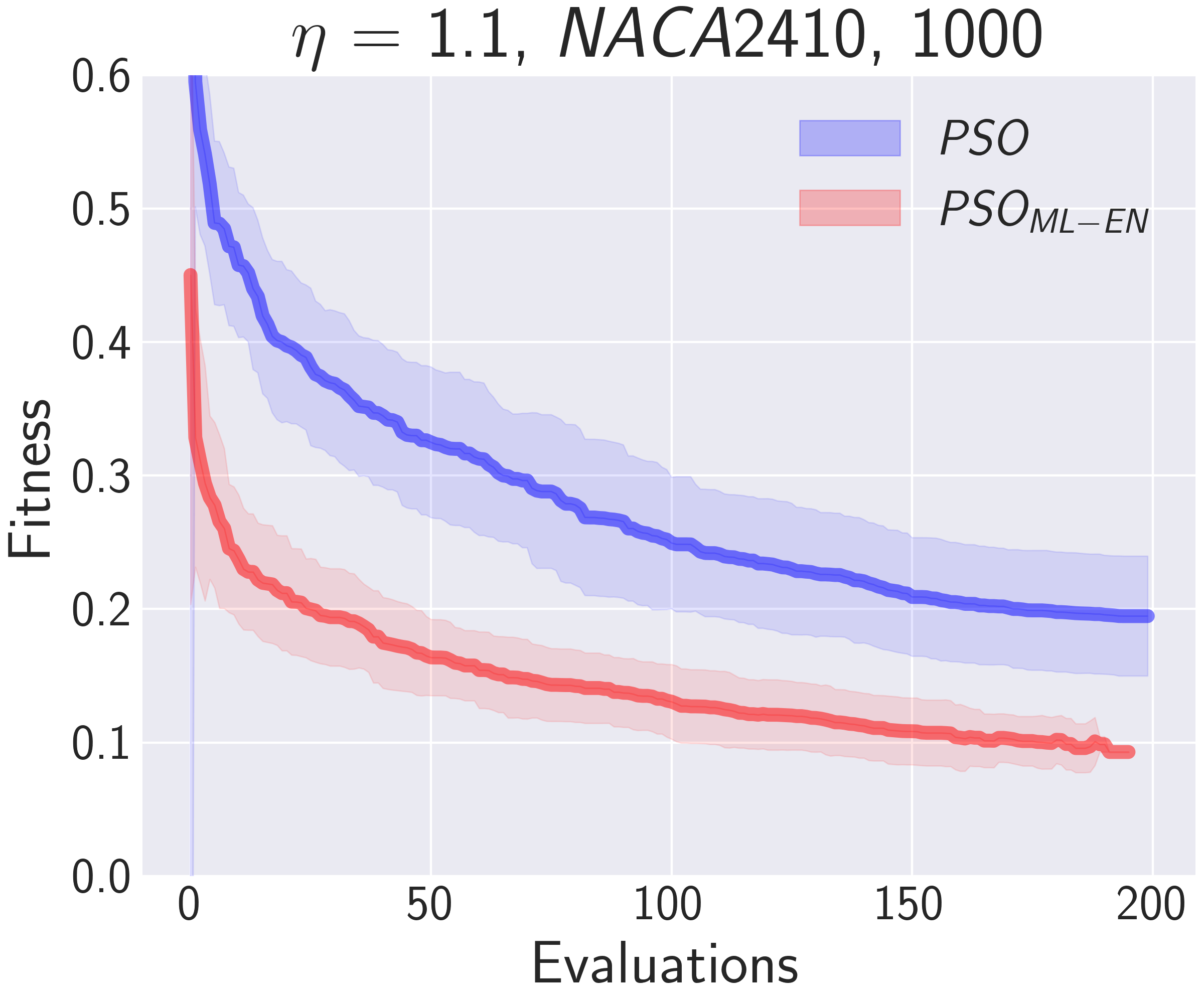}
		\caption{}
		\label{fig:enhpso_naca_f_comparison}
	\end{subfigure}
	
	\begin{subfigure}{0.3\textwidth}
		\centering
		\includegraphics[width=\linewidth]{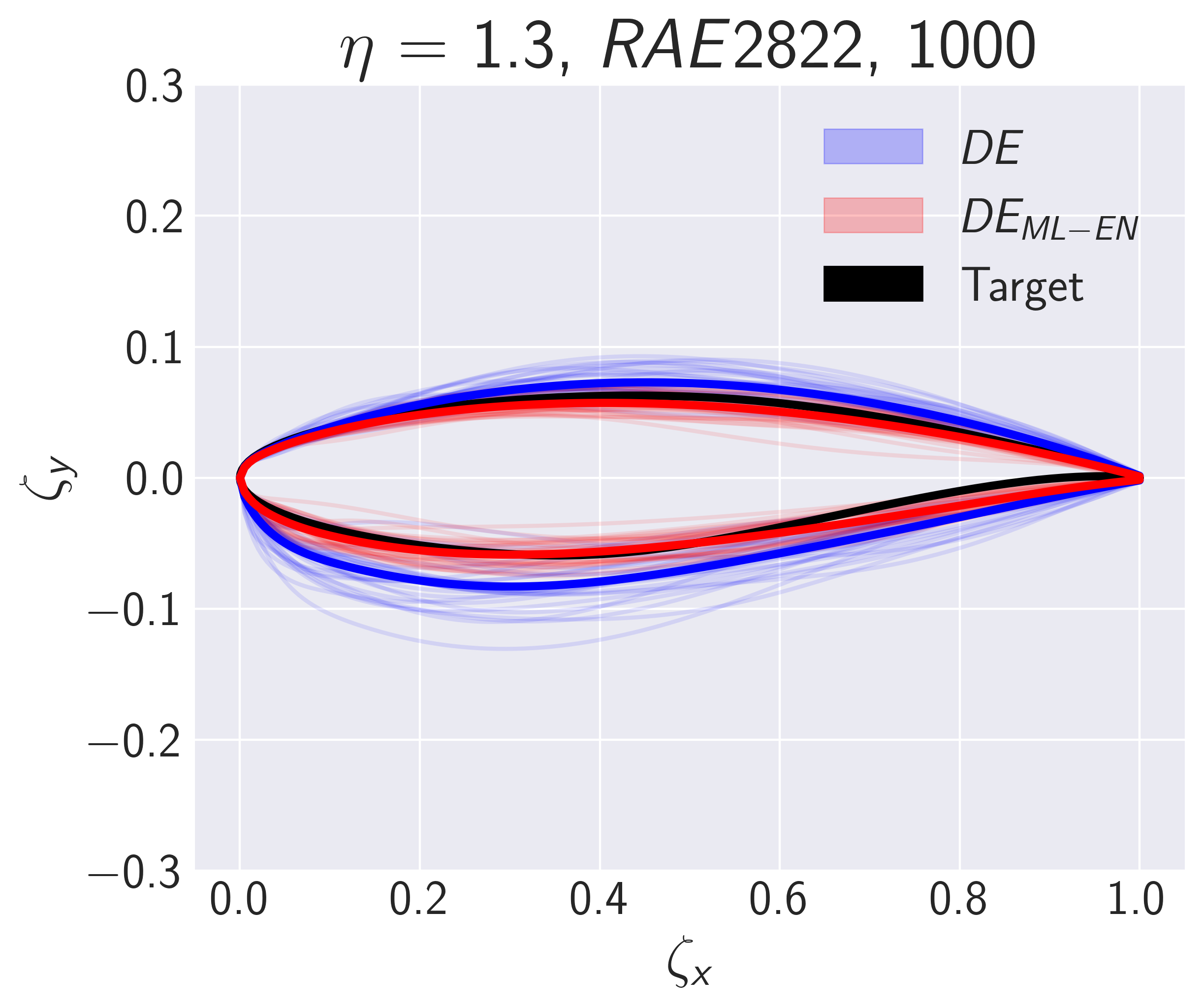}
		\caption{}
		\label{fig:enhde_rae_best_design_comparison}
	\end{subfigure}%
	\hfill
	\begin{subfigure}{0.3\textwidth}
		\centering
		\includegraphics[width=\linewidth]{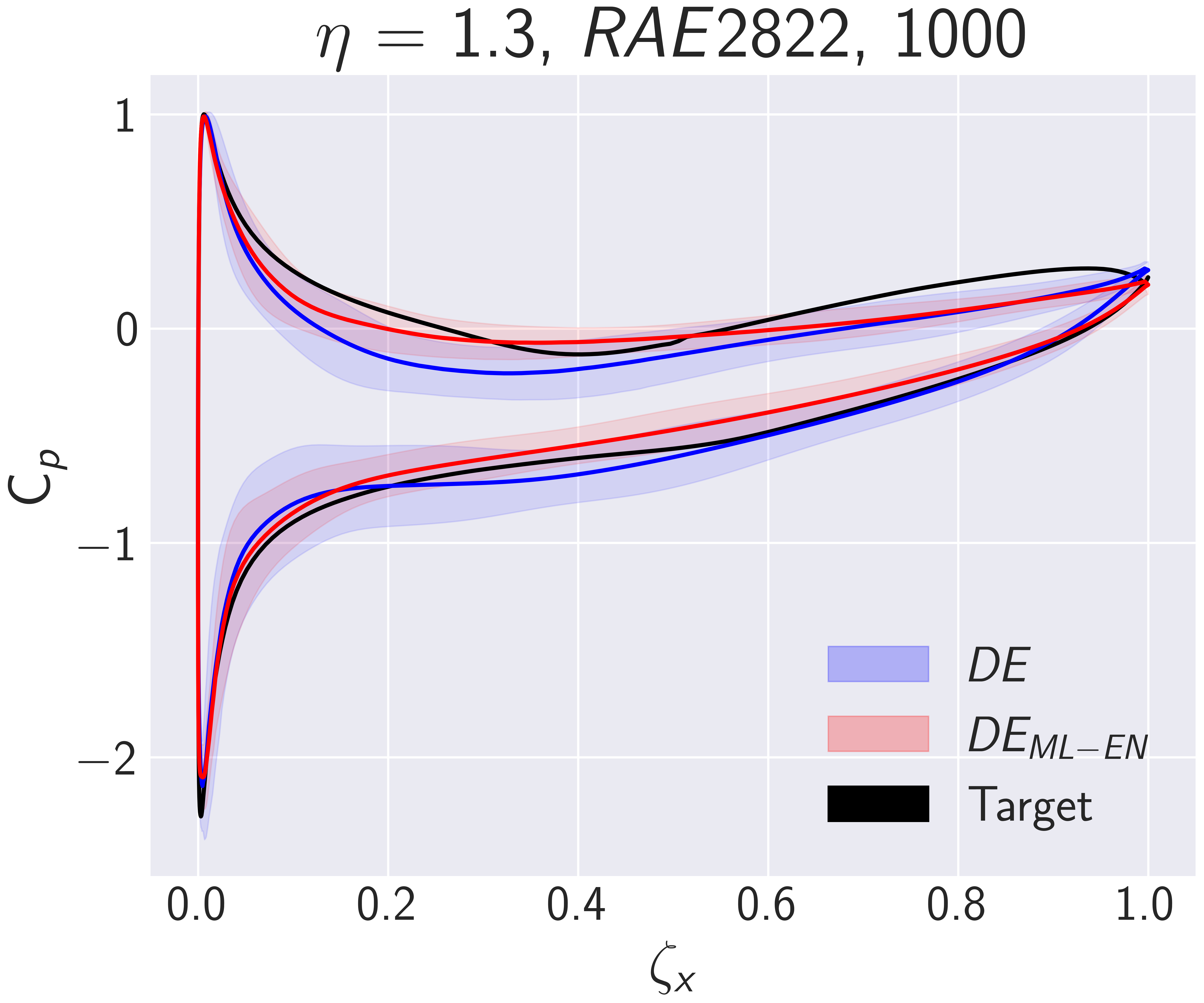}
		\caption{}
		\label{fig:enhde_rae_pressure_comparison}
	\end{subfigure}%
	\hfill
	\begin{subfigure}{0.3\textwidth}
		\centering
		\includegraphics[width=\linewidth]{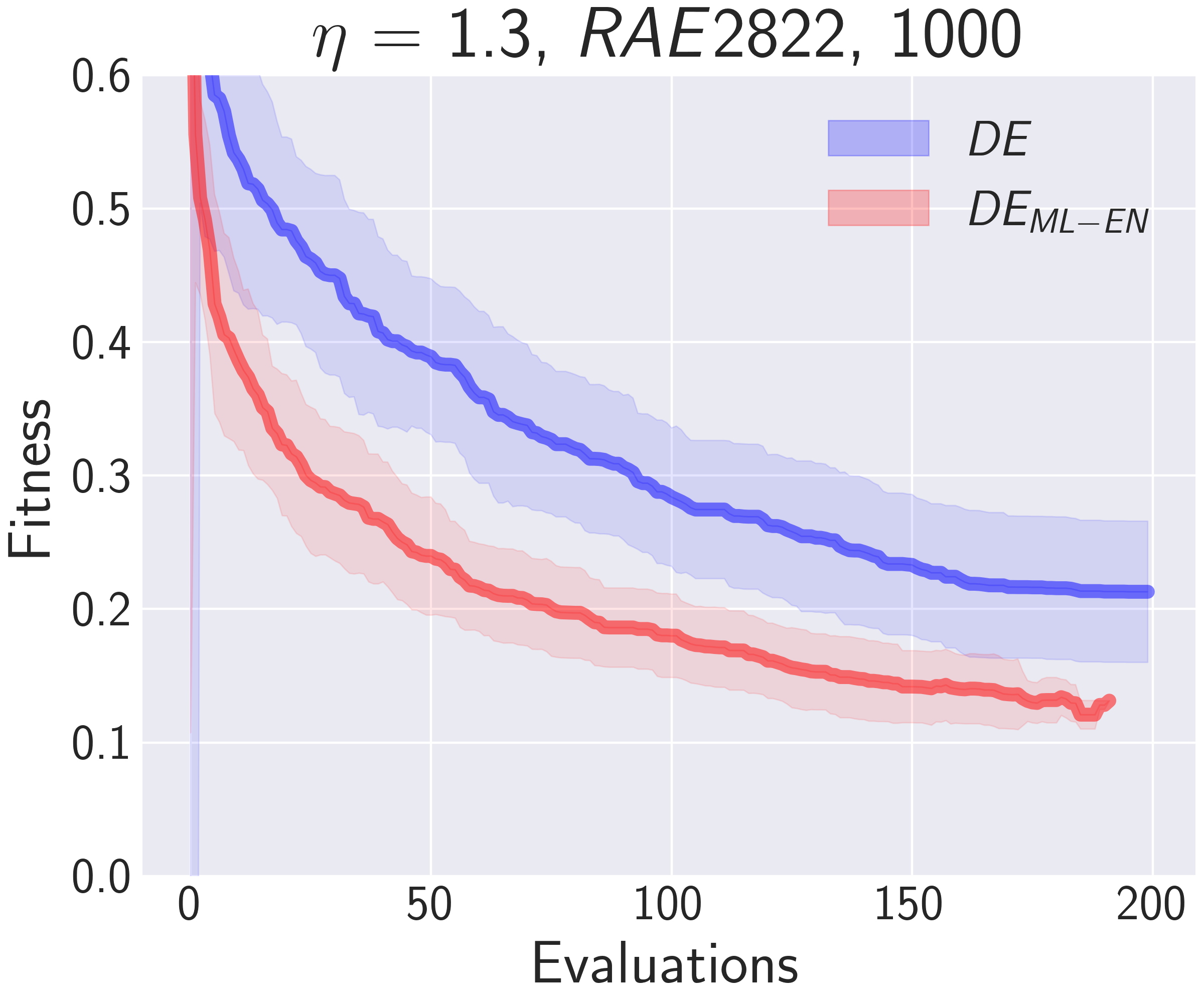}
		\caption{}
		\label{fig:enhde_rae_f_comparison}
	\end{subfigure}
	
	\caption{A comparative analysis of ML-enhanced algorithm configurations (dataset size = 1000 for both rows, $\eta$ = 1.1 for top and $\eta$ = 1.3 for bottom row, respectively) versus their unenhanced optimization equivalents: (a) Optimal achieved $NACA2410$ airfoil geometry for the PSO and $PSO_{ML-EN}$ (b) Optimal set of pressure coefficients for the same configuration juxtaposed with target values (c) Convergence of all 30 runs for PSO and $PSO_{ML-EN}$ (d) Optimal achieved $RAE2822$ airfoil geometry for DE and $DE_{ML-EN}$ (e) Corresponding optimal pressure coefficients set against target values for the $RAE2822$ airfoil (f) Average convergence and standard deviation of all 30 runs for both algorithm variants. The thicker lines in the first and second column represent the average optimized designs and pressure coefficients from the 30 runs, while the transparent shaded region in the second column depicts the standard deviation of the pressure coefficients.}
	\label{fig:naca_rae_convergence_comparison_pso_de}
\end{figure*}

Fig. \ref{fig:naca_rae_convergence_comparison_pso_de} offers a comparison between selected ML-enhanced algorithm configurations and their unenhanced optimization counterparts. The first column displays the optimal achieved airfoil geometry, while the second presents the optimal set of pressure coefficients, both set against the target values. The third column illustrates the convergence graphs of all 30 runs for both algorithm variants. The first row corresponds to the PSO algorithm and the $NACA2410$ airfoil, while the second shows an example of the DE algorithm and the $RAE2822$ airfoil. Considering all three visual metrics, both $DE_{ML-EN}$ and $PSO_{ML-EN}$ surpass their unenhanced counterparts. Yet, neither algorithm achieves an exact alignment with the target designs, in terms of geometry and pressure coefficient sets. This discrepancy arises because the framework is assessed under strict computational budgets, with a specific focus on only 200 HF simulations, however, further improvements for both approaches are likely with larger computational budgets.

\newpage
\clearpage

\newpage
\clearpage
\subsubsection{SFR Results}
\label{sec:sfr_results}

\begin{figure*}[h!]
	\centering
	\begin{subfigure}{0.45\textwidth}
		\centering
		\includegraphics[width=\linewidth]{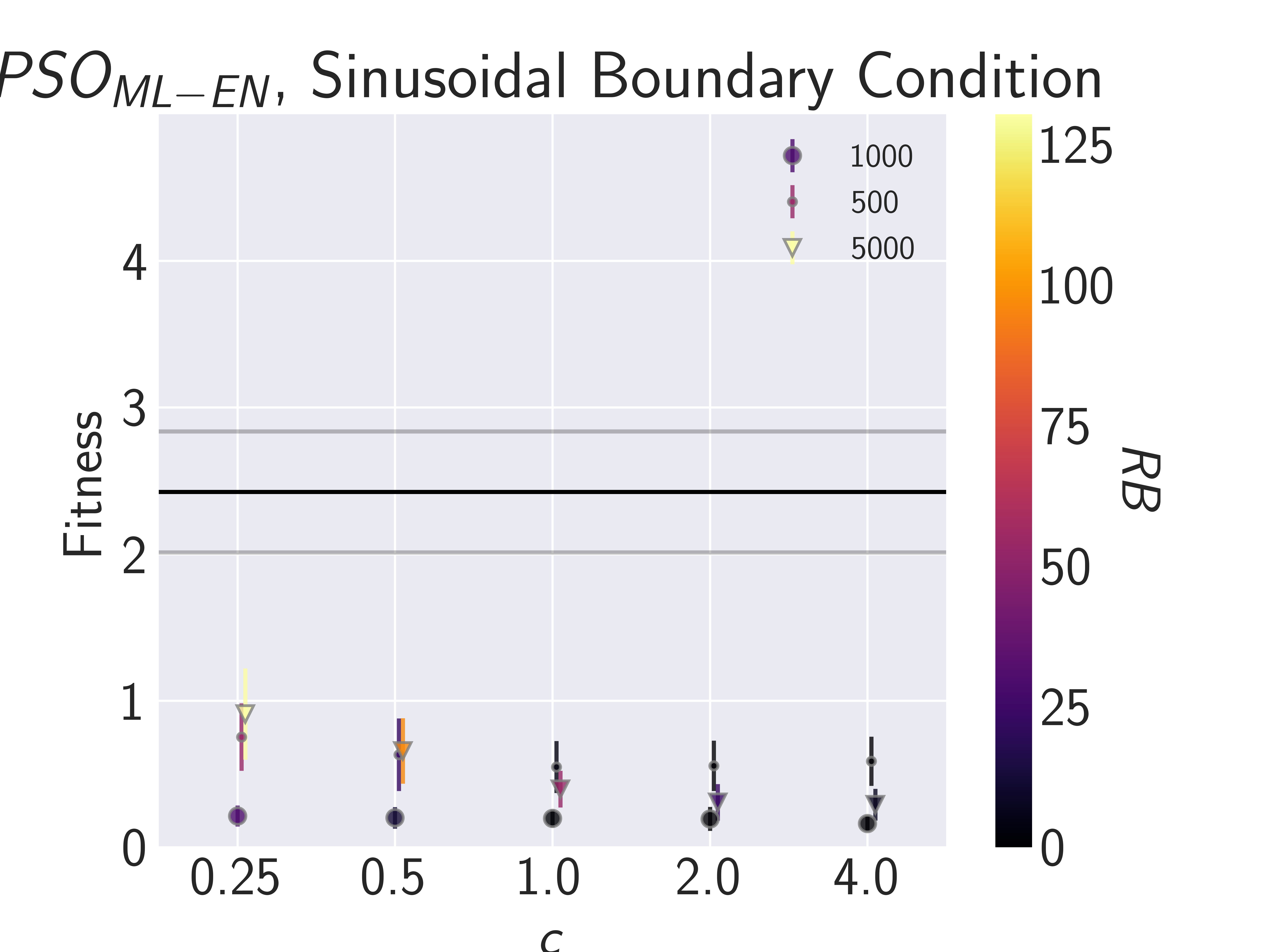}
		\caption{}
		\label{fig:pso_bc1}
	\end{subfigure}%
	\hfill
	\begin{subfigure}{0.45\textwidth}
		\centering
		\includegraphics[width=\linewidth]{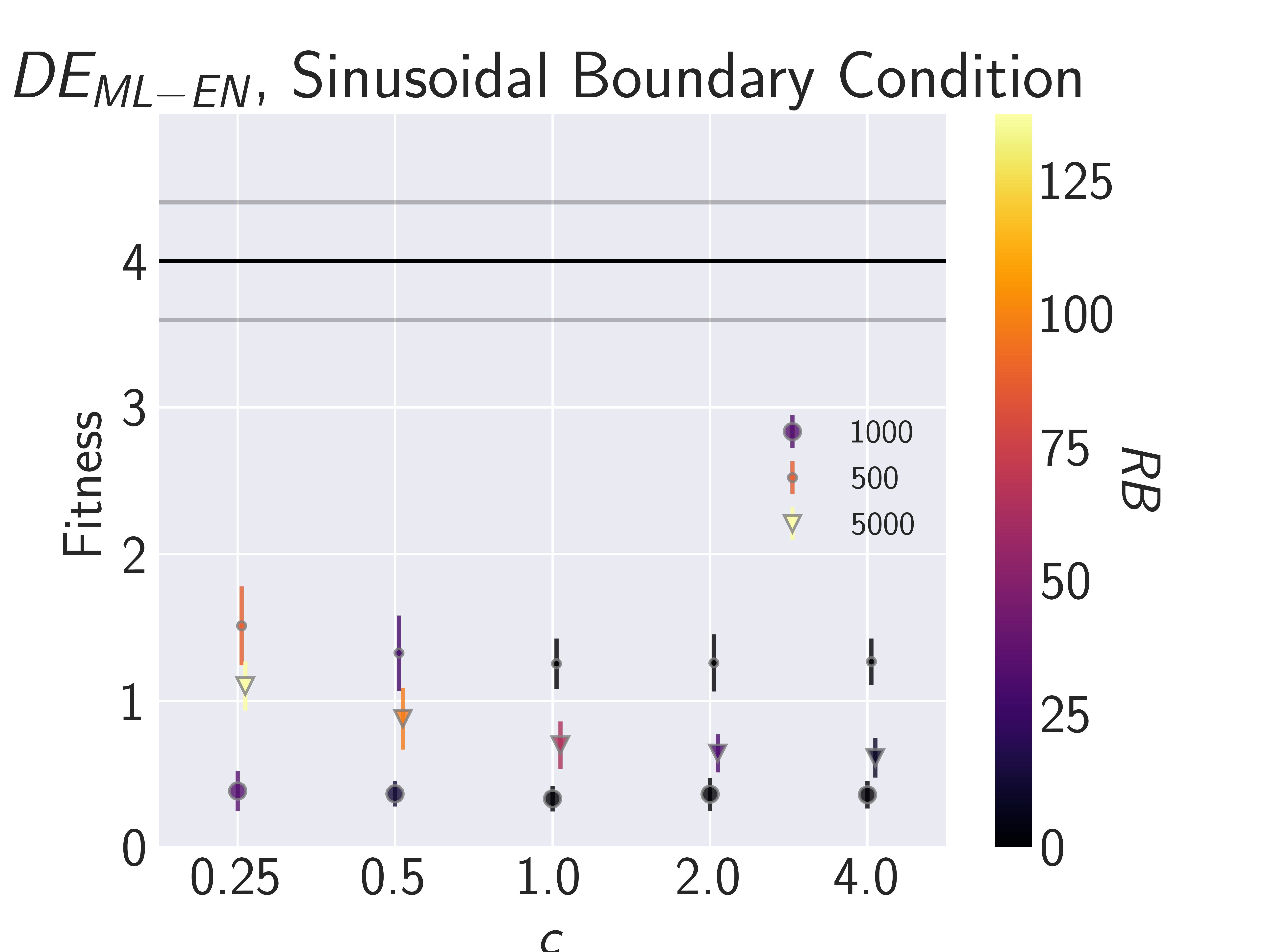}
		\caption{}
		\label{fig:de_bc1}
	\end{subfigure}
	
	\begin{subfigure}{0.45\textwidth}
		\centering
		\includegraphics[width=\linewidth]{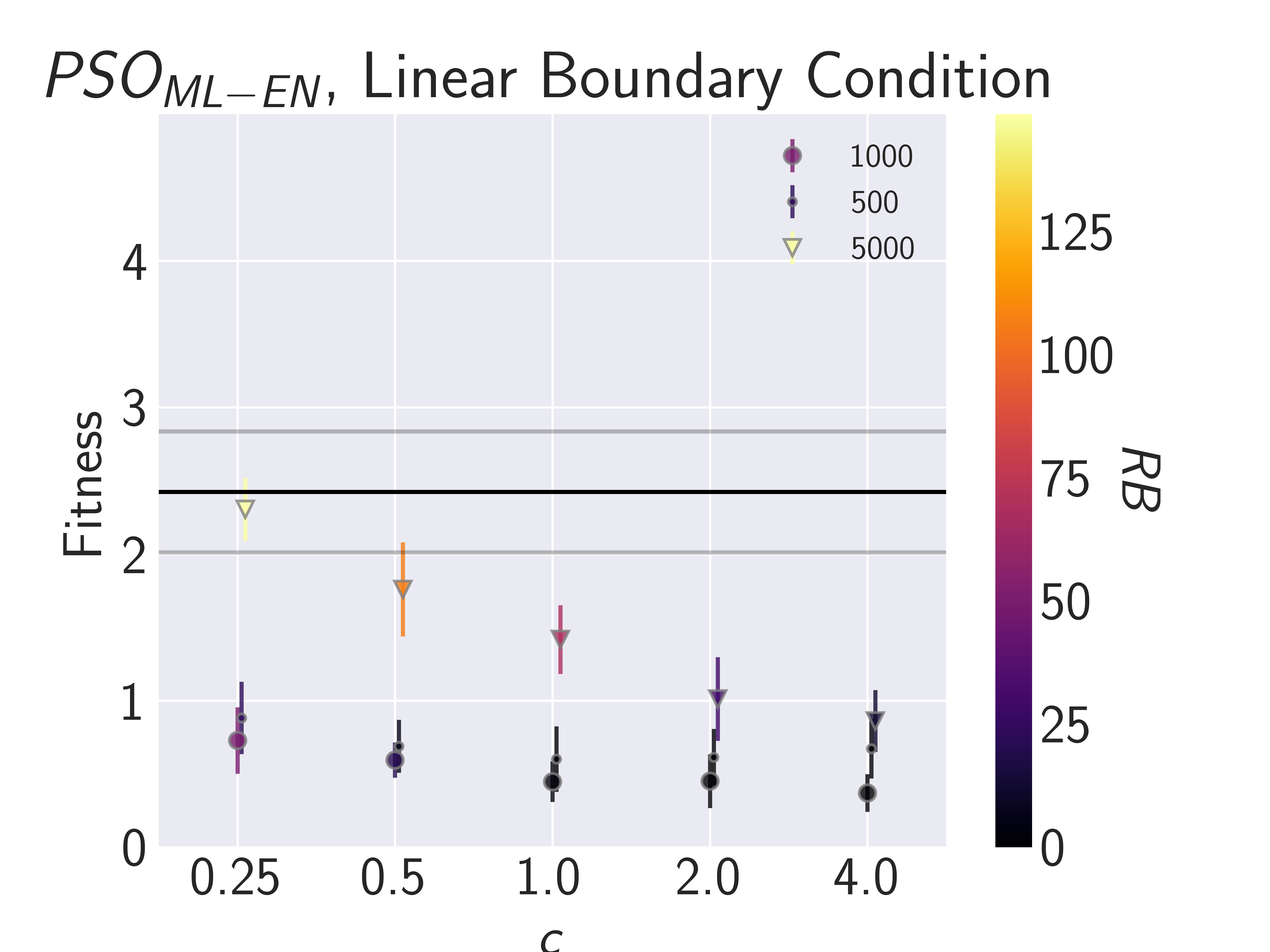}
		\caption{}
		\label{fig:pso_bc2}
	\end{subfigure}%
	\hfill
	\begin{subfigure}{0.45\textwidth}
		\centering
		\includegraphics[width=\linewidth]{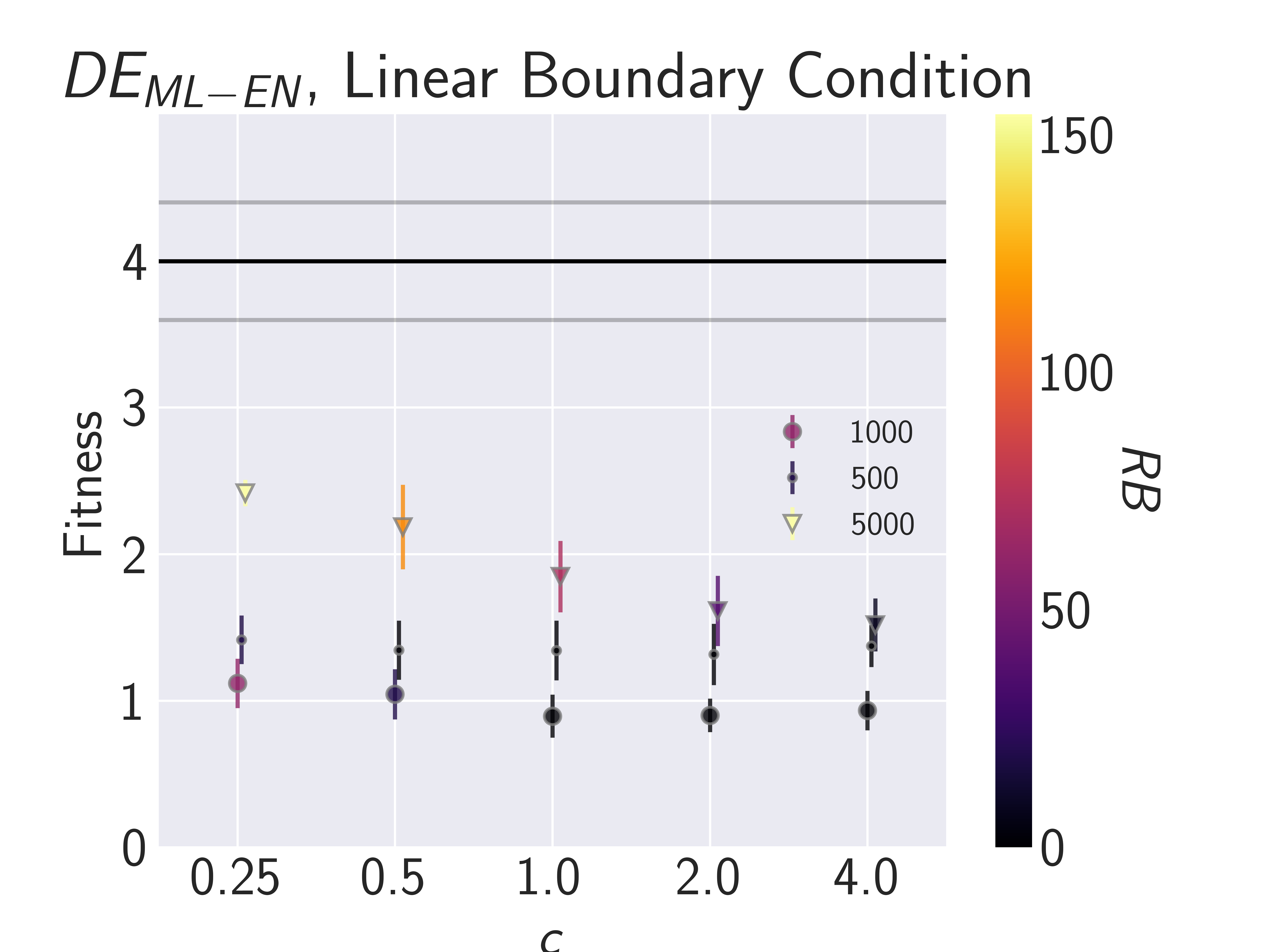}
		\caption{}
		\label{fig:de_bc2}
	\end{subfigure}
	
	\caption{ML-enhanced framework results for the surface field reconstruction problem: (a) $PSO_{ML-EN}$ for the sinusoidal BC (b) $DE_{ML-EN}$ for the sinusoidal BC (c) $PSO_{ML-EN}$ for the linear BC
		(d) $DE_{ML-EN}$ for the linear BC. The markers denote the different dataset sizes used to train the ML model, while the coloring of the markers represents the remaining simulation budget ($RB$) values. A higher $RB$ signifies greater savings in the computational budget, while low fitness values imply a better approximation of the target performance. The markers are slightly offset for each $c$ value to improve visibility.}
	\label{fig:bc1bc2_pso_de_hyperparameters}
\end{figure*}

Fig. \ref{fig:bc1bc2_pso_de_hyperparameters} presents the hyperparameter analysis for the ML-framework applied to the SFR problem. It also provides a comparison with the unenhanced algorithms showing the average and standard deviation of the fitness, indicated by the horizontal black and grey lines, respectively. The ML-enhanced algorithms consistently outshine their traditional counterparts. Drawing parallels with the AID problem, it is observed that while elevating the $c$ parameter allows the framework to focus on improving the target performance approximation (reducing the fitness value), it does so at the expense of fully utilizing the entire simulation budget.

The ML-enhanced optimizers with the MLP model trained on the dataset size  1000  exhibit superior performance in terms of fitness value compared to their counterparts trained on dataset sizes  500 and 5000, respectively. This difference can be attributed to the more effective boundary refinement achieved by the 1000-instance model, as evidenced by Fig. \ref{fig:bc1_ssr} and Fig. \ref{fig:bc2_ssr}. The applied boundary refinement notably contributes to reducing fitness uncertainty across all hyperparameter combinations as observed through the standard deviation lines corresponding to each marker. This advantage becomes even more pronounced when compared to  the standard deviation observed in the unenhanced algorithms.

For dataset sizes of 500 and 1000, a $c$ value of 1 or greater causes the ML-enhanced algorithms to consume the entire budget of HF simulations. Notably, when optimizers are paired with the MLP model trained on 5000 instances, the fitness scales almost linearly with the $c$ value. The 5000-instance trained ML-enhanced optimizers strike an good trade-off between achieving low fitness ($RMSE$) values and conserving HF simulations. Considering results from both BC scenarios, the hyperparameter settings that would achieve a trade-off between a good target performance approximation and simulation budget would be the 1000 dataset model at $c$ = 0.25 or the 5000 dataset model at $c$ = 1. If the goal is to substantially narrow down the design space, a noisy model, like the one trained on 500 instances, proves sufficient.

\begin{figure*}[h!]
	\centering
	\begin{subfigure}{0.45\textwidth}
		\centering
		\includegraphics[width=\linewidth]{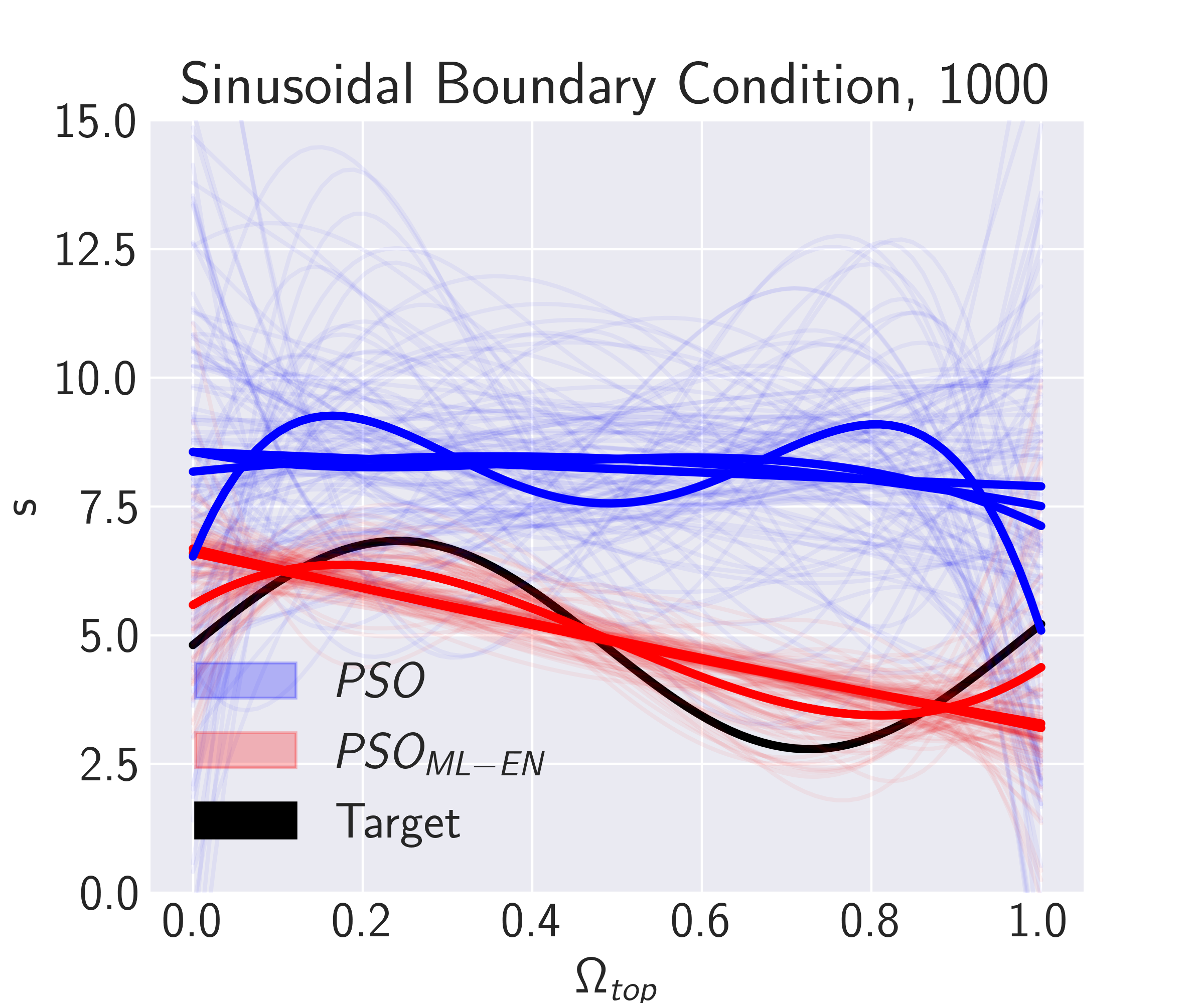}
		\caption{}
		\label{fig:pso_bc1}
	\end{subfigure}%
	\hfill
	\begin{subfigure}{0.45\textwidth}
		\centering
		\includegraphics[width=\linewidth]{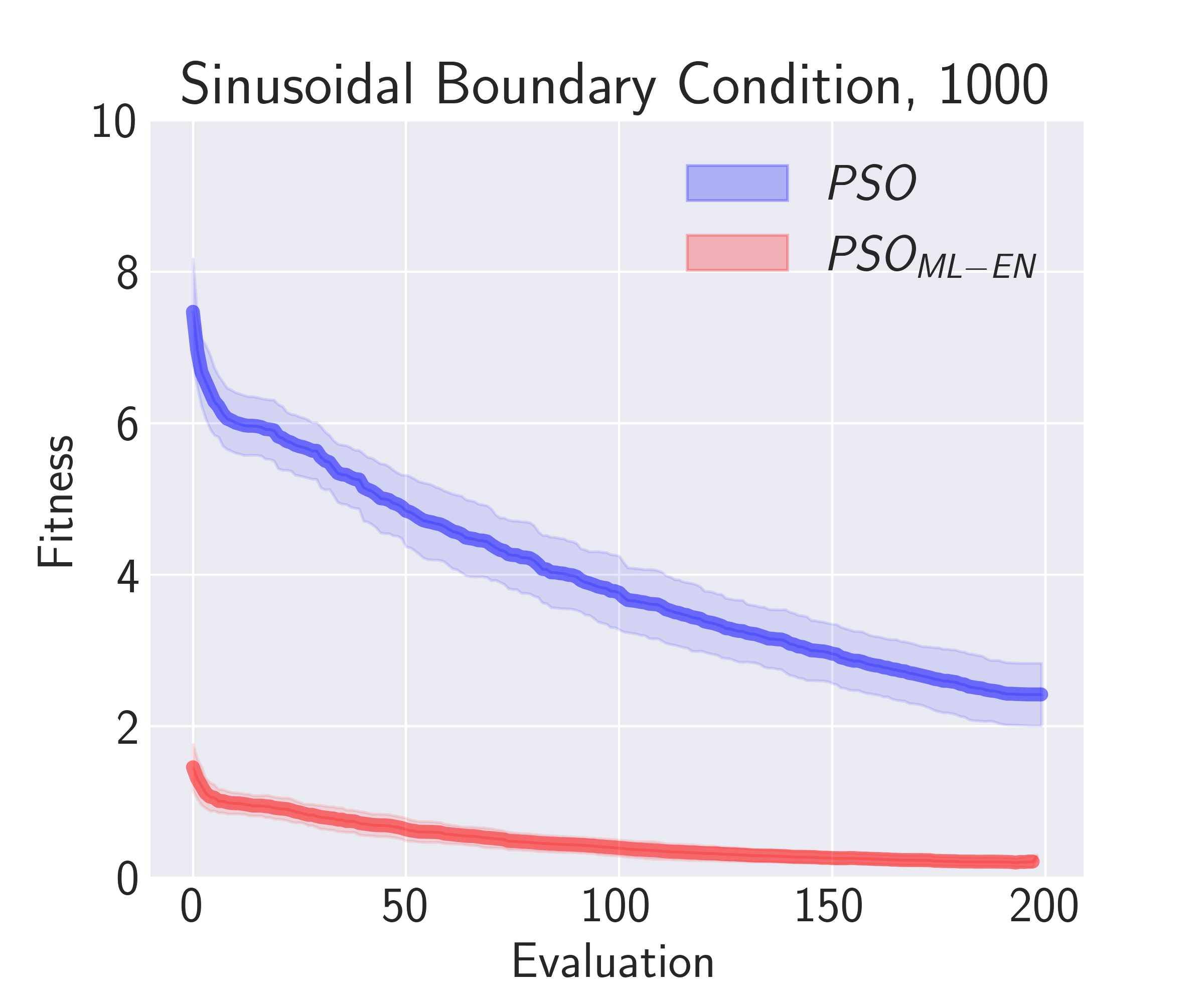}
		\caption{}
		\label{fig:de_bc1}
	\end{subfigure}
	
	\begin{subfigure}{0.45\textwidth}
		\centering
		\includegraphics[width=\linewidth]{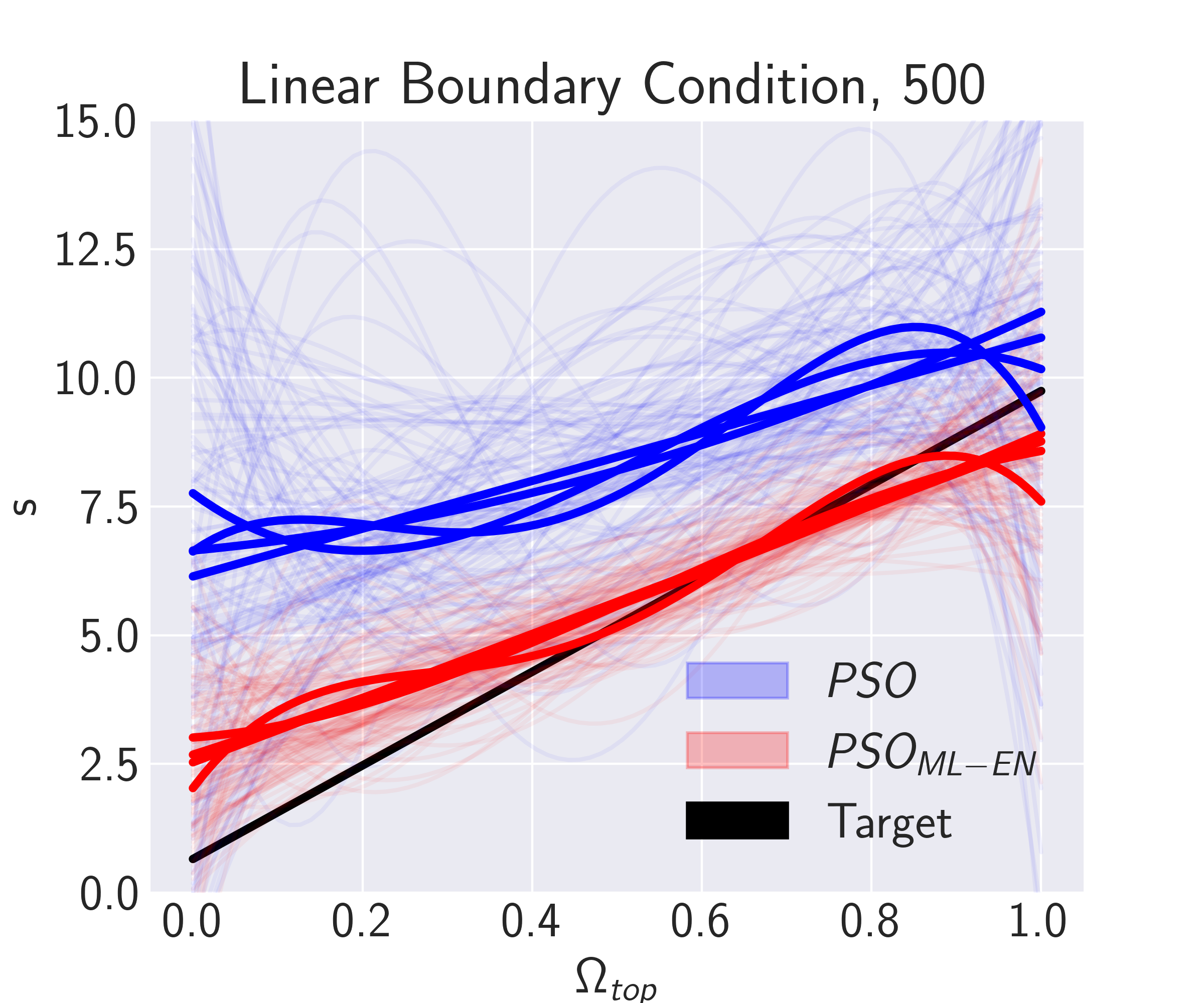}
		\caption{}
		\label{fig:pso_bc2}
	\end{subfigure}%
	\hfill
	\begin{subfigure}{0.45\textwidth}
		\centering
		\includegraphics[width=\linewidth]{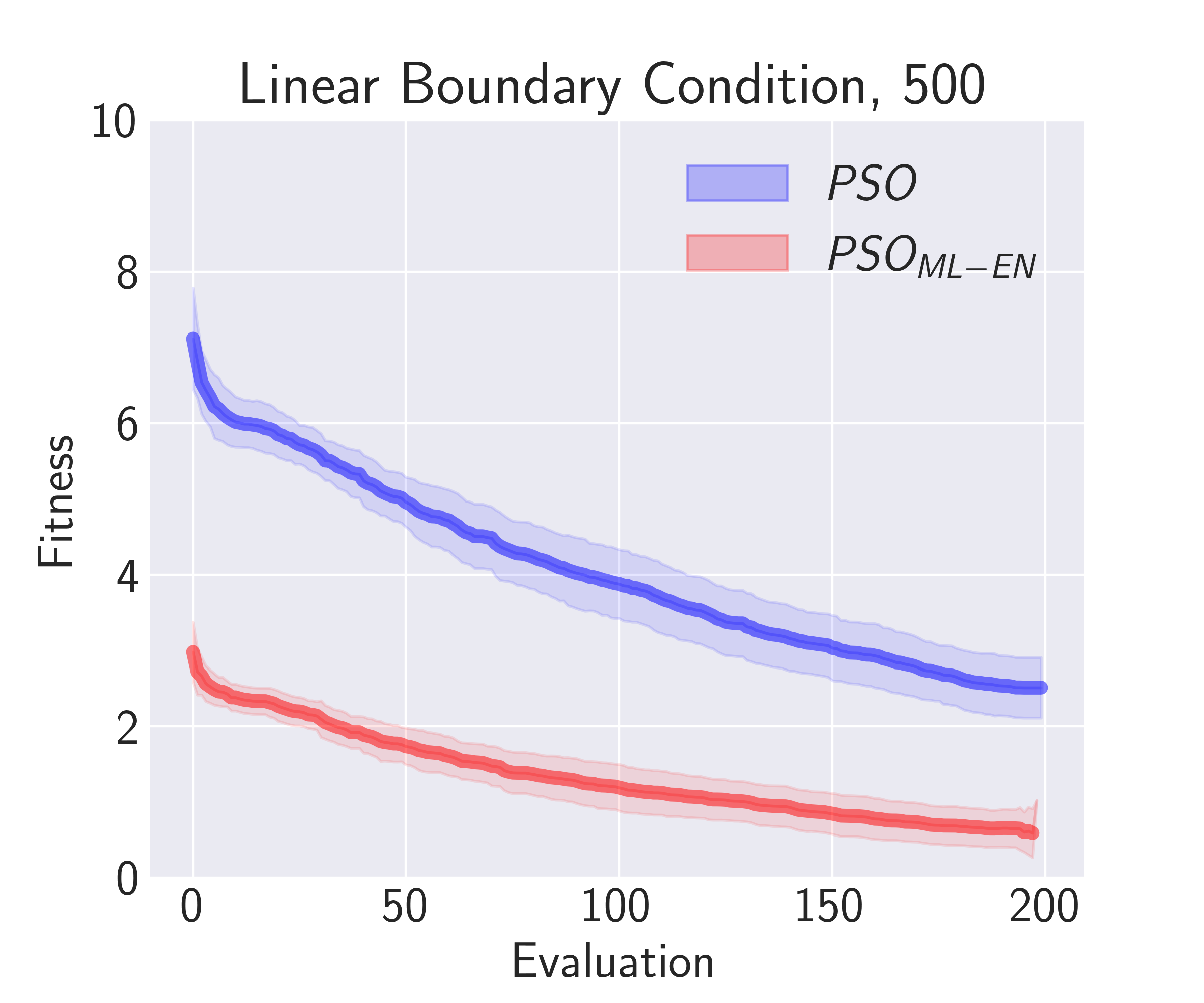}
		\caption{}
		\label{fig:de_bc2}
	\end{subfigure}
	
	\caption{A comparative analysis of ML-enhanced PSO algorithm configurations with $c$ = 1 (MLP training dataset size = 1000 for top row, and 500 for bottom row) versus their unenhanced optimization equivalents: (a) PSO and $PSO_{ML-EN}$ comparison of the reconstructed sinusoidal BC juxtaposed with the true BC (b) Sinusoidal BC case 30 runs average convergence and standard deviation plot of PSO and $PSO_{ML-EN}$ (c) PSO and $PSO_{ML-EN}$ comparison of the reconstructed linear boundary condition juxtaposed with the true BC (d) Linear BC case 30 runs average convergence and standard deviation plot of PSO and $PSO_{ML-EN}$. The thicker lines in the first column represent the average BCs of the 30 runs. There are four thicker lines shown since the solutions were approximated with four different degrees of regression.}
	\label{fig:bc1bc2_comparison_design_f}
\end{figure*}

Fig. \ref{fig:bc1bc2_comparison_design_f} displays examples of the optimized BCs for both test instances. The results from PSO$_{ML-EN}$ for the sinusoidal BC employed an MLP trained on 1000 instances with $c$ = 1, while for the linear BC an MLP trained on 500 instances with $c$ = 1 was used. Different reconstructed averaged BCs are depicted for both instances and algorithms. This variety arises because the final optimized average design vectors, which contained raw scalar values for each $\Omega_{top}$ coordinate, underwent regression model fitting ranging from degrees 1 to 4, described in Sect. \ref{sec:sfr_ssr}. In both cases presented in Fig. \ref{fig:bc1bc2_comparison_design_f}, both the average reconstructed BCs (for all regression model degrees) and the convergence plot clearly demonstrate the superiority of PSO$_{ML-EN}$ over PSO.

\begin{figure*}[ht]
	\centering
	\begin{subfigure}{0.33\textwidth}
		\centering
		\includegraphics[width=\linewidth]{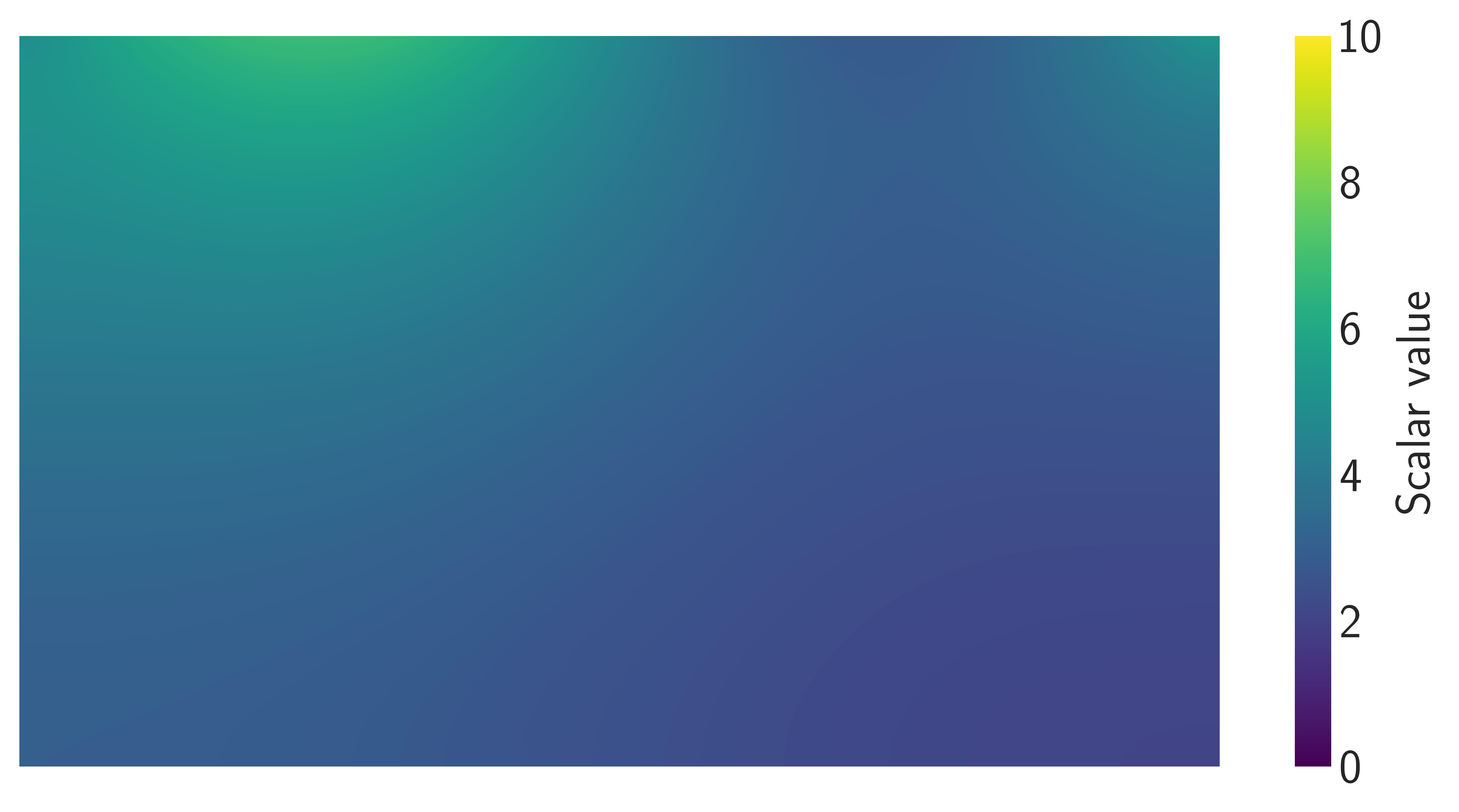}
		\caption{}
		\label{fig:sub1}
	\end{subfigure}%
	\hfill
	\begin{subfigure}{0.33\textwidth}
		\centering
		\includegraphics[width=\linewidth]{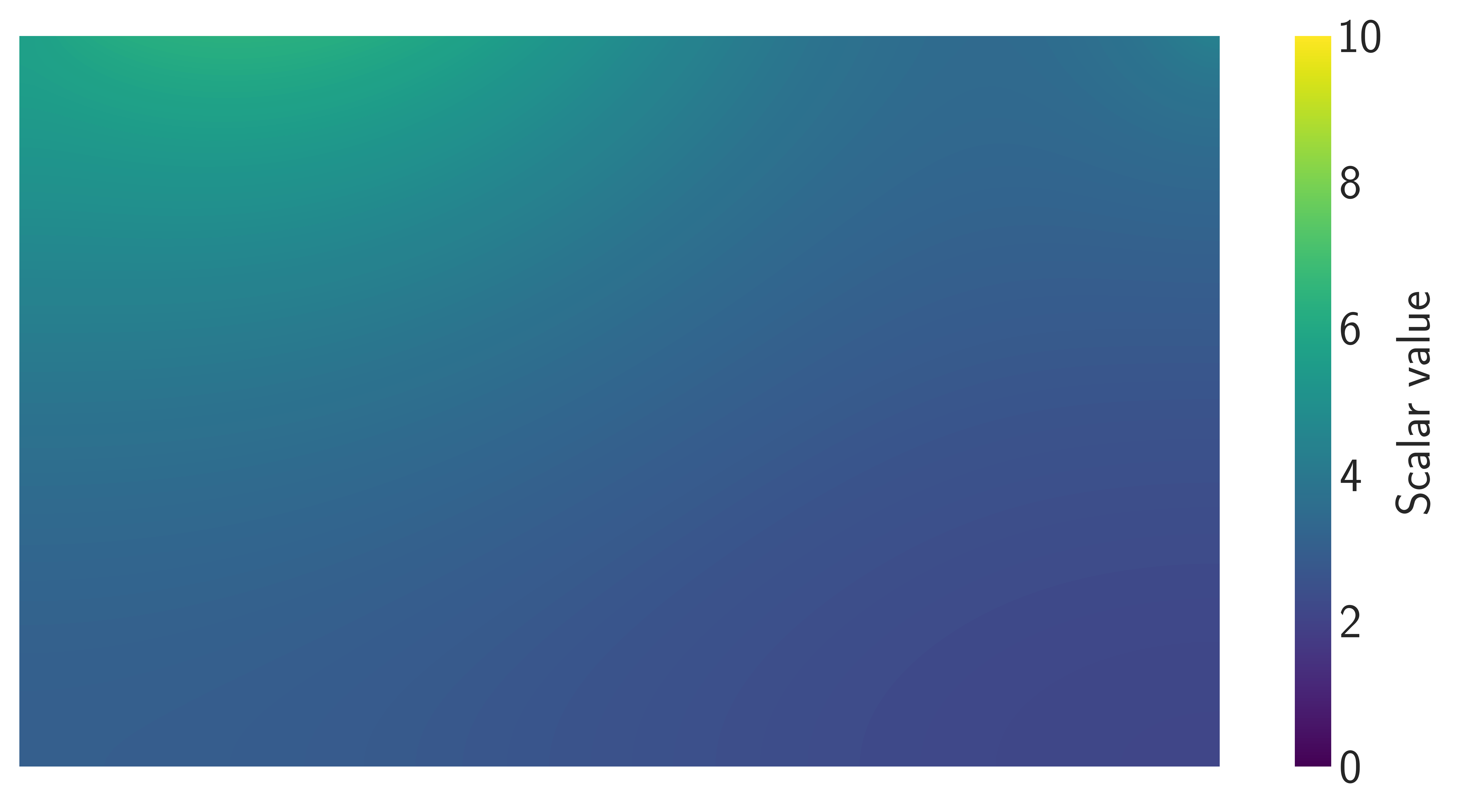}
		\caption{}
		\label{fig:sub2}
	\end{subfigure}%
	\hfill
	\begin{subfigure}{0.33\textwidth}
		\centering
		\includegraphics[width=\linewidth]{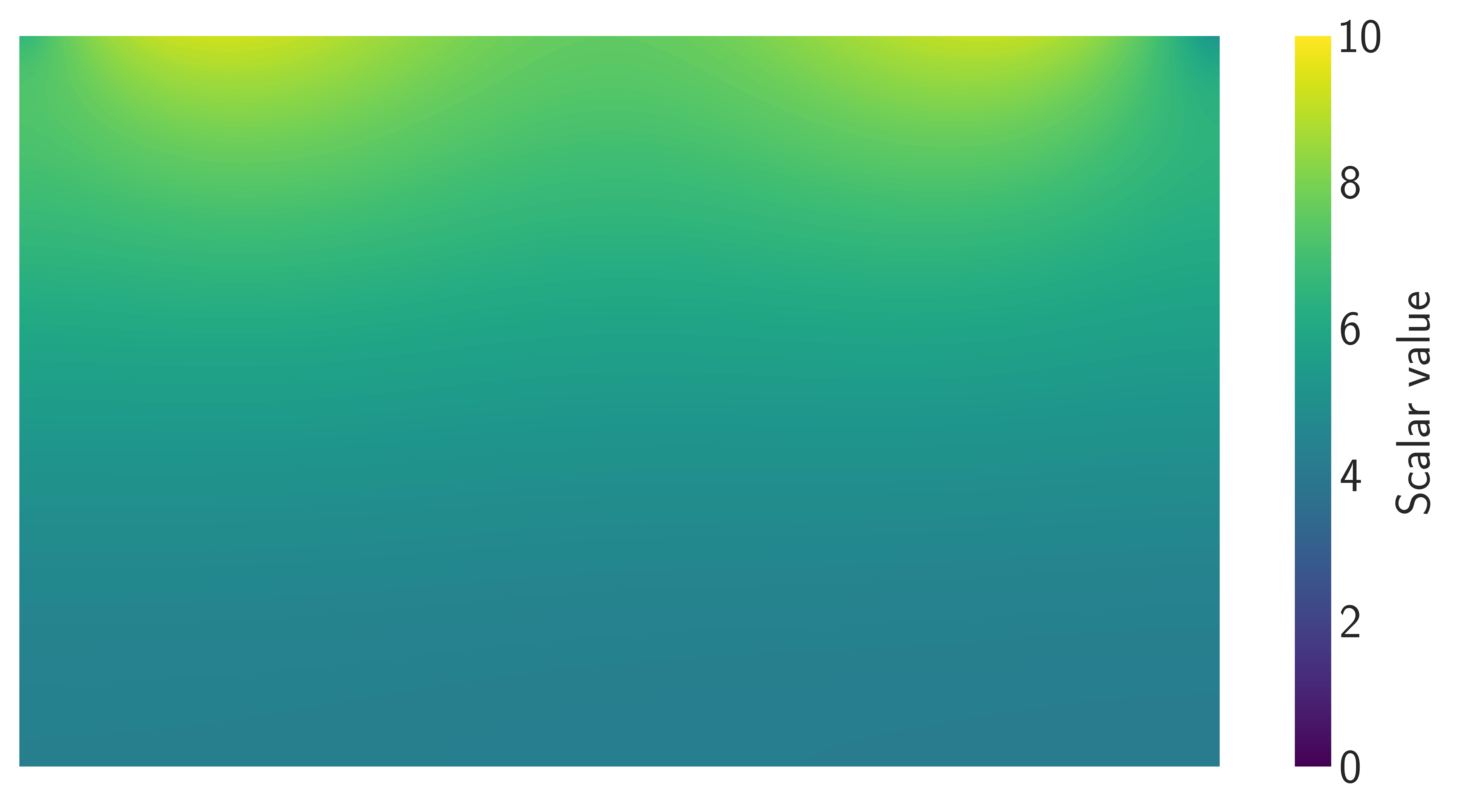}
		\caption{}
		\label{fig:sub3}
	\end{subfigure}
	
	\begin{subfigure}{0.33\textwidth}
		\centering
		\includegraphics[width=\linewidth]{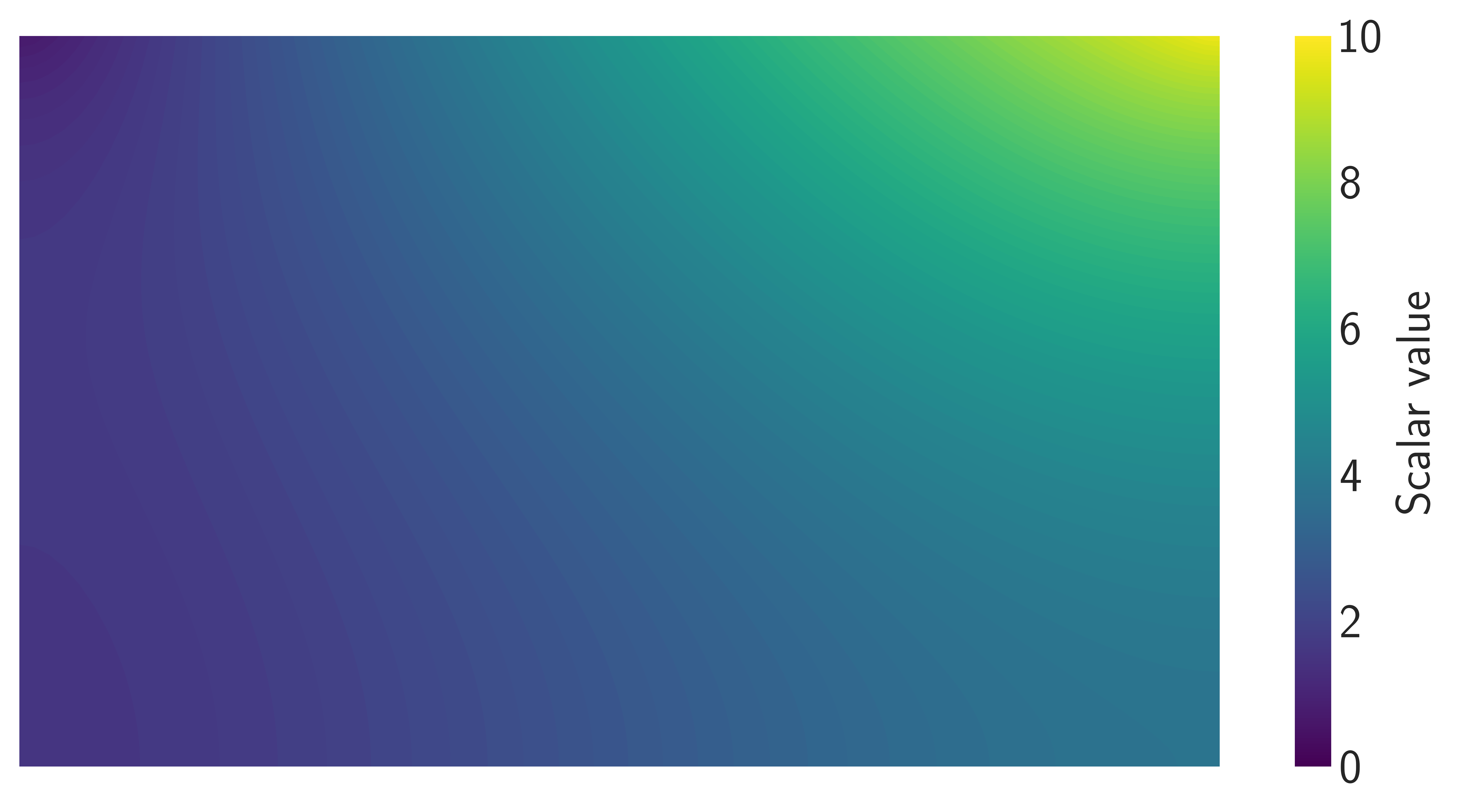}
		\caption{}
		\label{fig:sub4}
	\end{subfigure}%
	\hfill
	\begin{subfigure}{0.33\textwidth}
		\centering
		\includegraphics[width=\linewidth]{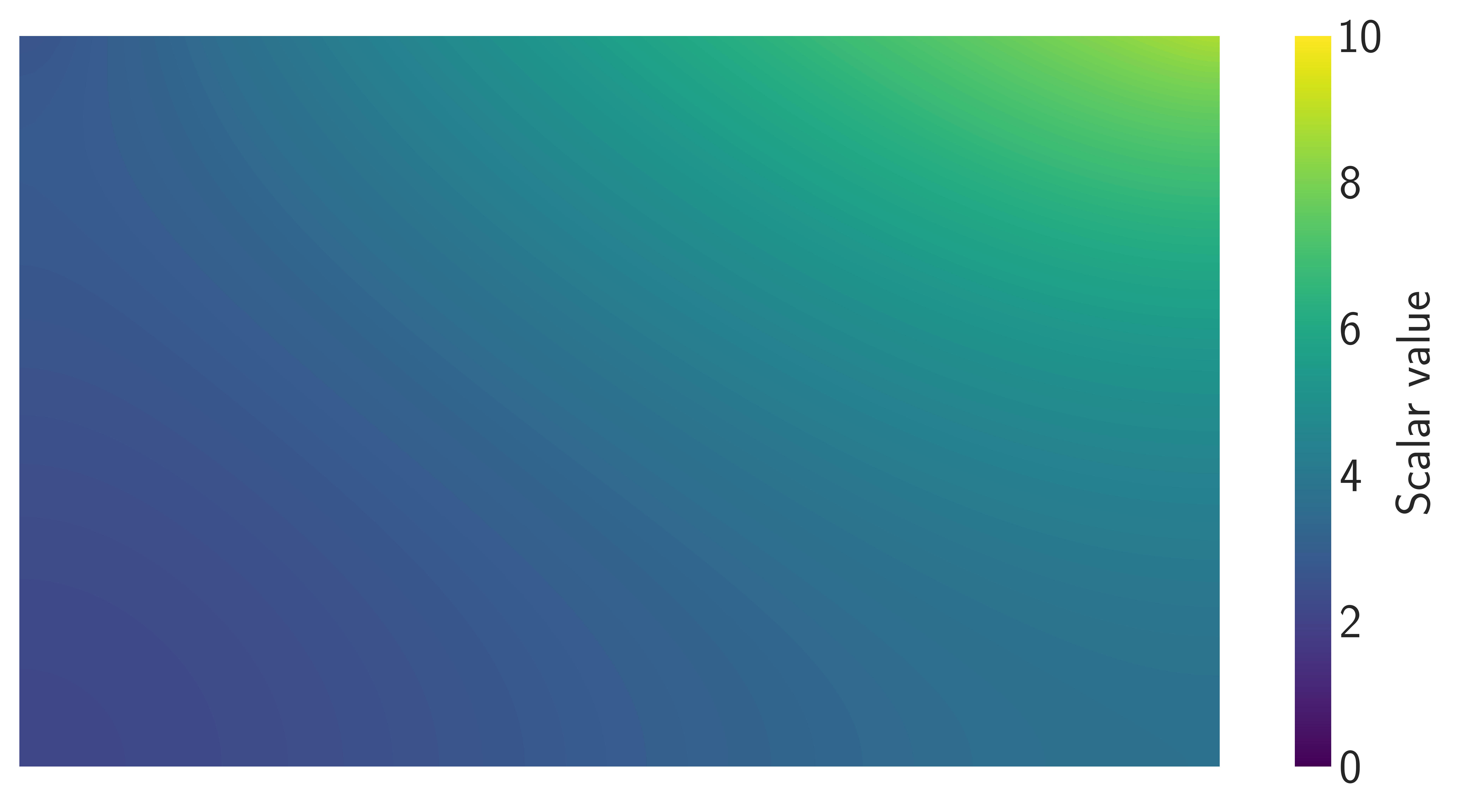}
		\caption{}
		\label{fig:sub5}
	\end{subfigure}%
	\hfill
	\begin{subfigure}{0.33\textwidth}
		\centering
		\includegraphics[width=\linewidth]{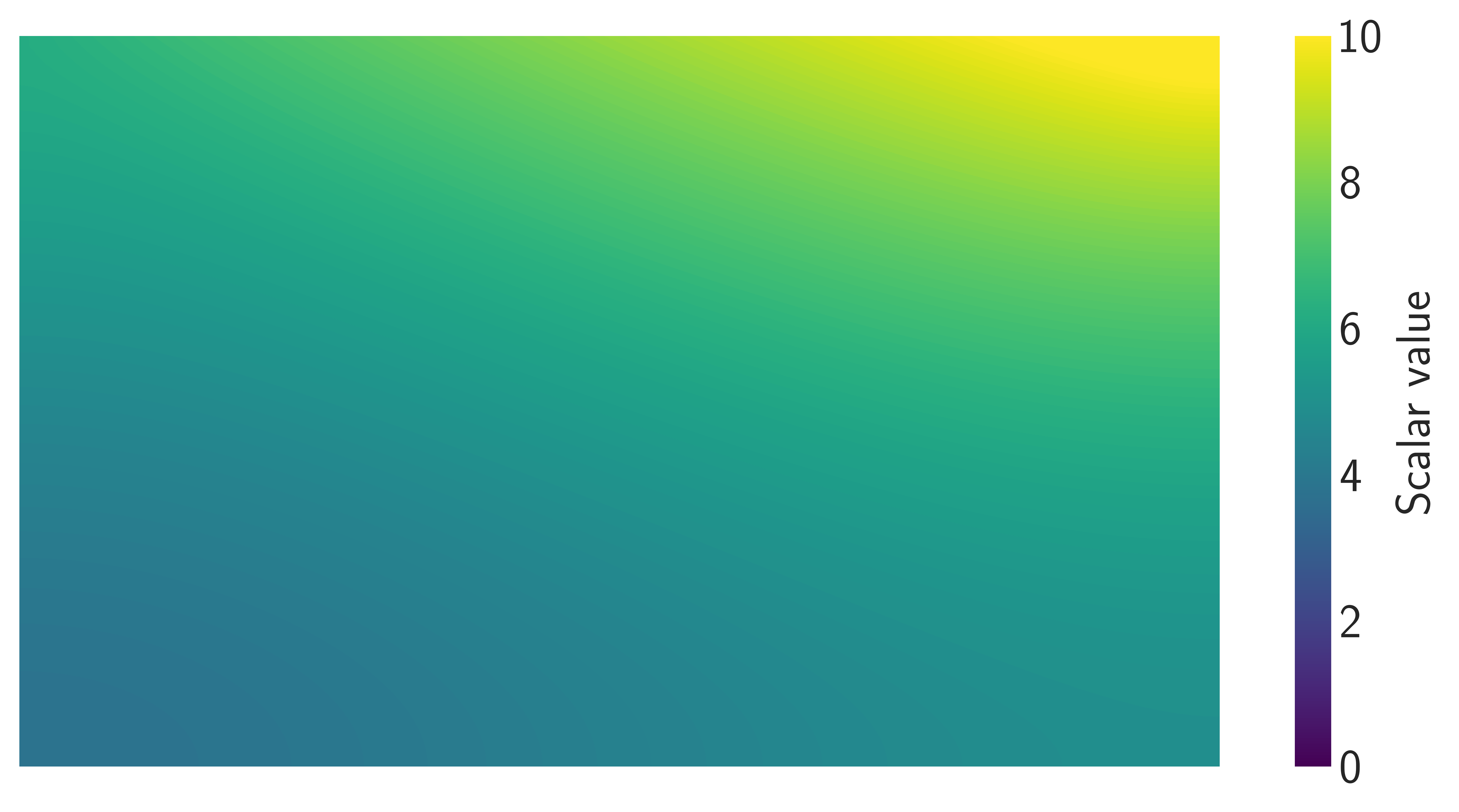}
		\caption{}
		\label{fig:sub6}
	\end{subfigure}
	
	\caption{Reconstructed scalar fields with the average optimized design vectors presented in Fig. \ref{fig:bc1bc2_comparison_design_f}: (a) True sinusoidal boundary condition (b) $PSO_{ML-EN}$ reconstructed sinusoidal boundary condition (4-th degree regression model) (c) PSO reconstructed sinusoidal boundary condition (4-th degree regression model) (d) True linear boundary condition (e) $PSO_{ML-EN}$ reconstructed linear boundary condition (linear regression model) (f) PSO reconstructed linear boundary condition (linear regression model).}
	\label{fig:bc1b2_fields}
\end{figure*}

\begin{figure*}[h!]
	\centering
	\begin{subfigure}{0.45\textwidth}
		\centering
		\includegraphics[width=\linewidth]{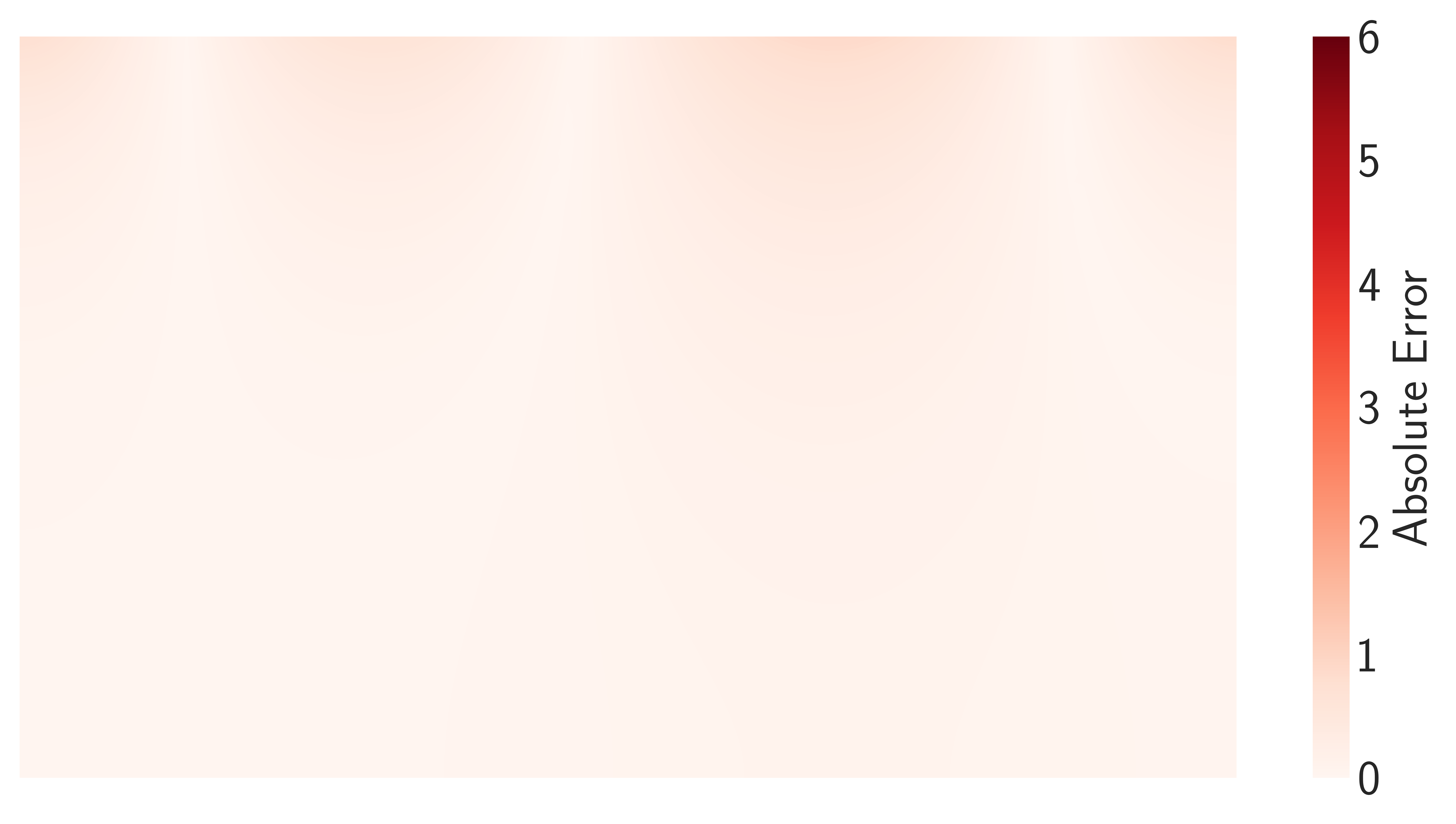}
		\caption{}
		\label{fig:enhpso_abs_bc1}
	\end{subfigure}%
	\hfill
	\begin{subfigure}{0.45\textwidth}
		\centering
		\includegraphics[width=\linewidth]{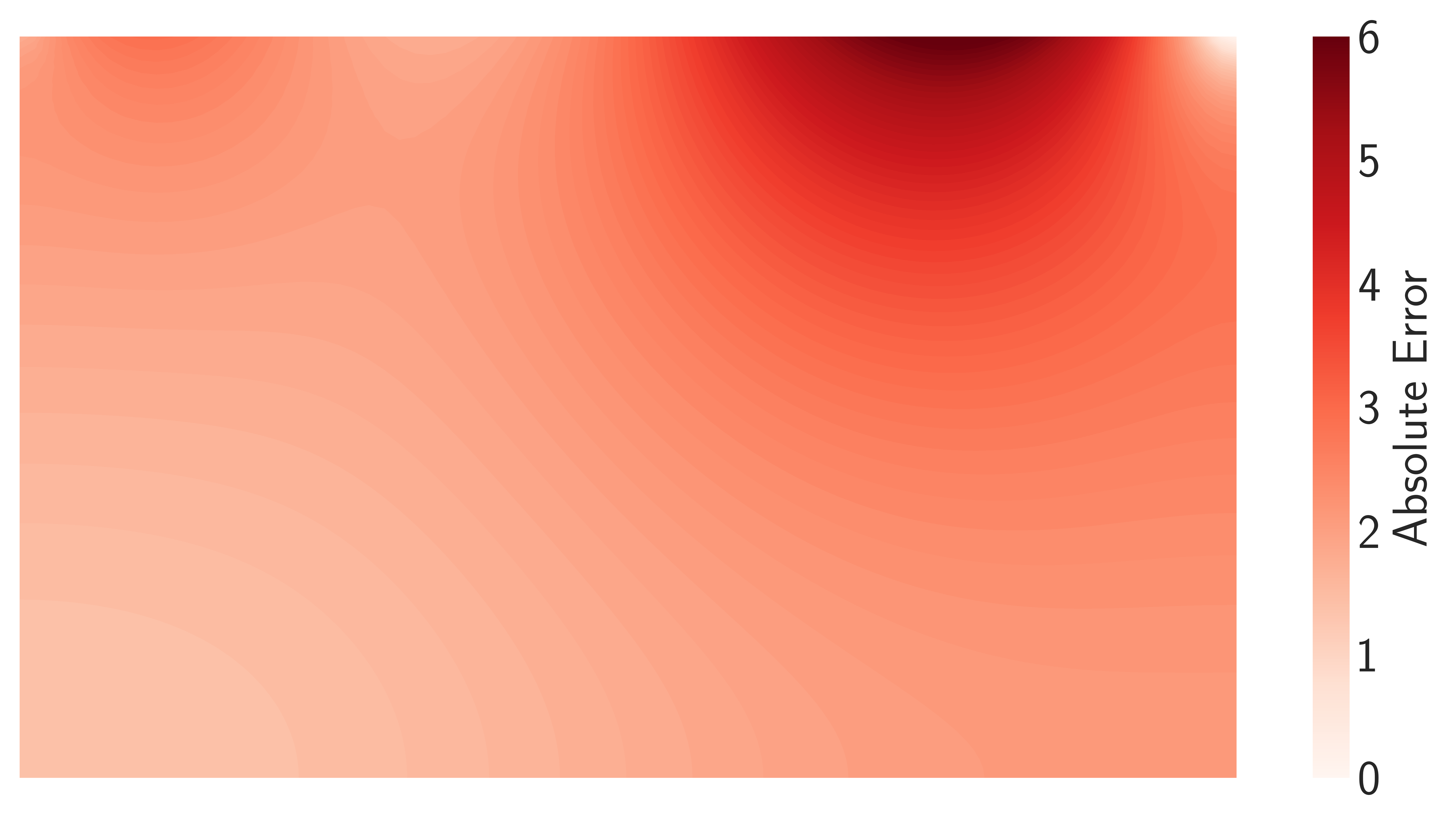}
		\caption{}
		\label{fig:vanpso_abs_bc1}
	\end{subfigure}
	
	\begin{subfigure}{0.45\textwidth}
		\centering
		\includegraphics[width=\linewidth]{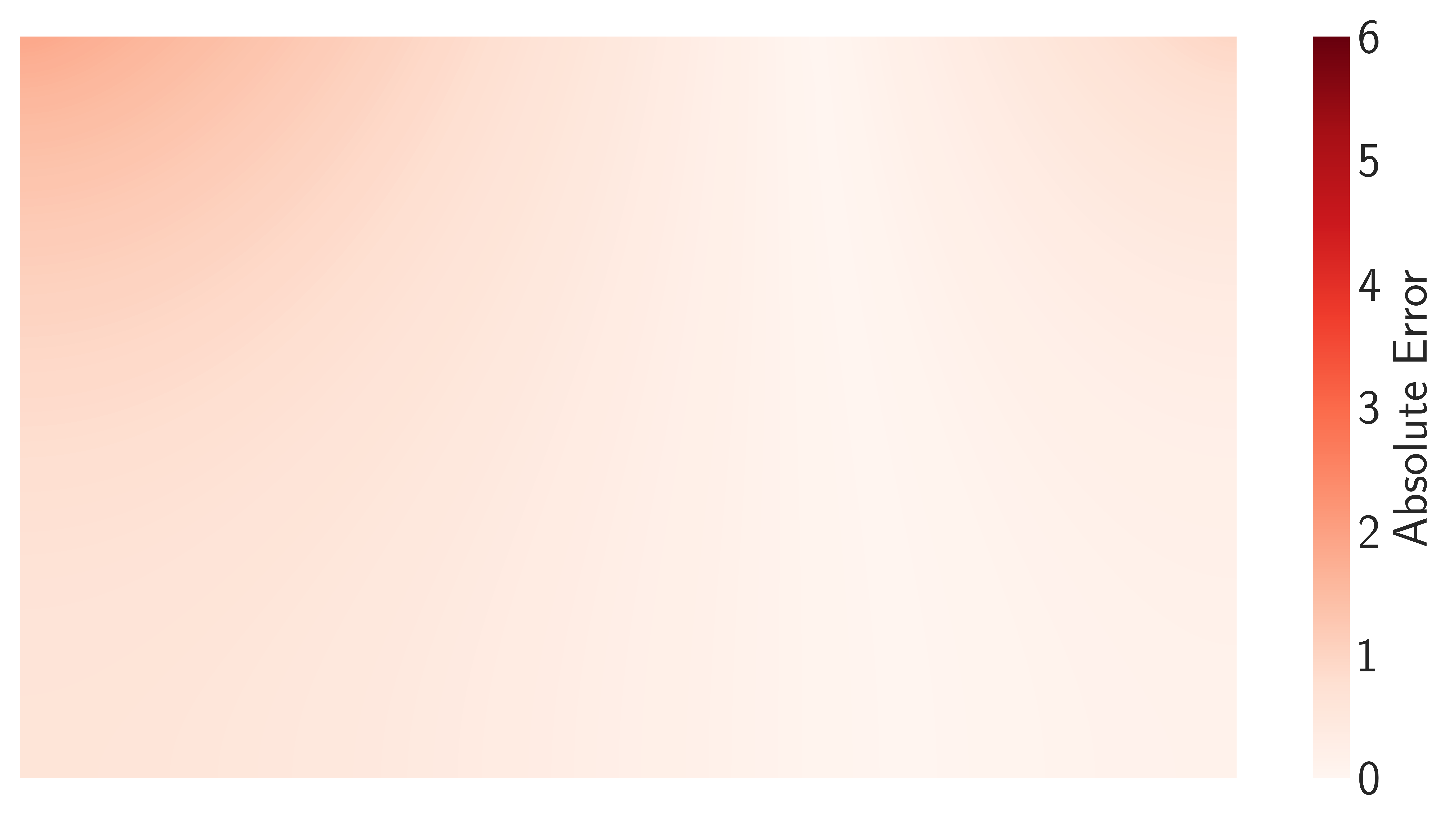}
		\caption{}
		\label{fig:enhpso_abs_bc2}
	\end{subfigure}%
	\hfill
	\begin{subfigure}{0.45\textwidth}
		\centering
		\includegraphics[width=\linewidth]{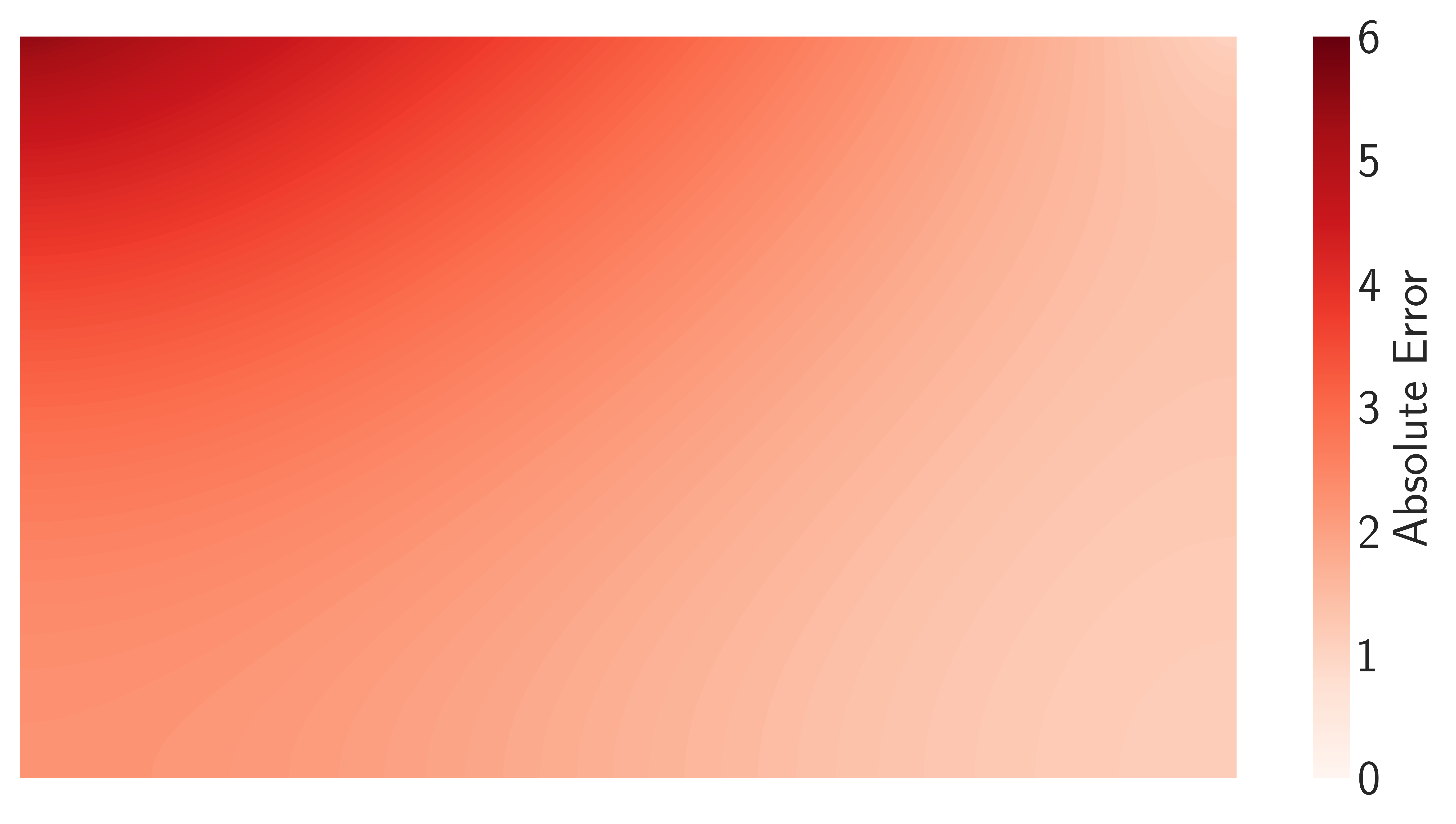}
		\caption{}
		\label{fig:vanpso_abs_bc2}
	\end{subfigure}
	
	\caption{Absolute error between the true scalar field and the scalar field obtained by the average optimized design vectors presented in Fig. \ref{fig:bc1bc2_comparison_design_f} for both BC cases: (a) True and $PSO_{ML-EN}$ scalar field absolute error for the sinusoidal BC (b) True and PSO scalar field absolute error for the sinusoidal BC (c) True and $PSO_{ML-EN}$ scalar field absolute error for the linear BC (d) True and PSO scalarfield absolute error for the linear BC.}
	\label{fig:abs_error_scalar_fields}
\end{figure*}

In Fig. \ref{fig:bc1b2_fields}, the reconstructed scalar fields generated by the BCs presented in Fig. \ref{fig:bc1bc2_comparison_design_f} are shown. The top row shows the fields for the sinusoidal BC, while the bottom row shows the fields for the linear BC. The first column shows the ground truth, while the second and third columns show the $PSO_{ML-EN}$ and PSO reconstructed scalar fields. It is obvious that the BCs generated by the ML-enhanced algorithm align much more closely with the true solution. Finally, Fig. \ref{fig:abs_error_scalar_fields} shows the absolute error between the true scalar fields (for both BC cases) and those obtained by $PSO_{ML-EN}$ and PSO-optimized boundary conditions. The absolute error was calculated for every point in the HF domain, and with the range of the absolute error being the same for both results shown, it is apparent that the $PSO_{ML-EN}$ generated BC is more accurate.

\subsection{Advantages and Limitations of the ML-enhanced Framework}

While the ML-enhanced inverse design method shows improved performance, it is not without limitations. Primarily, the framework requires a pre-trained ML model to estimate the M$_{info}$ value. To harness this model effectively for boundary refinement and to cut back on the number of HF simulations, it is vital to understand and determine the pertinent reduced-order information related to the optimization challenge. 

The main advantage of the proposed method is that an  ML model is trained independently of the optimization loop using LF data only, and can then be exploited for different inverse design instances of the same type of problem  (e.g., one ML model for airfoils enables the efficient  optimization of multiple different types of airfoils). Furthermore, the ML model does not have to be highly accurate as highlighted in the detailed hyperparameter analysis for both investigated problems, which is advantageous in cases where obtaining LF data is computationally non-trivial.

Another limitation of the framework lies in its dependence on multiple hyperparameters. Both the boundary refinement technique, as applied to AID, and the ML-enhanced framework itself require safety hyperparameters ($\eta$ and $c$ respectively). Although this study demonstrates that the $c$ value can correlate with the $RMSE$ of the model suggesting values of $c \in \{0.5, 1, 2\}$ for both problem a more exhaustive analysis encompassing a broader set of similar problems is essential. However, pinpointing the appropriate $c$ parameter could be accomplished through an exploratory analysis leveraging an ML model and exclusively LF simulations.

\section{Conclusion}

The paper presents an ML-enhanced inverse design framework for problems with stringent simulation budgets. This framework, applied to two distinct engineering challenges--AID and SFR--leveraged a pre-trained ML model. The goal was to reduce the size of the optimization design space and to decrease the need for costly HF simulations to arrive at an optimal design. In this ML-enhanced framework, both the DE and PSO optimization algorithms, which have an extensive demand for objective function evaluations, were enhanced with the ML model and contrasted with their conventional versions.

The main contributions of the study can be summarized in several points:

\begin{itemize}
	\item An ML model trained on a small set of LF data effectively narrows the optimization design space. This facilitates a better rate of convergence of both PSO and DE towards a better approximation of the target performance within a predefined HF simulation budget. 
	
	\item The ML-framework proves highly effective  for both minimizing the number of HF simulations and approximating user-defined target designs. A relationship between the ML model's error metric ($RMSE$) and the mechanism for minimizing HF simulations has been established and explored. For the AID and SFR problems, the hyperparameter $c$ which is used to multiply the $RMSE$, is recommended to be in the range $c \in \{0.5, 1, 2\}$.
	
	\item The solutions obtained with population-based stochastic global optimization algorithms, such as DE and PSO, can be significantly improved when guided by ML models.
	
	\item The effectiveness of the ML-enhanced inverse design framework was demonstrated on two  conceptually different engineering challenges. 
	
\end{itemize}

For the AID problem, future studies could delve into the integration of sophisticated computational fluid dynamics models like RANS or LES as the main HF simulators in the optimization loop, complemented by the ML model. Regarding the SFR problem, research emphasis should be on increasing the problem complexity, e.g.,  by employing a fully transient simulation model, integrating the diffusion coefficient value into both the ML model and inverse design, and potentially utilizing the RANS model for flow field reconstruction \cite{brunton2020machine}. Furthermore, an analysis of the influence of the number of field measurements should be conducted.

Generally, the ML-enhanced framework proposed here could find application in any problem where meaningful reduced-order information can be obtained and approximated using an ML model. Multiple scientific applications fall into this problem category including  simulations in climate and combustion that can be run with different grid resolutions and time step sizes.  The proposed framework could be implemented within a larger hybrid metaheuristic-Bayesian optimization framework to further minimize the number of HF function evaluations, and it could be further investigated with other derivative-free optimization algorithms.

\bmhead{Acknowledgments}

This work was supported by the Laboratory Directed Research and Development Program of Lawrence Berkeley National Laboratory under U.S. Department of Energy Contract No. DE-AC02-05CH11231. M\"uller's time was supported under U.S. Department of Energy Contract  No. DE-AC36-08GO28308. Funding for math developments was provided by U.S. Department of Energy Office of Science, Office of Advanced Scientific Computing Research, Scientific Discovery
through Advanced Computing (SciDAC) program through the FASTMath Institute. Funding for analysis of applications was provided by  the Laboratory Directed Research and Development Program of the National Renewable Energy Laboratory.

\section*{Declarations}

The authors have no relevant financial or non-financial interests to disclose.

\section*{Data availability}

All data required to reproduce the study is available in an online public repository: https://github.com/lukagrbcic/MLInverseDesignFrameworkData

\appendix

\section{AID Numerical Experiments and Dataset}
\label{app:airfoil_experiments_dataset}

This section details the target airfoil designs, flow conditions, and the flow simulation solver. Additionally, it describes the dataset used to train the ML models for the AID problem.

\subsection{Airfoils and Aerodynamic Flow Analysis}
\label{sec:airfoil_flow_analysis}

\begin{figure}[!h]
	\centering
	\begin{subfigure}[b]{0.45\textwidth}
		\includegraphics[width=\linewidth]{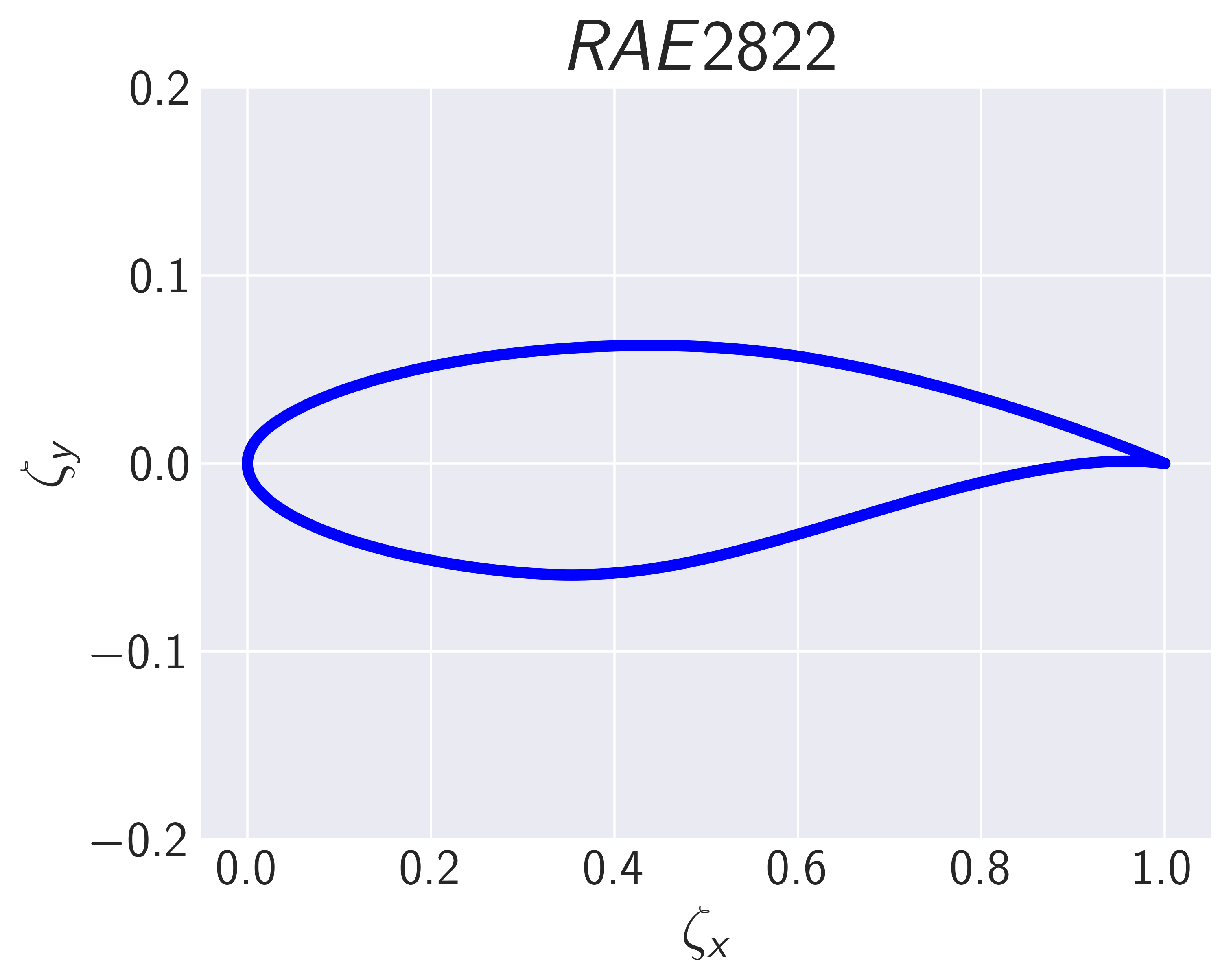}
		\caption{}
		\label{fig:rae2822} 
	\end{subfigure}
	\begin{subfigure}[b]{0.45\textwidth}
		\includegraphics[width=\linewidth]{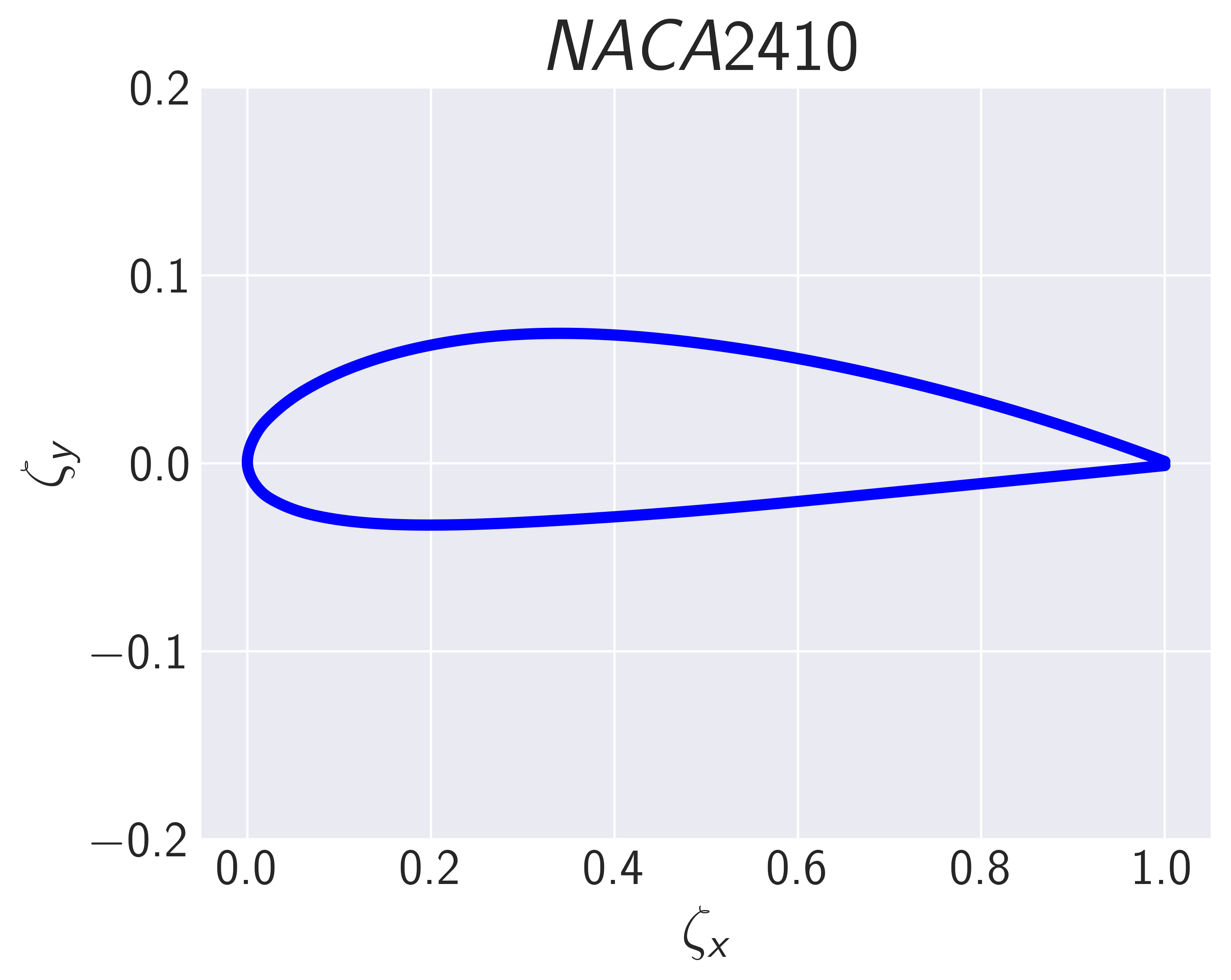}
		\caption{}
		\label{fig:naca2410}
	\end{subfigure}
	\begin{subfigure}[b]{0.45\textwidth}
		\includegraphics[width=\linewidth]{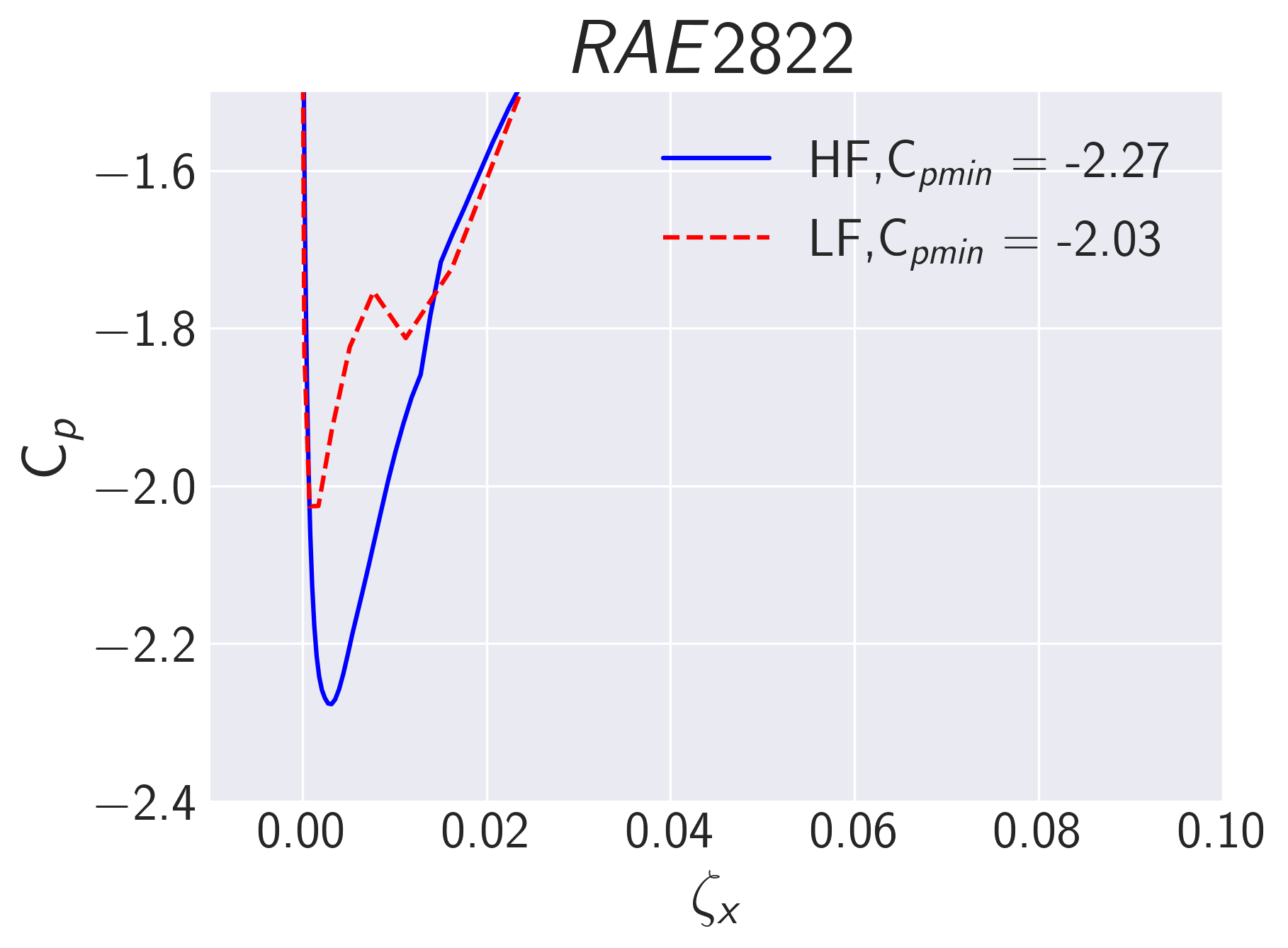}
		\caption{}
		\label{fig:rae2822_fidelity}
	\end{subfigure}
	\begin{subfigure}[b]{0.45\textwidth}
		\includegraphics[width=\linewidth]{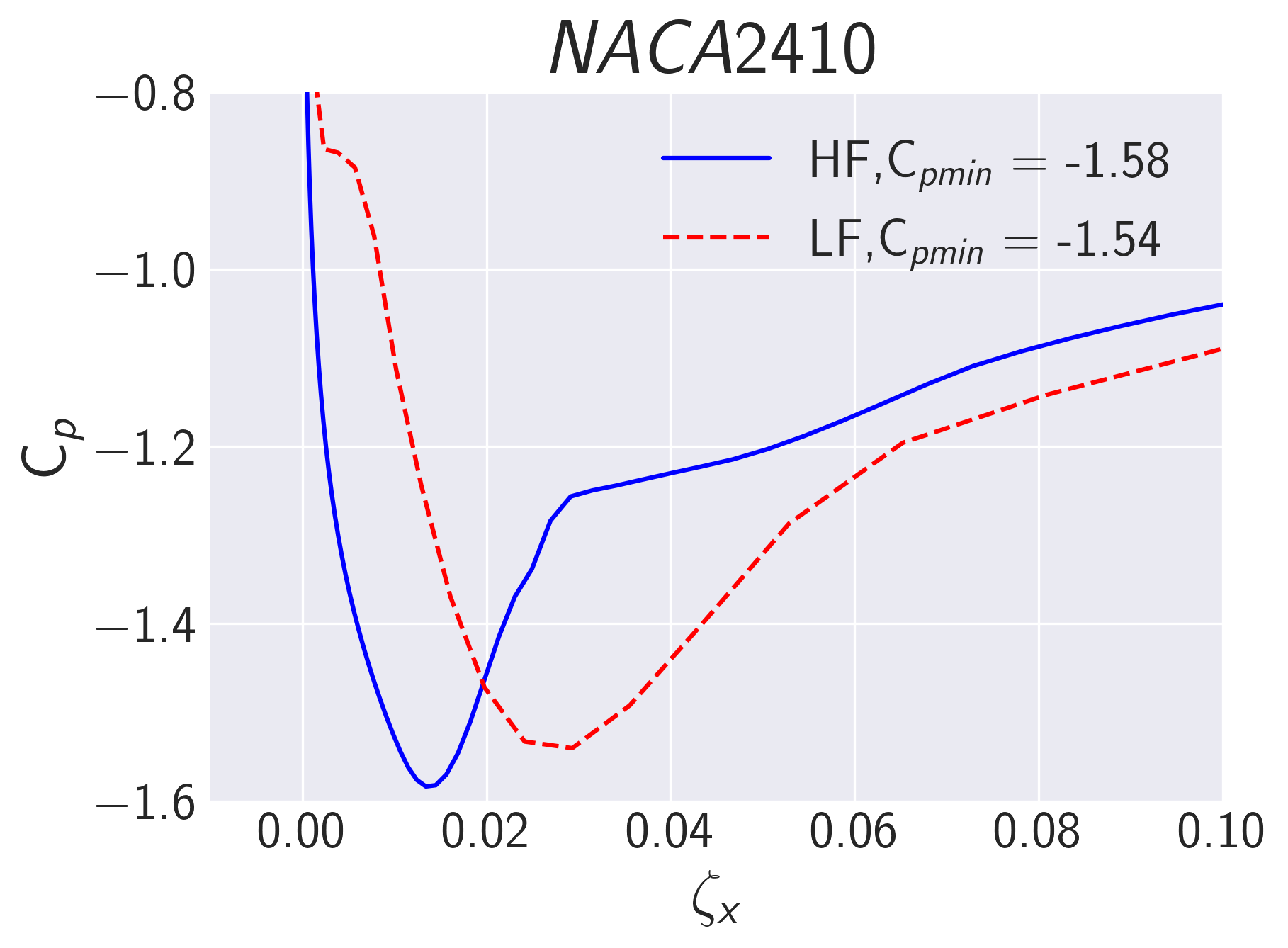}
		\caption{}
		\label{fig:naca2410_fidelity}
	\end{subfigure}
	\caption[Two airfoils]{Investigated airfoil geometries (top row) and the HF vs LF pressure coefficient distribution around the target $C^T_{p_{min}}$ area (bottom row): (a) $RAE2822$ airfoil (b) $NACA2410$ airfoil (c) $RAE2822$ LF vs HF pressure distribution (d) $NACA2410$ LF vs HF pressure distribution. Due to the optimization constraints being generated based on the $NACA0012$ airfoil as well as the ML model dataset (Sect. \ref{sub:AIDdataset}), the $RAE2822$ shape provides a harder optimization challenge due to its larger deviation from the camber line. 
	}
	\label{fig:shapes}
\end{figure}

The airfoils $RAE2822$ and $NACA2410$ are used for the numerical analysis. They  both exhibit asymmetry along the chord line unlike the base $NACA0012$ airfoil which was used to construct the original lower and upper boundaries of the decision vector. The $NACA2410$ airfoil is a member of the same family as the $NACA0012$, which serves as a reference for defining the B-Spline coefficient constraints. The $RAE2822$ airfoil is one of the most widely used benchmark airfoils in the field of aerodynamic shape optimization and inverse design \cite{li2022machine, deng2023fast, han2013improving, li2023efficient}. The shapes of both airfoils are shown in the top row of Fig.~\ref{fig:shapes}, and the difference between the HF and LF simulation results for both airfoils through the $C_p$ distribution graph are shown in the bottom row of Fig.~\ref{fig:shapes}.

For both investigated airfoils, the flow simulation parameters--Reynolds number (Re), Angle of Attack (AoA), and Mach number (Ma)--were set to 5$\cdot10^7$, 4, and 0, respectively. The target pressure coefficients were derived from these flow conditions and the specific airfoils, including the $C^T_{p_{min}}$ values. Specifically, for the $RAE2822$,    $C^T_{p_{min}}=-2.27$ and  for the $NACA2410$, $C^T_{p_{min}}=-1.58$. The aerodynamic flow analysis was conducted using XFOIL. This software package, specifically designed for subsonic airfoil analysis, served as the primary tool for assessing the pressure coefficients around the airfoil \cite{drela1989xfoil}. The Python wrapper for XFOIL simulations -- xfoil 1.1.1 \cite{xfoilpython} was utilized. 

XFOIL operates on a numerical panel method, which is integrated with a boundary layer model, facilitating accurate predictions of flow behavior around an airfoil. Through an iterative process, XFOIL effectively solves the potential flow equation for inviscid flows and the integral boundary layer equations for momentum and energy in viscous flows. It is optimally designed to accommodate incompressible flow scenarios with a Reynolds number between $10^6$ and $10^8$. The number of discretization panels used for XFOIL simulations determines the fidelity of the simulation. It has been shown by \citet{morgado2016xfoil} that XFOIL is more accurate than other methods for high lift low Reynolds number airfoils. In HF simulations, the discretization panel value is set to 300, while in LF simulations it is reduced to 100.

For each analysis, XFOIL takes as input the airfoil design, which is represented by the coordinates generated by the optimization variables--B-Spline coefficients, as well as B-Spline degree, and knots. Additional parameters, such as Re, AoA, and Ma must be specified for each simulation. Each XFOIL evaluation outputs the pressure coefficients measured around the airfoil which are compared with the target pressure coefficients. The number of iterations was set to 400 for every simulation, while the panel bunching parameter was set to 1, the trailing and leading edge density ratio was set to 0.15, and the refined-area-leading edge panel density ratio was set to 0.2.

While the proposed inverse design framework can leverage various computational fluid dynamics (CFD) analysis tools, XFOIL has been selected for its computational efficiency and as a proof-of-concept. The difference in execution time between the HF and LF simulations generated by XFOIL is not significant, however, the quality of solution does differ (shown in Fig.~\ref{fig:shapes}). In the future, this methodology can easily be expanded to incorporate more sophisticated approaches, such as Reynolds-averaged Navier--Stokes (RANS) or Large Eddy Simulation (LES), which both have significantly  higher computational demands.

\subsection{Minimum Pressure Coefficient Dataset}
\label{sub:AIDdataset}

In order to train the ML model, a suitable dataset must be generated. As defined in Table \ref{tab:airfoil_parameters}, the $M_{info}$ value corresponds to the minimum pressure coefficient, denoted $C_{p_{\text{min}}}$, necessitating the simulation of aerodynamic properties across a wide array of geometries and their mapping to respective $C_{p_{min}}$ values.
This dataset  was assembled utilizing the LHS design of experiment technique. Input features for training the ML model were generated using LHS as B-Spline coefficients ($\mathbf{c_u}$ and $\mathbf{c_l}$ in Eq. (\ref{eqn:bspline_variables})). Each B-spline coefficient was subsequently transformed into an airfoil geometry to obtain the corresponding $C_{p_{min}}$ value. All data were generated utilizing LF simulations, employing 100 discretization panels, with the flow parameters defined in Sect. \ref{sec:airfoil_flow_analysis}. A total of 15000 LF simulations were conducted, meaning a total of 15000 B-Spline and C$_{p_{min}}$ pairs were generated.

\section{SFR Numerical Experiments and Dataset}
\label{app:sfr_experiments_dataset}

This section provides details on the inverse design targets and the solver used for simulating the scalar diffusion process. It also includes information on the scalar measurement locations and the random generator algorithm for the scalar boundary values. Additionally, it describes the specifics of the ML dataset generated for the SFR problem.

\subsection{Scalar Diffusion Boundary Conditions and Solver}
\label{sec:sd_solver}

\begin{figure}[!h]
	\centering
	\begin{subfigure}[b]{0.6\textwidth}
		\includegraphics[width=1\linewidth]{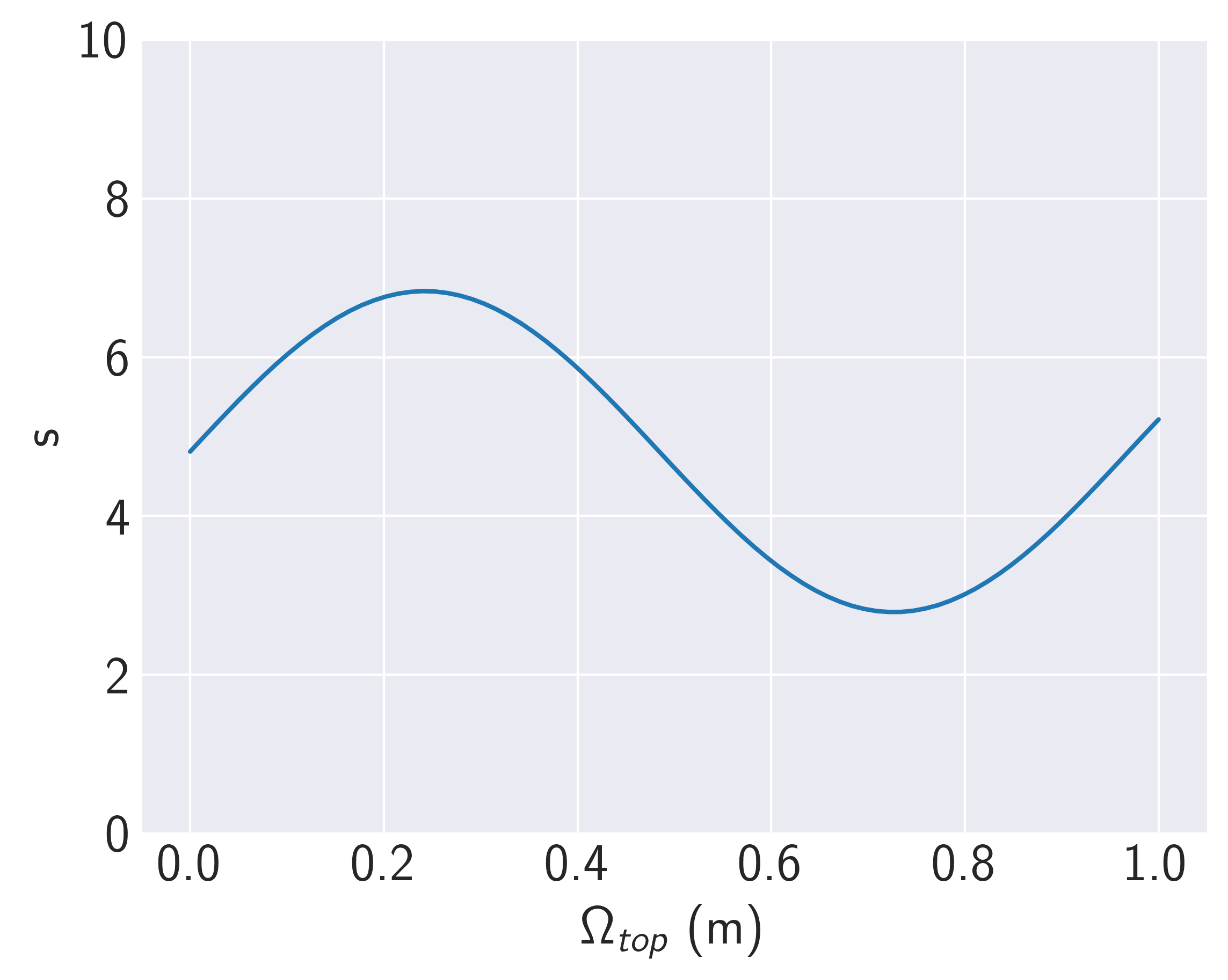}
		\caption{}
		\label{fig:bc1} 
	\end{subfigure}
	\begin{subfigure}[b]{0.6\textwidth}
		\includegraphics[width=1\linewidth]{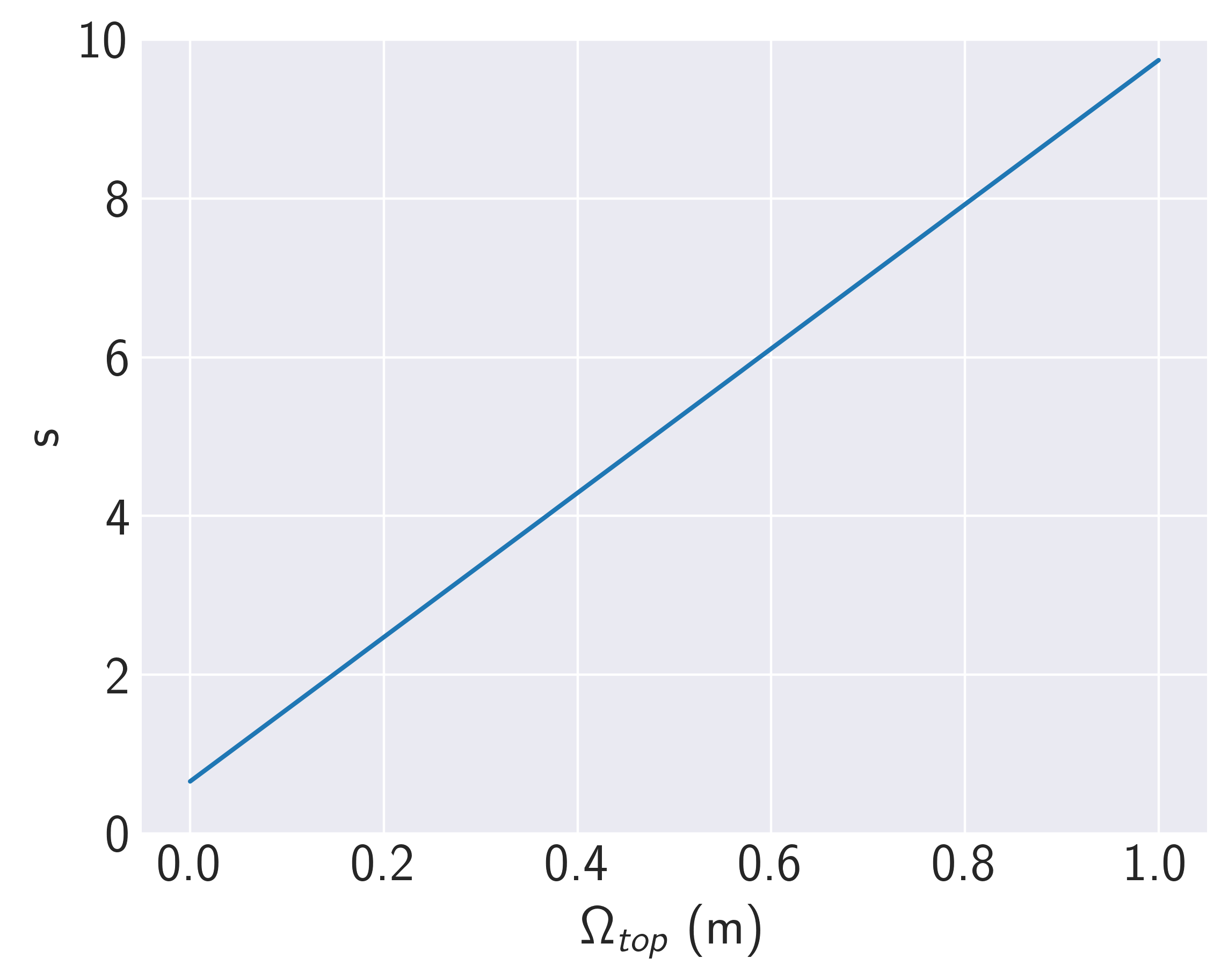}
		\caption{}
		\label{fig:bc2}
	\end{subfigure}
	\caption[Two bcs]{Two investigated boundary condition cases: (a) Sinusoidal boundary condition (b) Linear boundary condition. The x-axis defines the top boundary domain which ranges from 0 to 1, while the y-axis is the scalar value for each $\Omega_{top}$.}
	\label{fig:boundary_conditions}
\end{figure}

Two distinct boundary conditions (BC) were investigated to demonstrate the versatility of a single ML model across various scenarios. As depicted in Fig. \ref{fig:boundary_conditions}, one BC exhibits a sinusoidal pattern, whereas the other adheres to a linear trend. Both BCs were used to generate $\mathbf{s}^{T}$ arrays. The values were measured at locations given in Sect. \ref{sec:probes} (Fig. \ref{fig:probe_locations} and Table \ref{tab:probe_locations_table}). The $s^T_{{max}}$ ($T_{info}$) value for the sinusoidal BC (Fig. \ref{fig:bc1}) was 5.67, while for the linear BC (Fig. \ref{fig:bc2}) it was 9.1.

To simulate the BCs over the domain, the open source computational fluid dynamics library OpenFOAM 9 was used \cite{jasak2007openfoam}. More specifically, the \textit{laplacianFoam} diffusion PDE solver was used. The details about the LF and HF domains are presented in Table \ref{tab:lg_hf_mesh_of}, as well as the difference in the LF and HF modeled values of $s^T_{{max}}$. The LF scalar diffusion equation is solved on a computational grid that is 16 times smaller than the HF computational grid in terms of total finite volume cells. Moreover, both LF and HF OpenFOAM simulation execution times are similar, however, due to a difference between the obtained results, the cases can be utilized for investigation as a proof-of-concept for the ML-enhanced framework.

\begin{table}[h]
	\caption{``Total cells''  refers to the number of cells  within the computational domain. ``Top BC cells''  signifies the quantity of cells along the $\Omega_{x}$-axis direction, where the Dirichlet boundary condition is applied. Scalar values at the boundary are set in the cell centers.  $\delta \Omega_{x}$ and $\delta \Omega_{y}$ represent the cell sizes in the $\Omega_{x}$ and $\Omega_{y}$ axis directions, respectively.}\label{tab:lg_hf_mesh_of}
	\begin{tabular}{@{}lllllll@{}}
		\toprule
		Type &Total cells& Top BC cells & $\delta \Omega_{x}$ & $\delta \Omega_{y}$ &$s^T_{{max}}$ Sinusoidal BC & $s^T_{{max}}$ Linear BC \\
		\midrule
		LF & 400 & 20 & 0.05 & 0.025 & 5.72 & 8.67 \\
		HF & 6400 & 80 & 0.0125 & 0.00625 & 5.67 & 9.10 \\
		\botrule
	\end{tabular}
	
\end{table}

\subsection{Maximum Scalar Dataset}
\label{sub:SFRdataset}

The $M_{info}$ for the SFR requires identification of the maximum scalar value, $s_{max}$ within the domain. To curate a dataset, variations were made in the BC scalar values. For every distinct scalar value set, a simulation was executed, capturing the corresponding $s_{max}$ value. An in-depth analysis of the methodology employed to generate diverse BCs for this reconstruction problem can be found in Alg. \ref{alg:bc_generator_algorithm} presented in Sect. \ref{sec:sc_bc_gen}. For dataset creation, 15000 LF simulations were executed using OpenFOAM 9, with full details provided in Sect. \ref{sec:sd_solver}.

\subsection{Scalar Field Domain Probe Locations}\label{sec:probes}

\begin{figure}[!h]
	\centering
	\includegraphics[width=0.75\linewidth]{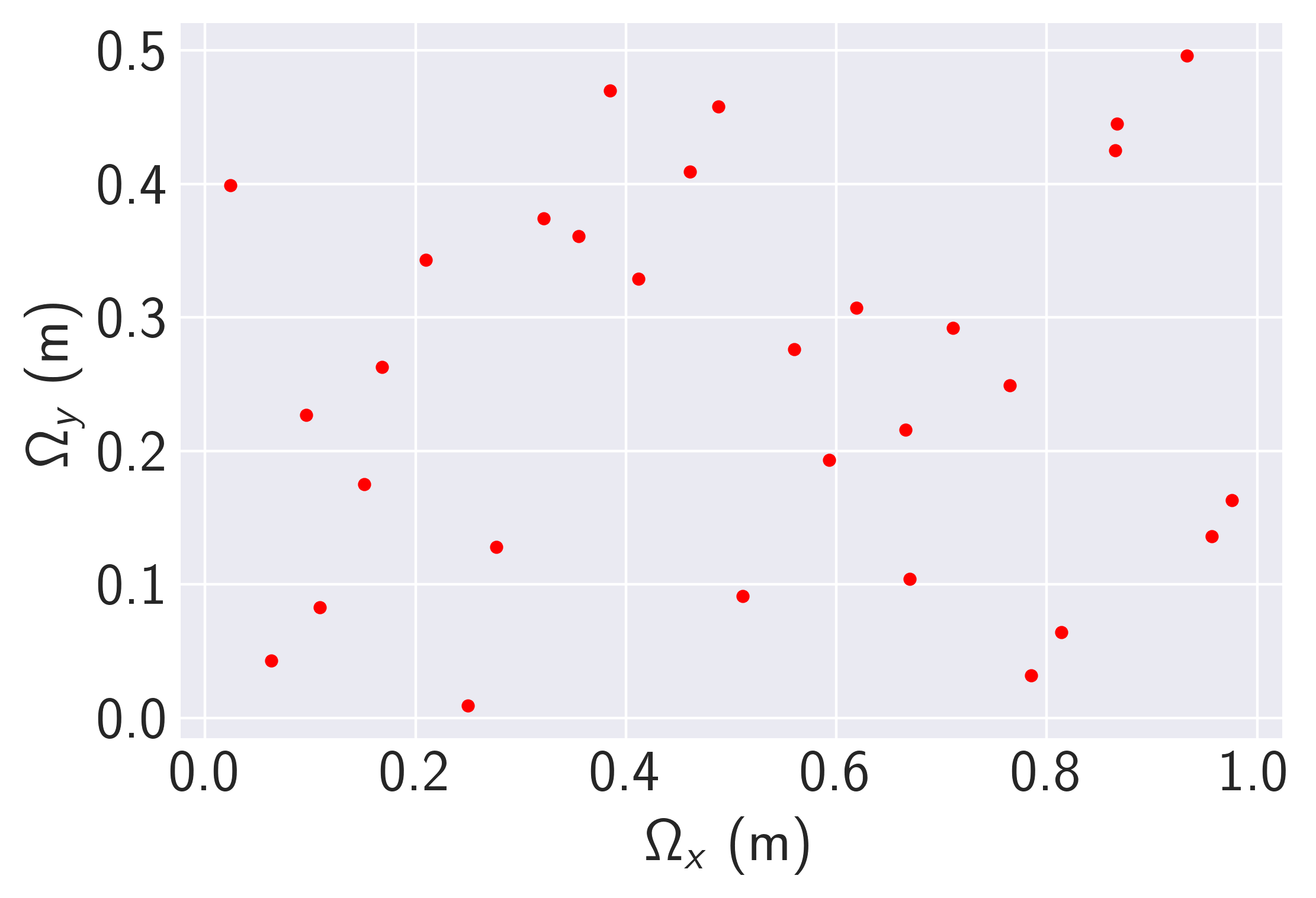}
	\caption{SFR problem domain $\Omega$ with probe locations marked as red points. The boundary condition values are given  at the top of the domain ($\Omega_y$ = 0.5 m).}
	\label{fig:probe_locations}
\end{figure}

In this section, the scalar measurement locations within the domain $\Omega$, used as postprocessing points for each OpenFOAM simulation, are presented. For the achievement of the target performance (scalar distribution) in both BC scenarios and for the training of ML models, the probe locations depicted in Fig. \ref{fig:probe_locations} and listed in Table \ref{tab:probe_locations_table} are to be used.

\begin{table}[h]
	\caption{$\Omega_x$-axis and $\Omega_y$-axis values for all probe locations within the domain $\Omega$ for the SFR problem.}\label{tab:probe_locations_table}
	\begin{tabular}{@{}lll@{}}
		\toprule
		Probe & $\Omega_x$&  $\Omega_x$ \\
		\midrule
		1 &0.168 & 0.263 \\
		2 &0.063  &0.043 \\
		3 &0.867 &0.445 \\
		4 &0.711 &0.292 \\
		5 &0.412& 0.329 \\
		6 &0.593& 0.193 \\
		7 &0.096& 0.227\\
		8 &0.670 &0.104 \\
		9 &0.814 &0.064\\
		10 &0.109 &0.083\\
		11 &0.666 &0.216 \\
		12 &0.024 & 0.399 \\
		13 &0.560 & 0.276 \\
		14 &0.322 & 0.374 \\
		15 &0.250 & 0.009 \\
		16 &0.210 & 0.343 \\
		17 &0.277 & 0.128 \\
		18 &0.957 & 0.136 \\
		19 &0.933 & 0.496 \\
		20 &0.151 & 0.175 \\
		21 &0.461 & 0.409\\
		22 &0.385 & 0.470 \\
		23 &0.785 & 0.032 \\
		24 &0.511 & 0.091 \\
		25 &0.488 & 0.458\\
		26 &0.619 & 0.307 \\
		27 &0.355 & 0.361 \\
		28 &0.865 &0.425 \\
		29 &0.976 &0.163 \\
		30 &0.765 &0.249 \\
		\botrule
	\end{tabular}
	
\end{table}

\subsection{Scalar Boundary Condition Generator}
\label{sec:sc_bc_gen}

Alg. \ref{alg:bc_generator_algorithm} illustrates the method used to generate the BC for SFR. The method begins with the values $LF_n$ and $s_{\Omega_{top}}$, which represent the number of LF discretization points at the top of the boundary and maximum scalar value that can be set a the top of the domain $\Omega$, respectively. The vector $\mathbf{BC}_{init}$ represents the initial BC. It consists of points that are equally spaced and sized $LF_n$. These points are derived from linear interpolation of values ranging from 1 to $LF_n$. A value $R$ is randomly chosen from a uniform distribution, representing one of three states that signify different BC variations: linear, parabolic, or sinusoidal. $\mathbf{G}_{noise}$ is a random vector, generated from a normal distribution, with a length of $LF_n$. Its standard deviation, $\sigma$, is drawn from a uniform distribution. This vector is added to the transformed $\mathbf{BC}$ to enhance model robustness, simulate real-world scenarios, and ensure better generalization in imperfect or noisy environments.

For each run, one of the three BC types is chosen and $\mathbf{BC}_{init}$ is transformed using the corresponding equation (linear, parabolic, or sinusoidal) incorporating randomly generated values $rand_1$, $rand_2$, and $rand_3$ from a uniform distribution. If the resultant $\mathbf{BC}$ with added noise $\mathbf{G}_{noise}$ has values exceeding $s_{max}$, they are substituted with a random value between 0 and $s_{max}$. To further diversify the generated BCs, if a random value between 0 and 1 is less than 0.5, the $\mathbf{BC}$ is reversed.

\begin{algorithm}
\begin{algorithmic}[1]
	\Procedure{GenerateBC}{$LF_n$, $s_{\Omega_{top}}$}
	\State \(\mathbf{BC}_{init} = \{bc_i\}_{i=1}^{LF_n} \)
	\State \(R \sim U(\{0, 1, 2\})\)
	\State \(\sigma \sim U(0, 100)\) \Comment{Scalar noise standard deviation}
	\State \(\mathbf{G}_{noise} \sim \mathcal{N}(0, \sigma )\) \Comment{Gaussian noise vector of length $LF_n$}
	\If{\(R == 0\)}
	\State \(rand_1, rand_2 \sim U(0, 10)\)
	\State \(\mathbf{BC} \gets rand_1 \times \mathbf{BC}_{init} + rand_2\) \Comment{Linear}
	\ElsIf{\(R == 1\)}
	\State \(rand_1, rand_2, rand_3 \sim U(0, 10)\)
	\State \(\mathbf{BC} \gets rand_1 \times \mathbf{BC}_{init}^2 + rand_2 \times \mathbf{BC}_{init} + rand_3\) \Comment{Parabolic}
	\Else
	\State \(rand_1, rand_2, rand_3 \sim U(0, 10)\)
	\State \(\mathbf{BC} \gets rand_1 \times \sin(rand_3 \times \mathbf{BC}_{init}) + rand_2\) \Comment{Sinusoidal}
	\EndIf
	\State \(\mathbf{BC} \gets \mathbf{BC} + \mathbf{G}_{noise} \)
	\State Replace all \(bc_i > s_{\Omega_{top}}\) with \(U(0, s_{\Omega_{top}})\)
	\If{\(U(0, 1) < 0.5\)}
	\State \(\mathbf{BC} \gets \) reverse of \(\mathbf{BC}\)
	\EndIf
	\State \Return \(|\mathbf{BC}|\)
	\EndProcedure
\end{algorithmic}
\caption{The algorithm for generating the scalar reconstruction problem BC for the purpose of ML model training dataset creation.}
\label{alg:bc_generator_algorithm}
\end{algorithm}

\newpage
\section{ML Algorithms Hyperparameters}
\label{app:ml_hyperparams}

In this section, the optimized hyperparameters of all investigated ML algorithms are presented. The best performing algorithm was further used as a part of the enhanced inverse design framework. All three ML algorithms were optimized using the Python framework for hyperparameter optimization Optuna 3.1.0. \cite{akiba2019optuna}. The number of trials for all three algorithms was 100, and Optuna-based hyperparameter optimization goal was to minimize the average $RMSE$ of a shuffled K-Fold ($k=3$) cross-validation procedure. The ML algorithms were separately tuned for both investigated problems/datasets (described in Sect. \ref{sub:AIDdataset} and Sect. \ref{sub:SFRdataset}), and 15000 LF data instances were used for optimization. The optimal set of hyperparameters was independently selected for each of the three investigated algorithms, based on the results from 100 trials conducted using Optuna. 

The XGB algorithm hyperparameters used are presented in Table \ref{tab:xgb_hyper}. The \textsf{max\_depth} parameter controls the depth of each tree, the \textsf{n\_estimators} defines the total number of gradient boosted trees in the model, \textsf{learning\_rate} scales the contribution of each tree when it is added to the ensemble of trees, \textsf{colsample\_bytree} and \textsf{subsample} parameters specify the fraction of the randomly sampled features and data instances used to construct each tree, respectively,  and  \textsf{gamma}, \textsf{reg\_alpha} and \textsf{reg\_lambda} are regularization parameters. The Python module xgboost 1.7.4 was used.

\begin{table}[h]
	\caption{XGB model hyperparameters tuned with the Optuna Python framework. The best solution  of 100 trials is shown. The first column denotes the names of the tuned hyperparameters, while the second column shows the values obtained for the AID problem, and the third column shows the parameter values for the SFR problem.}\label{tab:xgb_hyper}
	\begin{tabular}{@{}lll@{}}
		\toprule
		Hyperparameter&  AID &  SFR \\ 
		\midrule
		max\_depth & 4 & 5 \\
		n\_estimators & 500 & 500 \\
		learning\_rate &  0.07 & 0.06 \\
		colsample\_bytree  & 0.94  & 0.19 \\
		subsample &  0.54  & 0.36 \\
		gamma & 0.42 &  1.33  \\
		reg\_alpha & 2.44 & 0.51 \\ 
		reg\_lambda & 4.16 & 0.20 \\ 
		\botrule
	\end{tabular}
	
\end{table}

The LGB model hyperparameters are presented in Table \ref{tab:lgb_hyper}. The \textsf{num\_iterations} parameter controls the number of boosting iterations performed.  Each iteration builds a new tree that boosts the performance of the model. The \textsf{learning\_rate} scales the contribution of each tree when it is added to the model (similarly as XGB), \textsf{lambda\_l1} and \textsf{lambda\_l2} are L1 and L2 regularization parameters added in order to reduce overfitting. The parameter \textsf{num\_leaves} controls the complexity of the model, and  \textsf{min\_child\_samples} refers to the minimum number of data instances a leaf node must have after a split, as a form of regularization. The  \textsf{feature\_fraction} parameter defines tha fraction of features used at each training iteration, while \textsf{bagging\_fraction} determines the number of data instances used at each iteration. Both parameters are also used as a form of regularization. The Python module LightGBM 3.3.5 was used.

\begin{table}[h]
	\caption{LGB model hyperparameters tuned with the Optuna Python framework (the best solution out of 100 for each problem). The first column denotes the names of the tuned hyperparameters,  the second column shows the values obtained for the AID problem, and the third column shows the parameter values for the SFR problem.}\label{tab:lgb_hyper}
	\begin{tabular}{@{}lll@{}}
		\toprule
		Hyperparameter&  AID &  SFR \\ 
		\midrule
		num\_iterations & 2500 & 2500 \\
		learning\_rate &  0.0187 & 0.0190 \\
		lambda\_l1  & 1.78  & 0.40 \\
		lambda\_l2 &  7.43  & 6.71 \\
		num\_leaves & 41 &  68  \\
		min\_child\_samples & 37 & 99 \\ 
		feature\_fraction & 0.77 & 0.56 \\
		bagging\_fraction & 0.45 & 0.52\\ 
		\botrule
	\end{tabular}
	
\end{table}

The hyperparameters for the MLP are detailed in Table \ref{tab:dnn_hyper}. The \textit{LeakyReLU} activation function was applied to all hidden layers in the AID problem, whereas the SFR used \textit{ReLU}. Monte Carlo dropout layers were integrated into the architecture to reduce overfitting. During training for both problems, 30\% of the data was reserved for validation. An early stopping criterion with a patience value of 20 was set based on the validation loss to further combat overfitting. The MLP was implemented in  Tensorflow 2.11.0 \cite{abadi2016tensorflow}.

\begin{table}[h]
	\caption{MLP hyperparameters tuned with the Optuna Python framework (the best solution out of 100 trials is shown for each problem). The first column denotes the names of the tuned hyperparameters,  the second column shows the values obtained for the AID problem, and the third column shows the parameter values for the SFR problem.}\label{tab:dnn_hyper}
	\begin{tabular}{@{}lll@{}}
		\toprule
		Hyperparameter&  AID &  SFR \\ 
		\midrule
		Layers & 3 & 2 \\
		Neurons per layer &  92,116,34 & 388,322 \\
		Dropout per layer  & 0.1,0.1,0.0  & 0.1,0.0 \\
		Activation function &  \textit{LeakyReLU}  & \textit{ReLU} \\
		Optimizer & \textit{Adam} &  \textit{Adam}  \\
		Epochs & 100 & 500 \\ 
		Batch size & 128 & 64 \\
		Learning rate & 0.00083 & 0.00060 \\
		\botrule
	\end{tabular}
	
\end{table}

\newpage
\section{Boundary Refinement Convergence}
\label{app:ssr_convergence}

\begin{figure}
	\centering
	\begin{subfigure}[b]{0.5\textwidth}
		\includegraphics[width=1\linewidth]{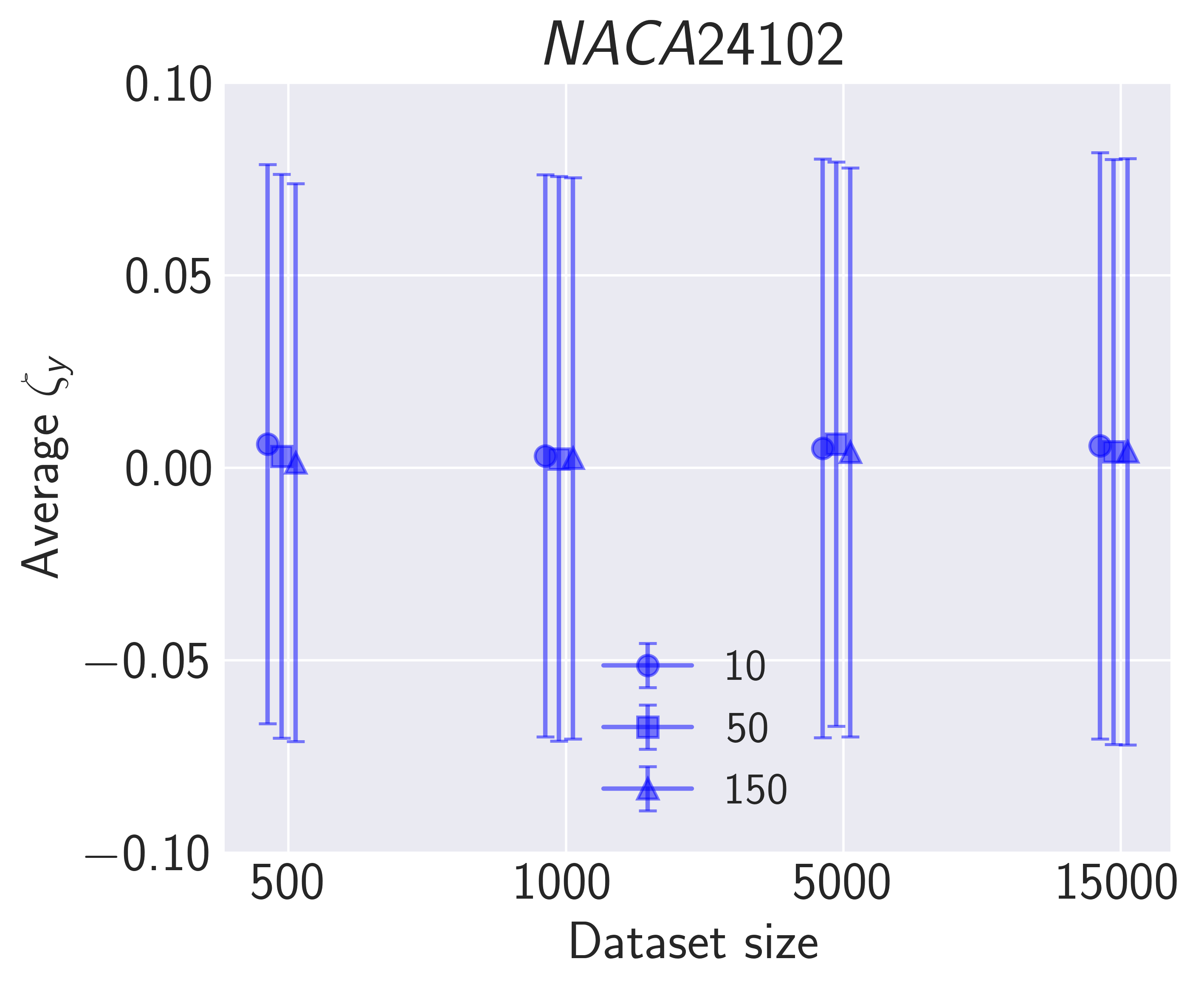}
		\caption{}
		\label{fig:naca2410_ssr_convergence} 
	\end{subfigure}
	
	\begin{subfigure}[b]{0.5\textwidth}
		\includegraphics[width=1\linewidth]{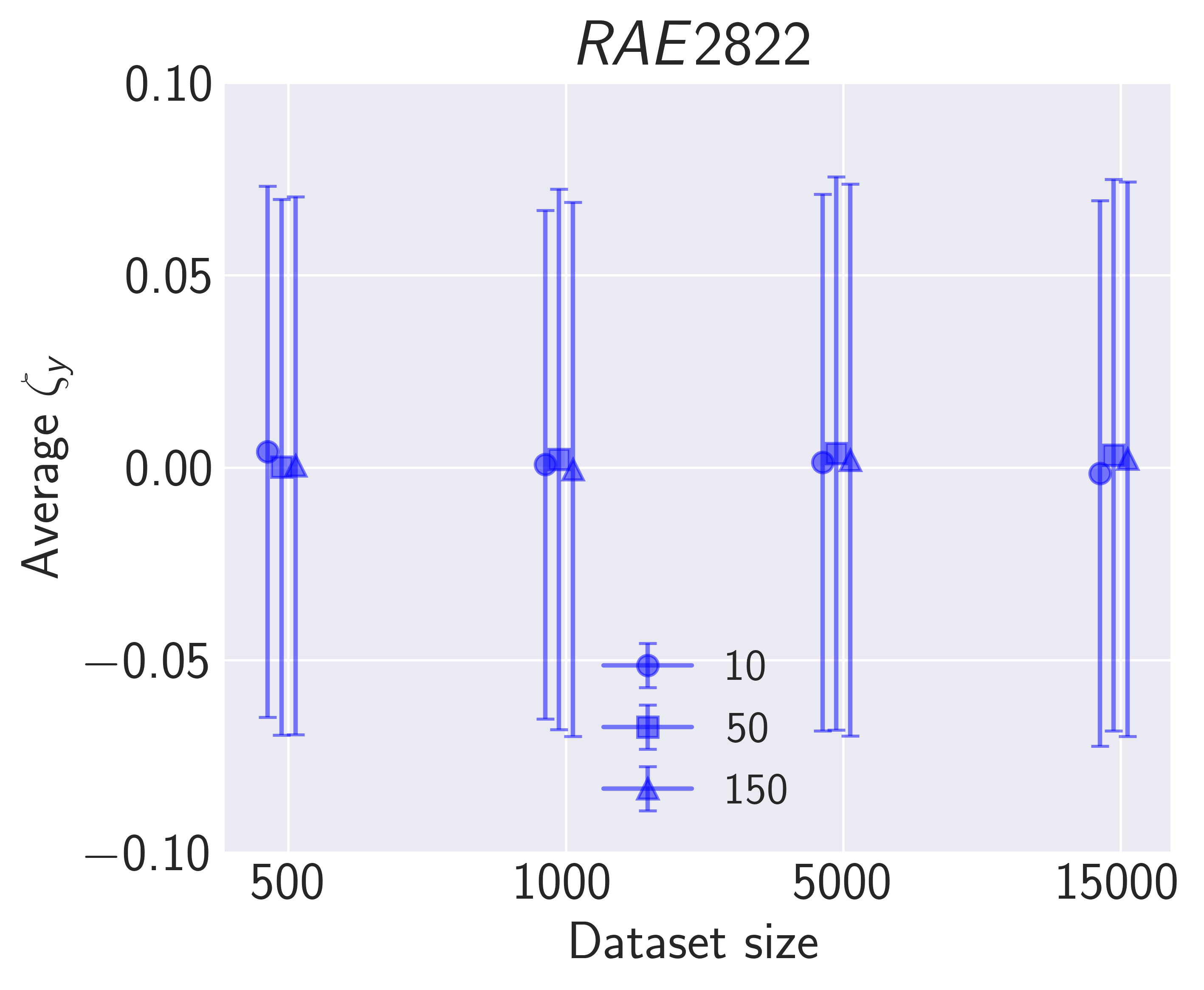}
		\caption{}
		\label{fig:rae2822_ssr_convergence}
	\end{subfigure}
	
	\caption[ssr conv]{Boundary refinement technique for different datasets sizes used to train the XGB model versus the average $\zeta_y$ for (a) $NACA2410$ airfoil and (b) $RAE2822$ airfoil. The number of solutions $N$ (denoted as runs) was varied to analyze how it affects the average $\zeta_y$. The markers are slightly offset for each $c$ value to improve visibility.}
	\label{fig:ssr_convergence_airfoil}
\end{figure}

Fig. \ref{fig:ssr_convergence_airfoil} illustrates the impact of both the dataset size used for training the XGB model and the number of solutions, $N$, derived from the boundary refinement technique on the formation of the new lower and upper boundaries, $\mathbf{lb}_R$ and $\mathbf{ub}_R$, respectively. Since these new boundaries can be interpreted as an airfoil shape, the effect of the dataset size and the $N$ value is articulated through the average and standard deviation of the $\zeta_y$ values (the chord length-normalized y-coordinates of the airfoil defined by $\mathbf{lb}_R$ and $\mathbf{ub}_R$).

For 10, 50, and 150 runs (or solutions), the average $\zeta_y$ and its standard deviation bandwidth exhibit only minor variations as the dataset size increases. This trend is discernible for both $NACA2410$ and $RAE2822$ in Fig. \ref{fig:naca2410_ssr_convergence} and Fig. \ref{fig:rae2822_ssr_convergence}. This implies that even a boundary refinement formulated by an XGB model trained with just 500 data instances and merely 10 repeated runs could be beneficial, as the boundaries remain relatively consistent despite increases in both parameters.

\begin{figure}
	\centering
	\begin{subfigure}[b]{0.5\textwidth}
		\includegraphics[width=1\linewidth]{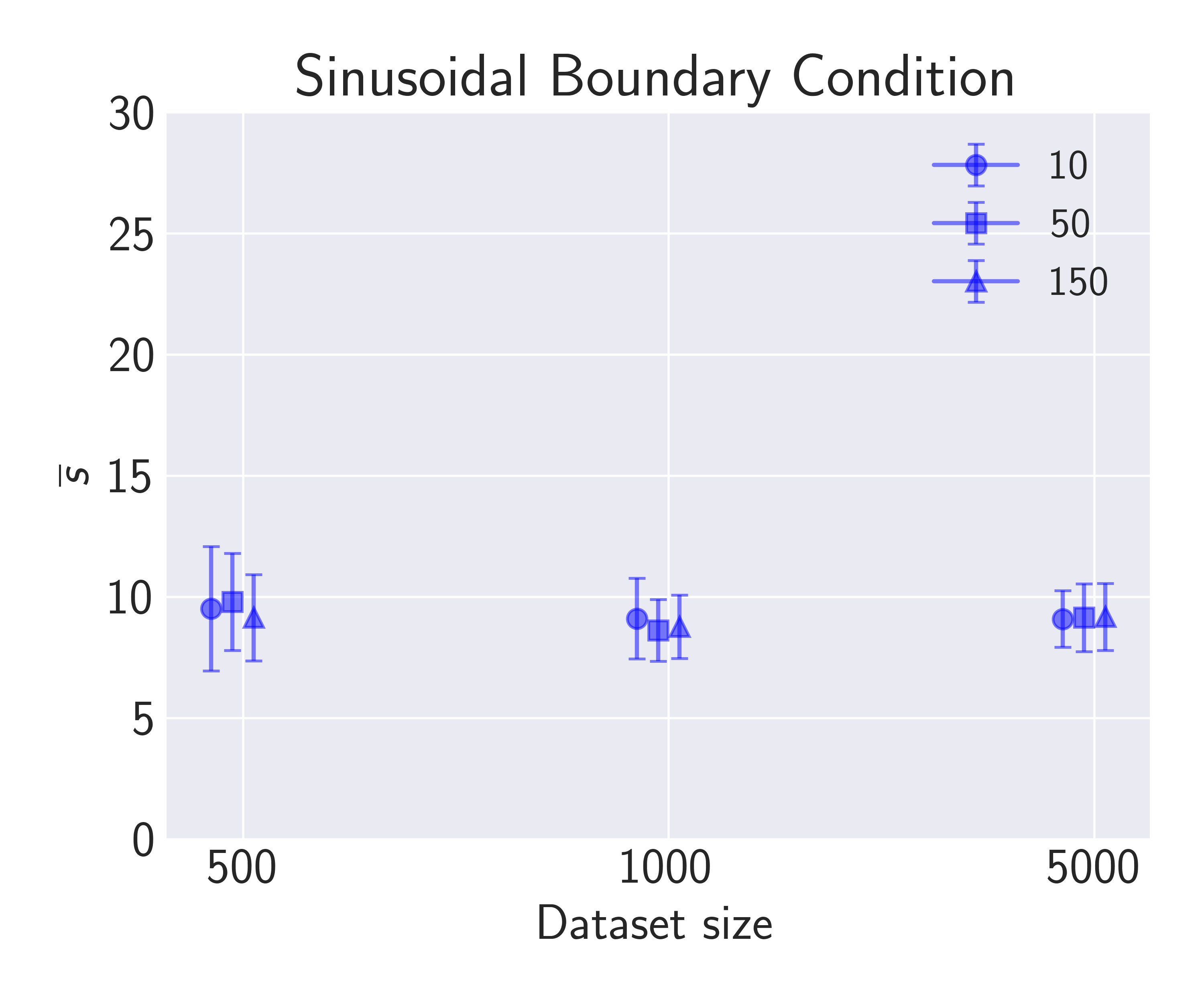}
		\caption{}
		\label{fig:case1_ssr_convergence} 
	\end{subfigure}
	
	\begin{subfigure}[b]{0.5\textwidth}
		\includegraphics[width=1\linewidth]{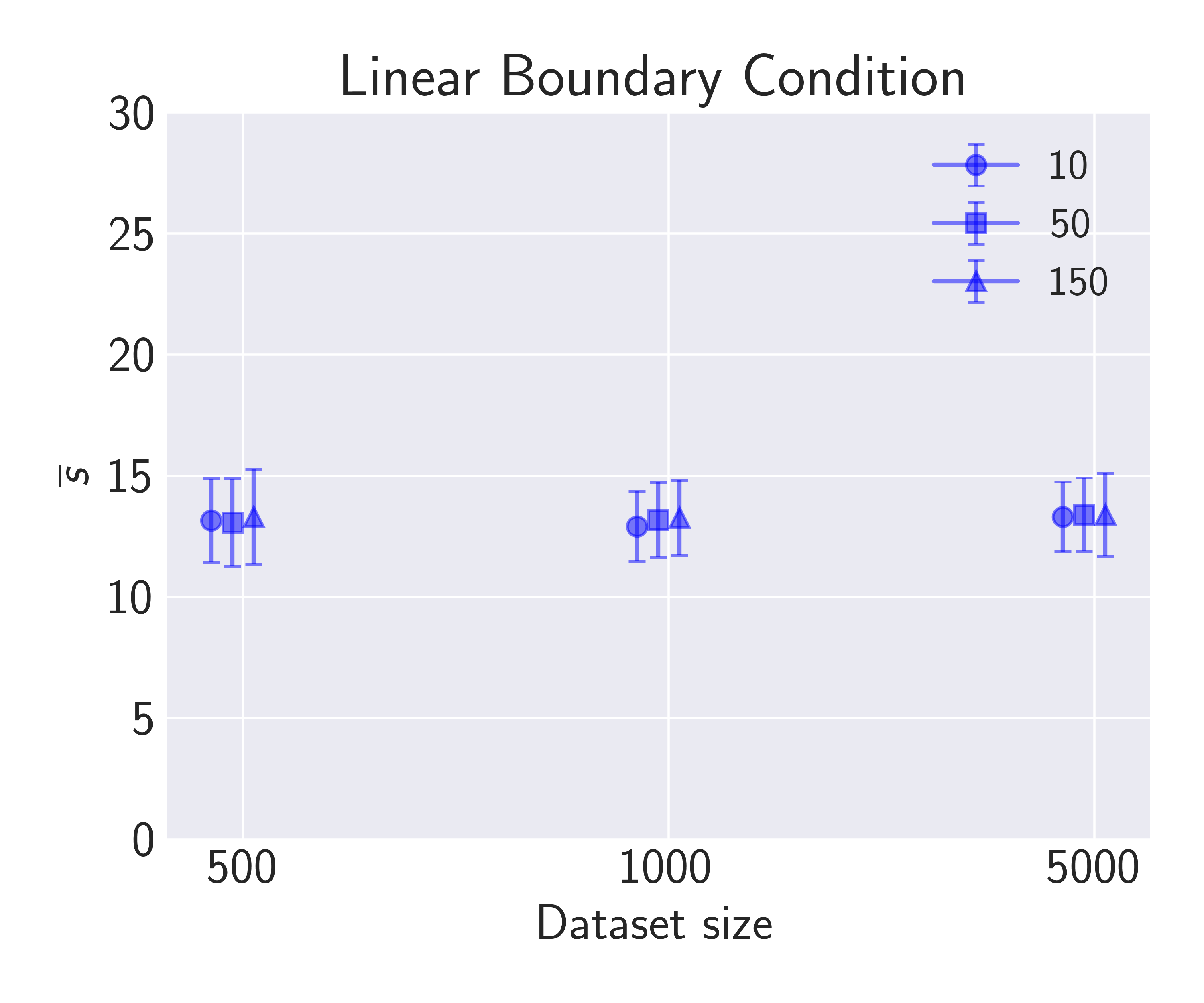}
		\caption{}
		\label{fig:case2_ssr_convergence}
	\end{subfigure}
	
	\caption[ssr conv]{Boundary refinement technique for different datasets sizes used to train the MLP model versus the average scalar value of the BC $\overline{s}$ for the (a) Sinusoidal BC and (b) Linear BC. The number of solutions $N$ (denoted as runs) was varied to observe its influence on the average $s$ value. The markers are slightly offset for each $c$ value to improve visibility.}
	\label{fig:ssr_convergence_bcs}
\end{figure}

Fig. \ref{fig:ssr_convergence_bcs} demonstrates the impact of the number of solutions, $N$, and the dataset size on the average scalar value $\overline{s}$ of the BC. Mirroring observations from the AID boundary refinement, neither the dataset size nor the number of solutions exert a significant effect on $\overline{s}$.

\newpage
\section{AID Results for $\eta$ = 1.1 and $\eta$ = 1.2}
\label{app:aid_full_analysis}

\begin{figure*}[h!]
	\centering
	\begin{subfigure}{0.45\textwidth}
		\centering
		\includegraphics[width=\linewidth]{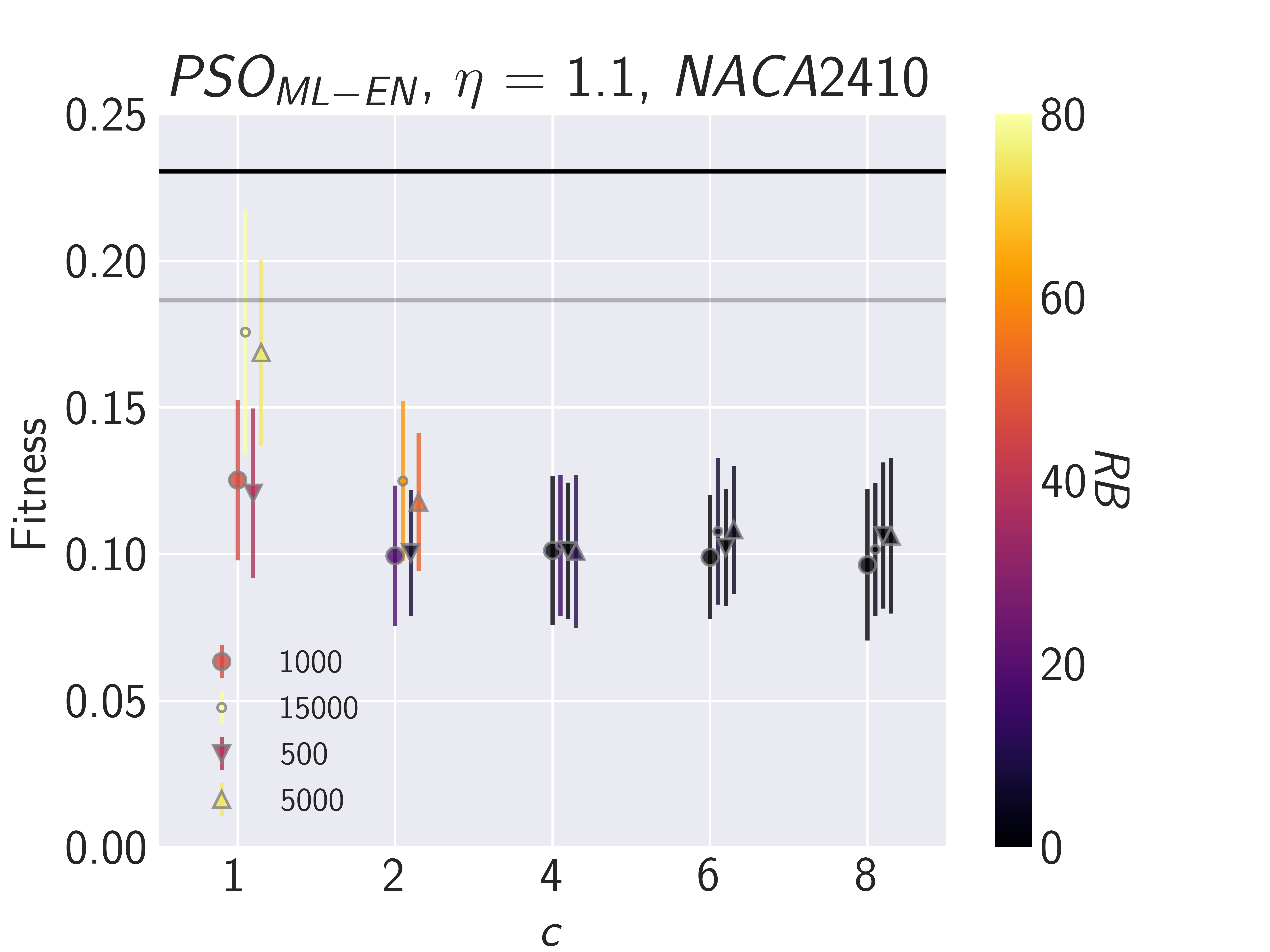}
		\caption{}
		\label{fig:naca2410_1_1_pso}
	\end{subfigure}%
	\hfill
	\begin{subfigure}{0.45\textwidth}
		\centering
		\includegraphics[width=\linewidth]{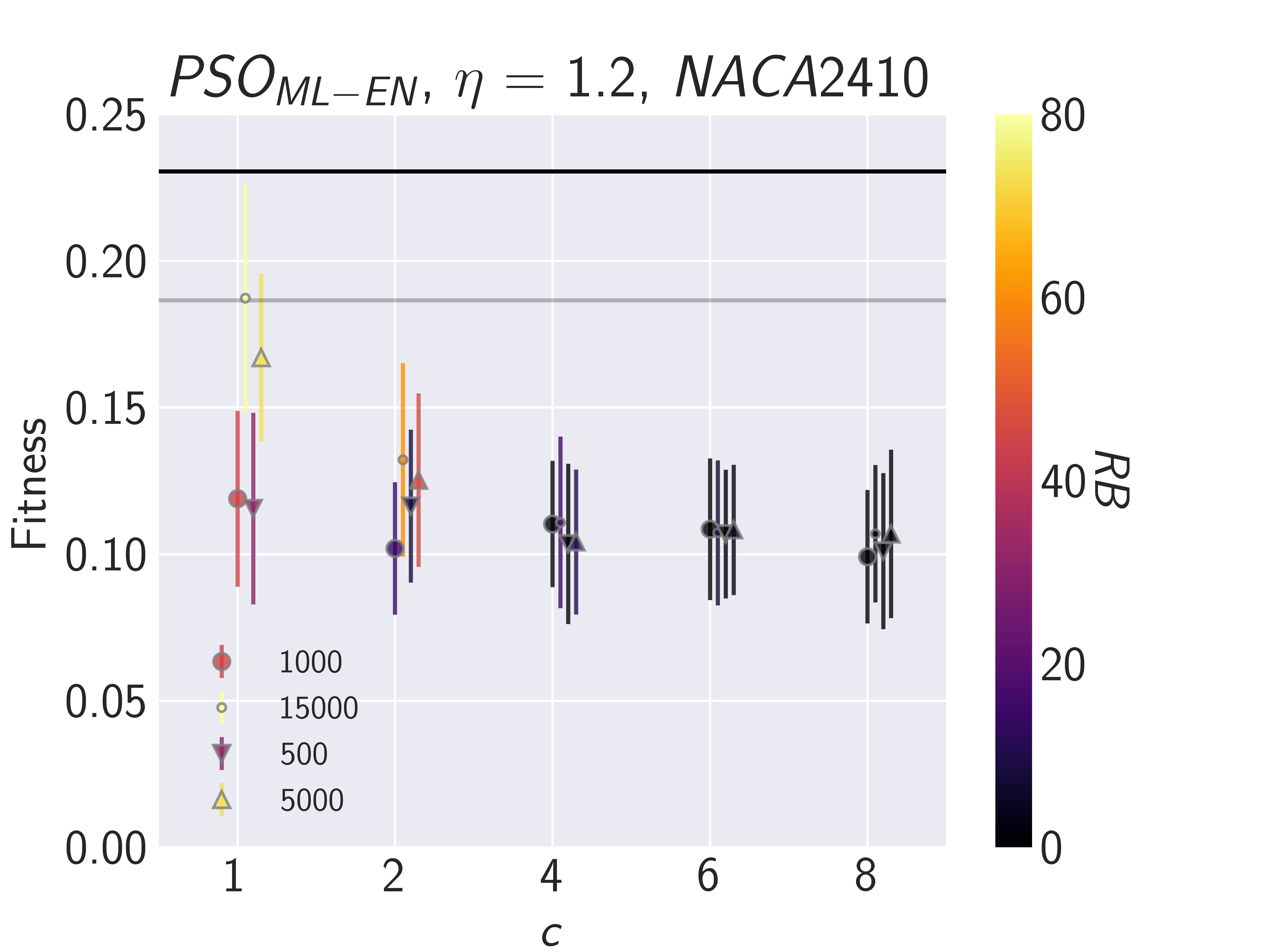}
		\caption{}
		\label{fig:naca2410_1_2_pso}
	\end{subfigure}
	
	\begin{subfigure}{0.45\textwidth}
		\centering
		\includegraphics[width=\linewidth]{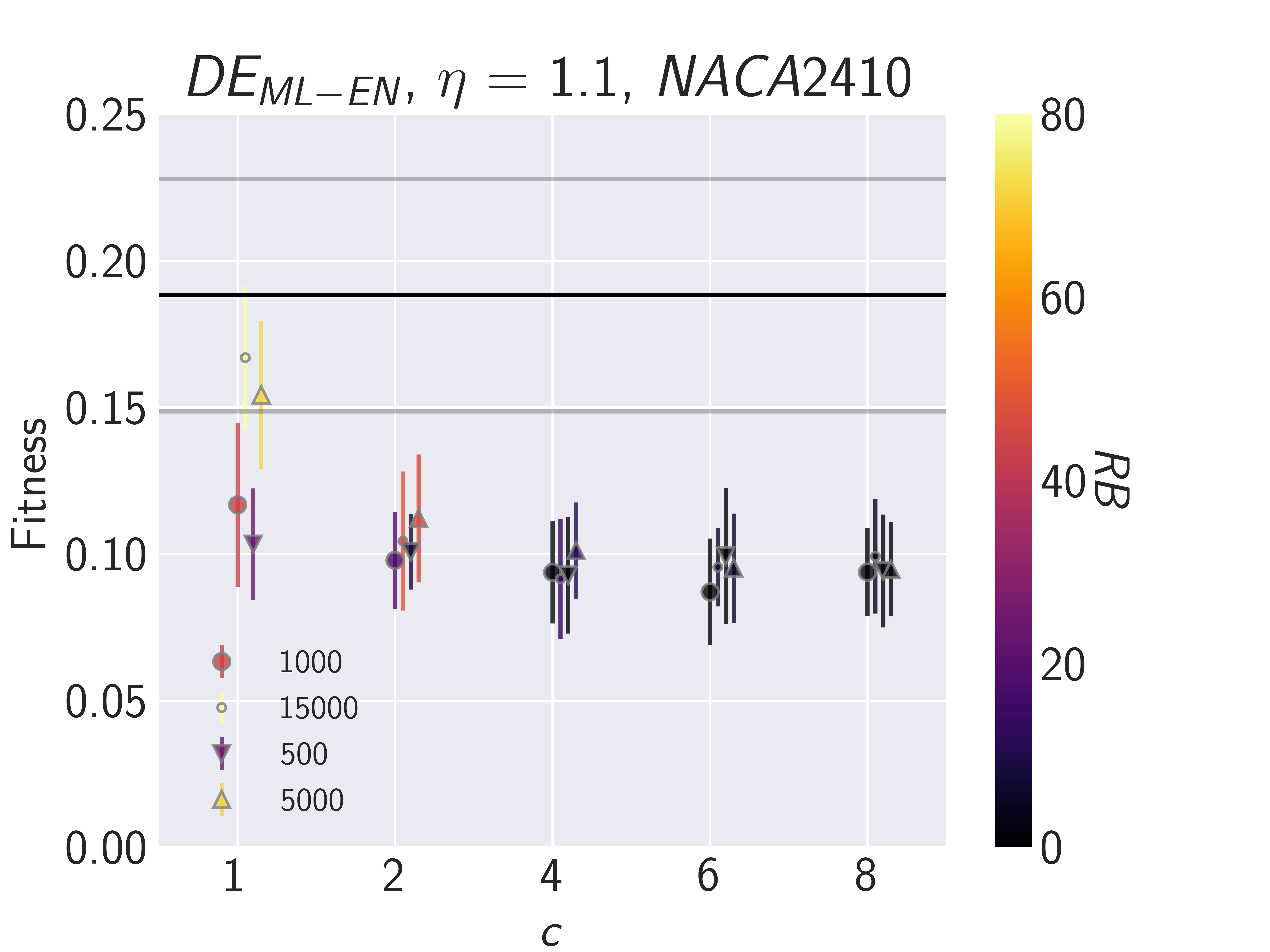}
		\caption{}
		\label{fig:naca2410_1_1_de}
	\end{subfigure}%
	\hfill
	\begin{subfigure}{0.45\textwidth}
		\centering
		\includegraphics[width=\linewidth]{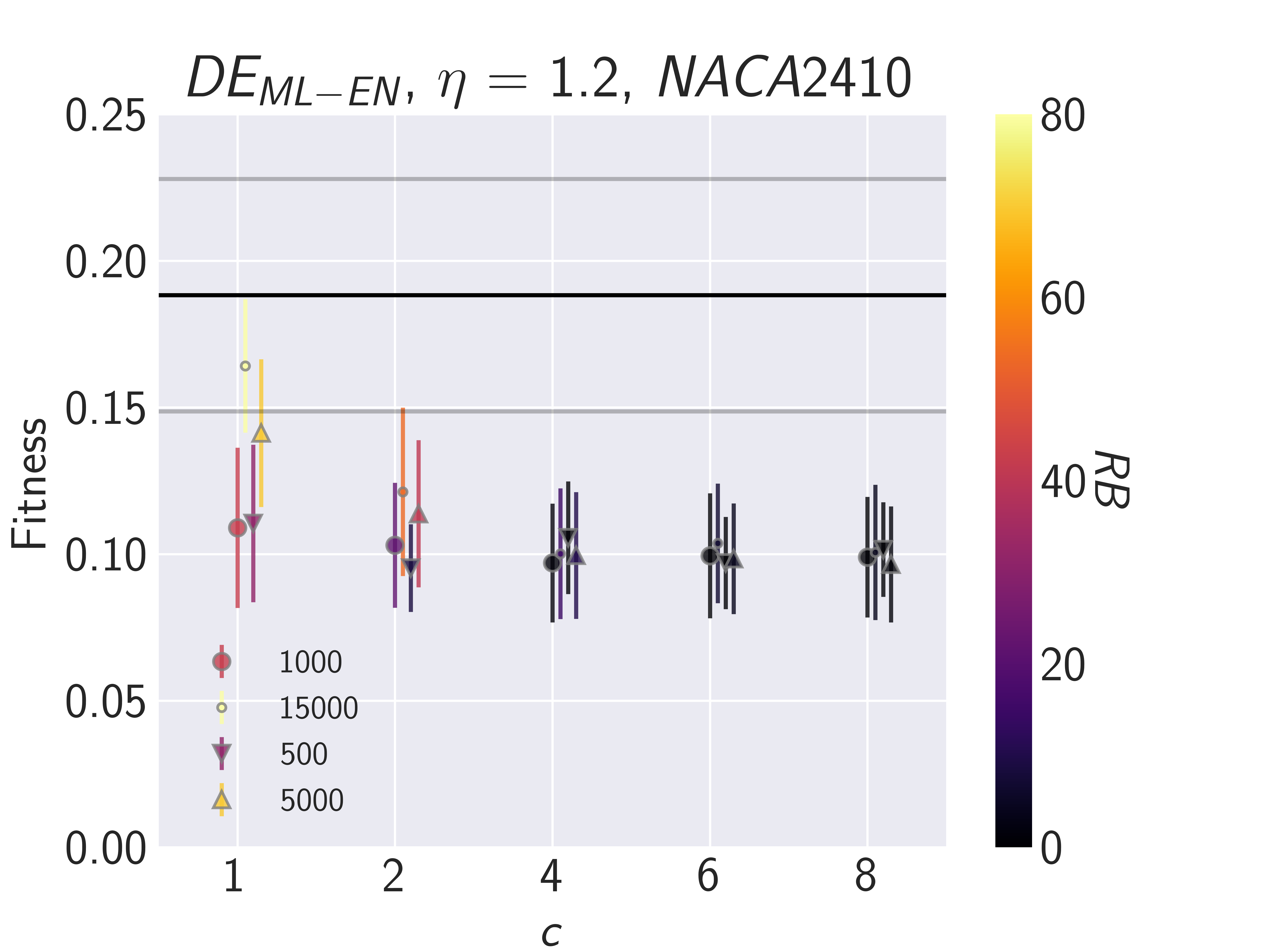}
		\caption{}
		\label{fig:naca2410_1_2_de}
	\end{subfigure}
	
	\caption{Results from the ML-enhanced framework for NACA2410 include the following boundary refinement configurations: (a) $PSO_{ML-EN}$, $\eta$ = 1.1 (b) $PSO_{ML-EN}$, $\eta$ = 1.2 (c) $DE_{ML-EN}$, $\eta$ = 1.1 (d) $DE_{ML-EN}$, $\eta$ = 1.2. The markers denote the different dataset sizes used to train the ML model, while the coloring of the markers represents the remaining simulation budget ($RB$) values. A higher $RB$ signifies greater savings in the computational budget, while low fitness values imply a better approximation of the target performance. The markers are slightly offset for each $c$ value to improve visibility.}
	\label{fig:naca2410_pso_de_hyperparameters_appendix}
\end{figure*}

\begin{figure*}[h!]
	\centering
	\begin{subfigure}{0.45\textwidth}
		\centering
		\includegraphics[width=\linewidth]{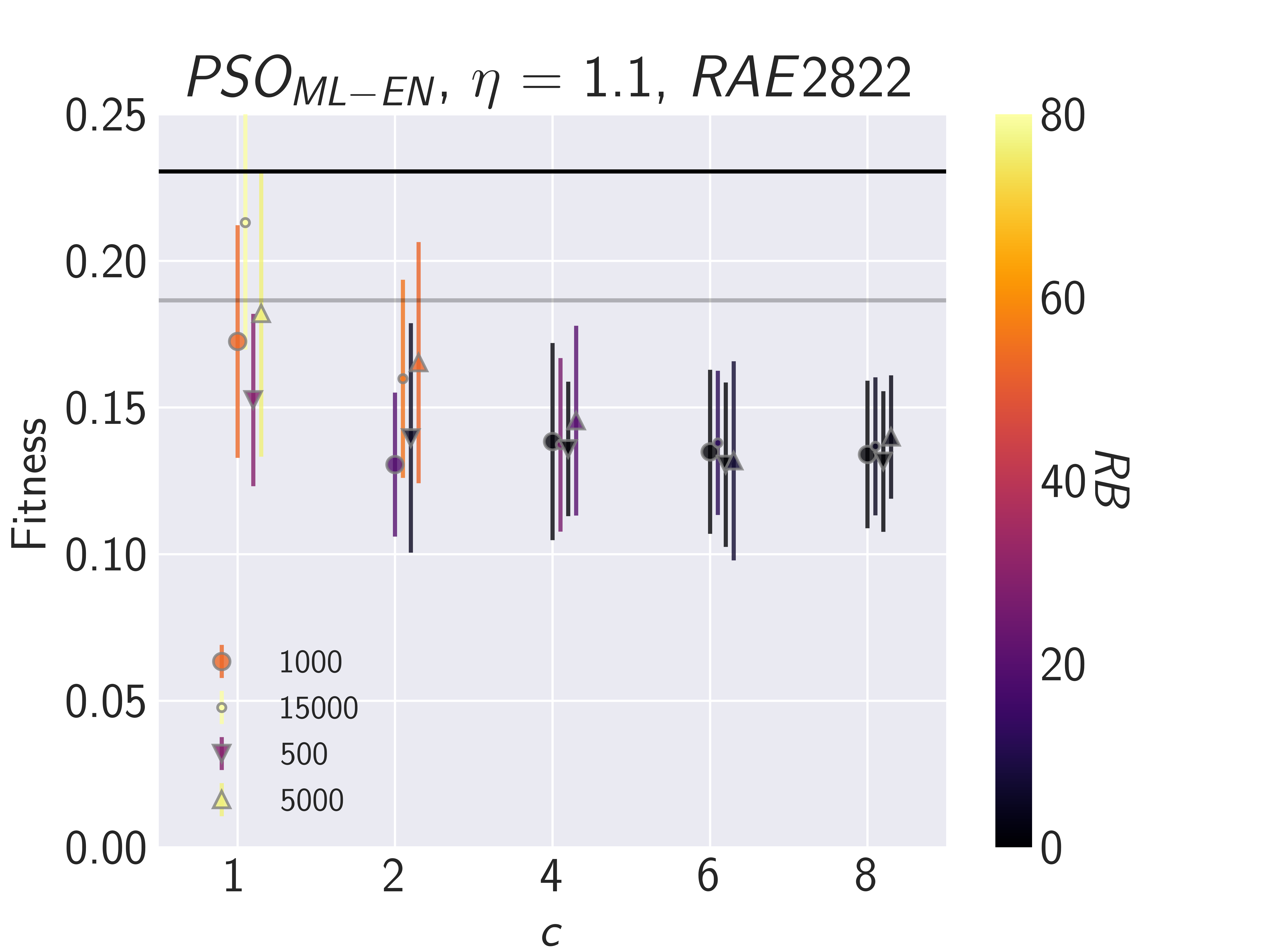}
		\caption{}
		\label{fig:rae2822_1_1_pso}
	\end{subfigure}%
	\hfill
	\begin{subfigure}{0.45\textwidth}
		\centering
		\includegraphics[width=\linewidth]{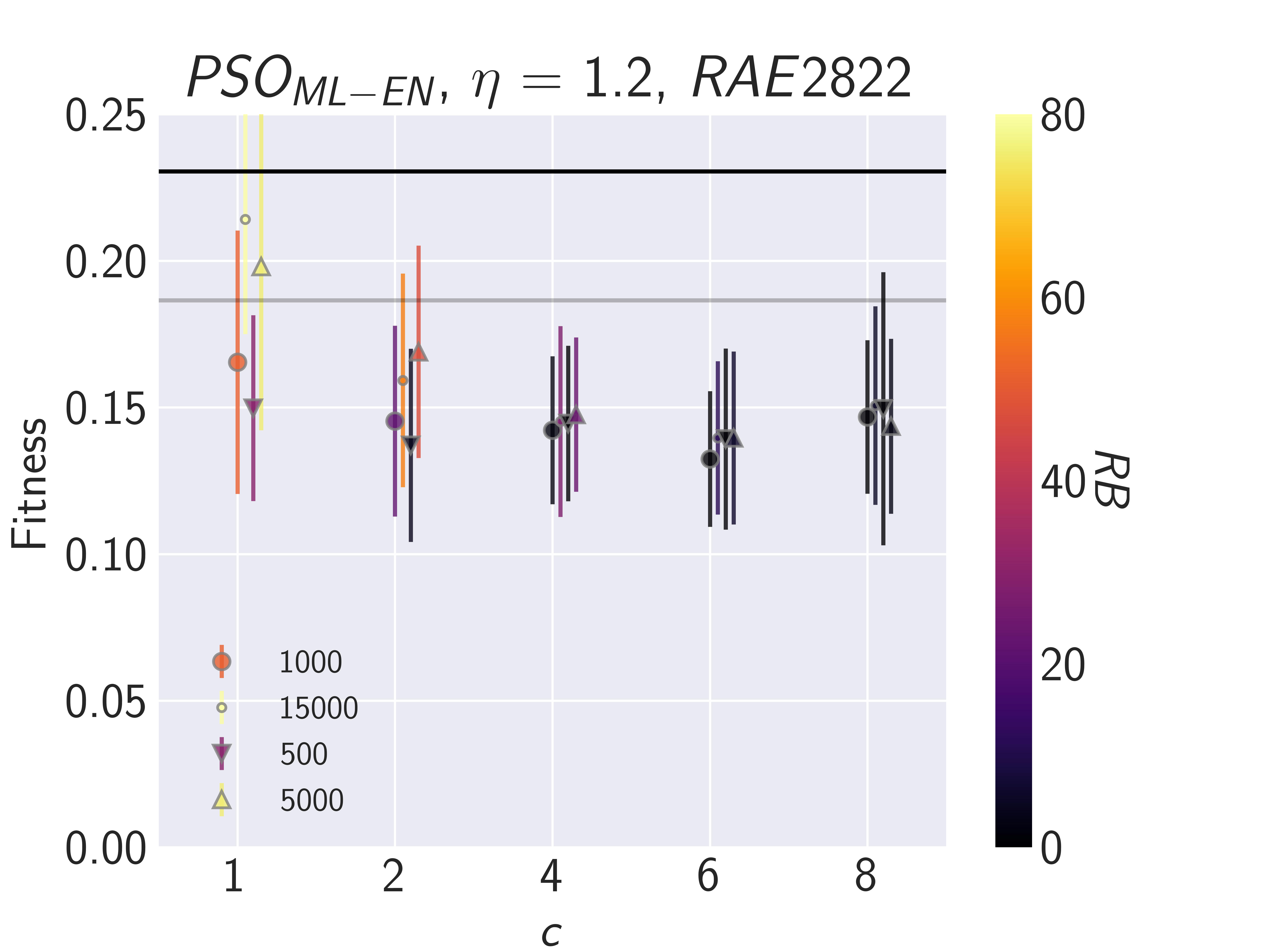}
		\caption{}
		\label{fig:rae2822_1_2_pso}
	\end{subfigure}
	
	\begin{subfigure}{0.45\textwidth}
		\centering
		\includegraphics[width=\linewidth]{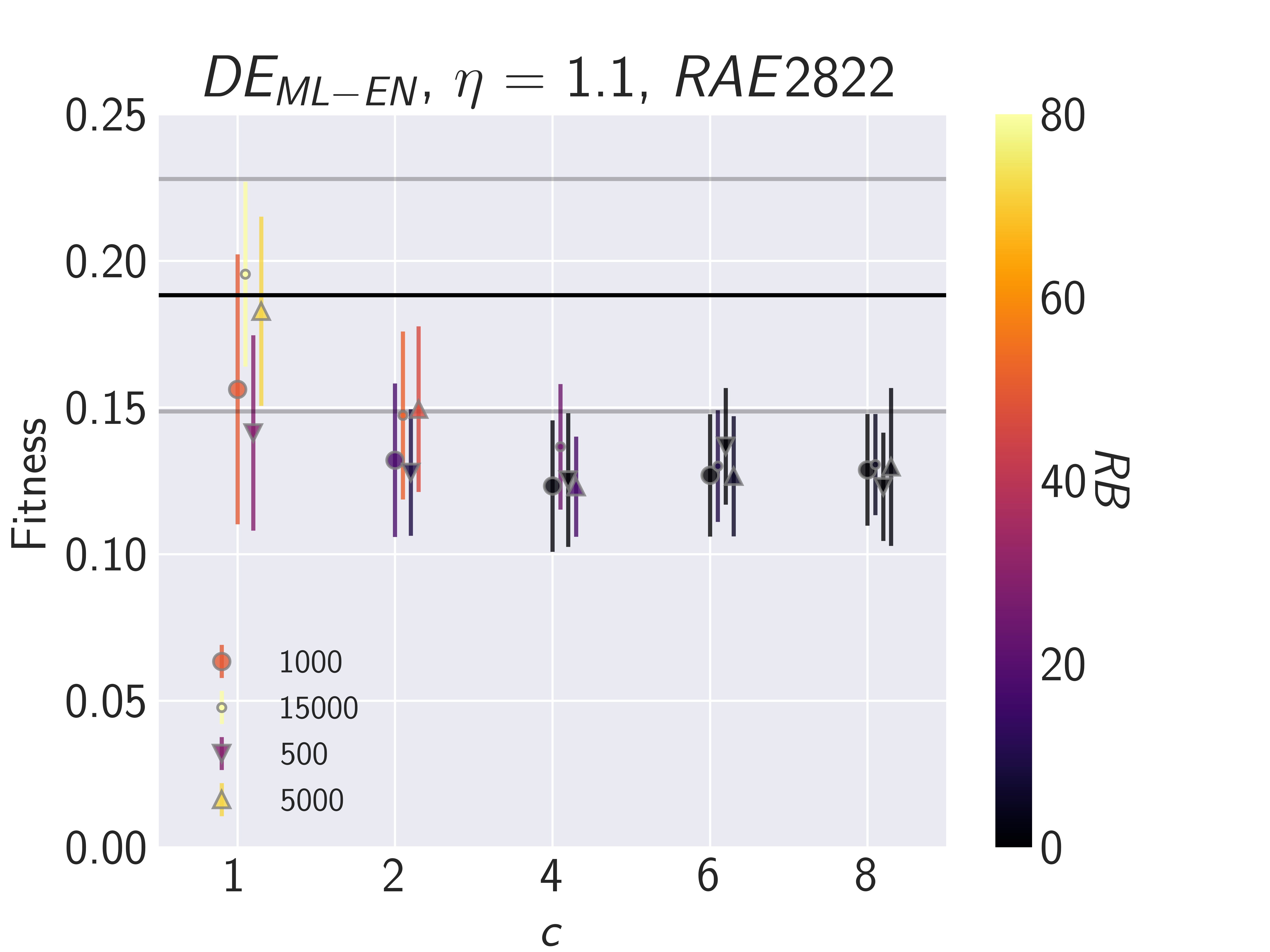}
		\caption{}
		\label{fig:rae2822_1_1_de}
	\end{subfigure}%
	\hfill
	\begin{subfigure}{0.45\textwidth}
		\centering
		\includegraphics[width=\linewidth]{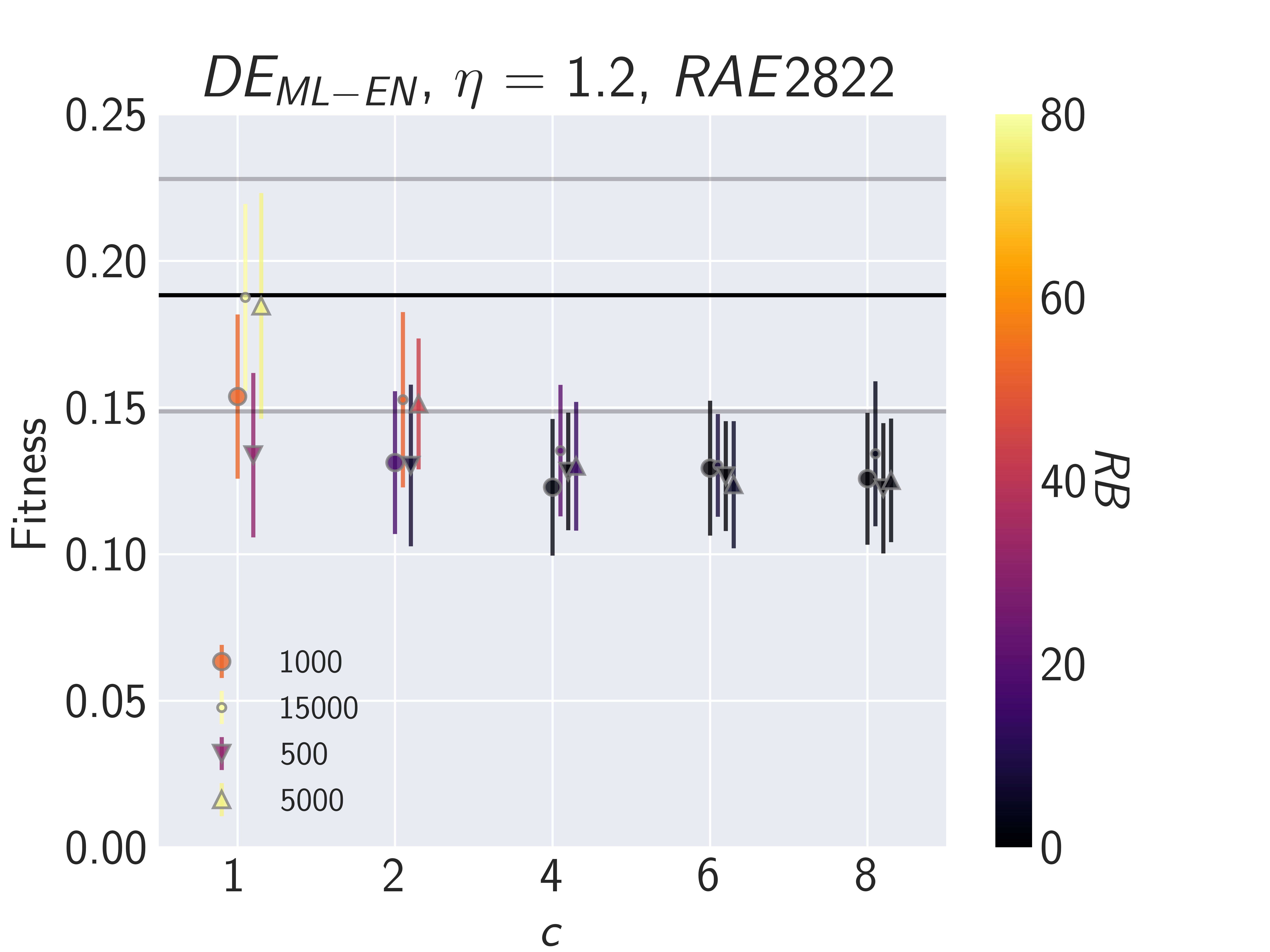}
		\caption{}
		\label{fig:rae2822_1_2_de}
	\end{subfigure}
	
	\caption{Results from the ML-enhanced framework for $RAE2822$ include the following boundary refinement configurations: (a) $PSO_{ML-EN}$, $\eta$ = 1.1 (b) $PSO_{ML-EN}$, $\eta$ = 1.2 (c) $DE_{ML-EN}$, $\eta$ = 1.1 (d) $DE_{ML-EN}$, $\eta$ = 1.2. The markers denote the different dataset sizes used to train the ML model, while the coloring of the markers represents the remaining simulation budget ($RB$) values. A higher $RB$ signifies greater savings in the computational budget, while low fitness values imply a better approximation of the target performance. The markers are slightly offset for each $c$ value to improve visibility.}
	\label{fig:rae2822_pso_de_hyperparameters_appendix}
\end{figure*}

The results of the ML-enhanced inverse design framework utilizing the XGB model and the boundary refinement technique ($\eta$ = 1.1 and $\eta$ = 1.2) applied to the AID problem for the $NACA2410$ and $RAE2822$ airfoils are shown in Fig. \ref{fig:naca2410_pso_de_hyperparameters_appendix} and Fig. \ref{fig:rae2822_pso_de_hyperparameters_appendix}.

\newpage
\clearpage






\bibliography{sn-bibliography}

\end{document}